# The Collisional Evolution of the Primordial Kuiper Belt, Its Destabilized Population, and the Trojan Asteroids

William F. Bottke[1,2], David Vokrouhlický[3], Raphael Marshall[4], David Nesvorný[2], Alessandro Morbidelli[4], Rogerio Deienno[2], Simone Marchi[2], Luke Dones[2], Harold F. Levison[2]

**Abstract**

The tumultuous early era of outer solar system evolution culminated when Neptune migrated across the primordial Kuiper belt (PKB) and triggered a dynamical instability among the giant planets. This event led to the ejection of ~99.9% of the PKB (here called the destabilized population), heavy bombardment of the giant planet satellites, and the capture of Jupiter's Trojans. While this scenario has been widely tested using dynamical models, there have been fewer investigations into how the PKB, its destabilized population, and the Trojans experienced collisional evolution. Here we examined this issue for all three populations with the code Boulder. Our constraints included the size-frequency distributions (SFDs) of the Trojan asteroids and craters on the giant planet satellites. Using this combination, we solved for the unknown disruption law affecting bodies in these populations. The weakest ones, from an impact energy per mass perspective, were diameter $D \sim 20$ m. Overall, collisional evolution produces a power-law-like shape for multikilometer Trojans and a wavy-shaped SFD in the PKB and destabilized populations. The latter can explain (i) the shapes of the ancient and younger crater SFDs observed on the giant planet satellites, (ii) the shapes of the Jupiter family and long-period comet SFDs, which experienced different degrees of collision evolution, and (iii) the present-day impact frequency of superbolides on Jupiter and smaller projectiles on Saturn's rings. Our model results also indicate that many observed comets, most which are $D < 10$ km, are likely to be gravitational aggregates formed by large-scale collision events.

[1] Corresponding author william.bottke@swri.org
[2] Southwest Research Institute, 1050 Walnut St, Suite 300, Boulder, CO, USA
[3] Institute of Astronomy, Charles University, V Holešovičkách 2, CZ-18000, Prague 8, Czech Republic
[4] Observatoire de la Côte d'Azur, CNRS, Laboratoire Lagrange, Nice, France

# 1. Introduction

The early migration of the giant planets, and their influence on the objects residing in the primordial Kuiper belt (PKB) between ~24 and ~50 au, played a foundational role in sculpting our solar system (Fig. 1). Their co-dependent evolution has been extensively explored using dynamical models. They show that after a giant planet instability takes place, giant planet encounters, both with one another and with destabilized primordial Kuiper belt objects (KBOs), can potentially explain numerous small body reservoirs (e.g., the observed Kuiper belt, Oort cloud, scattered disk, irregular satellites of the giant planets, Jupiter and Neptune Trojans, Hilda asteroids, and D- and P-type asteroids captured in the main asteroid belt (e.g., Tsiganis et al. 2005; Morbidelli et al. 2005; Levison et al. 2008a; Nesvorný 2011; Nesvorný and Morbidelli 2012; Batygin et al. 2012; Nesvorný and Vokrouhlický 2016; 2019; Nesvorný et al. 2007; 2013; 2016; 2018; 2020; 2021; Kaib and Sheppard 2016; Clement et al. 2018; Vokrouhlický et al. 2016; 2019; Lawler et al. 2019; see Nesvorný 2018 for a review).

Moreover, the objects placed on planet-crossing orbits by the dispersal of the PKB, characterized here as the destabilized population (Fig. 1), are not only responsible for most of the larger craters formed on giant planet satellites over the last 4.5 Gyr (e.g., Zahnle et al. 2003; Wong et al. 2019; 2021), but also include the long-lived survivors that make up our ecliptic and nearly-isotropic comet populations (e.g., Nesvorný et al. 2017; Vokrouhlický et al. 2019). They may even explain many of the craters formed on prominent worlds like Pluto-Charon and Arrokoth, all which reside in various parts of the Kuiper belt (Singer et al. 2019; Spencer et al. 2020; Morbidelli et al. 2021; Robbins and Singer 2021).

PLACE FIGURE 1 HERE

Overall, dynamical models of the giant planet instability and its aftermath tell a compelling story of outer solar system evolution. Nevertheless, our understanding of what happened at ancient times is still incomplete because we have yet to adequately test the predictions of these models against bombardment and collisional evolution constraints.

The answers to a number of intriguing questions hang in the balance. For example:

- Various planetesimal formation models have been proposed to explain the sizes of KBOs in the Kuiper belt (and PKB) (e.g., see review by Johansen and Lambrechts 2017), but our ability to test these results depends on the unknown nature of the PKB's initial size-frequency distribution (SFD). For example, did the SFDs of the PKB follow a cumulative power law slope of $q \sim -2$ for bodies that are diameter $D <$ 100 km, the same slope observed today among the Jupiter Trojans (Yoshida and Nakamura 2005; Wong and Brown 2015; Yoshida and Terai 2017), or did the SFDs start with more shallow slopes (e.g., Napier et al. 2023)?

- Destabilized PKB objects that enter the giant planet zone to strike the giant planet satellites should presumably have a SFD with a shape similar to the one found among the Jupiter Trojans, since both come from the same source population. Instead, as we will show below, they appear to have different shapes. Understanding how this could have happened can help us probe the physical properties of comets, Trojans, and KBOs.

- Analysis of images from the New Horizons mission indicate there is an extreme paucity of $D_{\text{crat}} < 10$ km craters on Charon (Singer et al. 2019; Robbins and Singer 2021). This crater deficit is not unique; a shallow crater SFD can also be found on Europa, Enceladus, Miranda, Hyperion, Phoebe, and other giant planet satellites whose terrains are not dominated by secondary and sesquinary craters (e.g., Zahnle et al. 2003; Porco et al. 2005; Bierhaus et al. 2012; Thomas et al. 2007; 2013; Kirchoff et al. 2022). The explanation for this deficit may be found in how KBOs respond to disruptive collisions (Morbidelli et al. 2021).

- Ground-based observations of flashes on Jupiter indicate it is being hit by a substantial population of small bolides (Hueso et al. 2013; 2018). Approximately 10–65 bodies strike Jupiter per year that are $D > 5$–20 m or larger. Curiously, this rate is ~1 to ~1.5 orders of magnitudes larger than expectations based on an extrapolation of ~1 to 20 km craters (i.e., approximately ~0.05 to 1 km projectiles) found on Europa (Zahnle et al. 2003; Schenk et al. 2004). This mismatch suggests we are missing something crucial in our understanding of the small body comet flux hitting outer solar system worlds.

- Multikilometer-sized comets, such as 67P/Churyumov–Gerasimenko, often have irregular shapes, high porosities, and structures that suggest limited tensile strength (e.g., El-Maarry et al. 2017). From these and other properties, some argue that comets this size were formed directly by planetesimal formation processes (e.g., Davidsson et al. 2016), while others argue they are mainly fragments produced by large PKB objects bashing into one another (e.g., Morbidelli and Rickman 2015; Jutzi et al. 2016). Our interpretation of comet shapes, compositions, and physical properties, as well as the nature of returned samples, depends on how well we understand the origin of these bodies.

Our ultimate goal is to address these questions, and many others, by modeling the collisional evolution of the PKB and all of its daughter populations. To date, most groups have thus far concentrated on collisional evolution of the Kuiper belt alone (e.g., Davis and Farinella 1997; Stern and Colwell 1997; Kenyon and Bromley 2004; 2020; Pan and Sari 2005; Krivov et al. 2005; Charnoz and Morbidelli 2007; Levison et al. 2009; Schlichting et al. 2013; Morbidelli and Rickman 2015; Brunini and Zanardi 2016; Jutzi et al. 2017; Nesvorný et al. 2018; Kenyon and Bromley 2020; Benavidez et al. 2022). Calculating compelling solutions, however, is challenging for several reasons. For example:

1. **Dynamical evolution model of the PKB**. A prerequisite for any good collisional model of the PKB is the inclusion of the dynamical history of both the giant planets and the PKB after the dissipation of the solar nebula (e.g., Nesvorný et al. 2018). Such runs are needed in order to calculate collision probabilities, impact velocities, and depletion factors that control the collision rates between small bodies in the PKB and daughter populations. The lack of such information can be an obstacle to modelers, who are forced to guess several critical parameters (e.g., timing and nature of PKB's excitation; the evolution of its daughter populations).

2. **Limited observational constraints on the Kuiper Belt size distribution.**, Ground-based and space-based observations have difficulties detecting KBOs smaller than a few tens of km in diameter (e.g., Bernstein et al. 2004; Fuentes et al. 2009; Fraser et al. 2014; Parker et al. 2021; Kavelaars et al. 2021). Accordingly, direct constraints of the Kuiper belt SFD at small sizes are limited. One potential way to overcome this problem is to interpret the crater SFDs found on Pluto, Charon, and Arrokoth (Singer et al. 2019; Spencer et al. 2020; Morbidelli et al. 2021). These data are new, however, and only a few groups have tried to include them as modeling constraints to date (e.g., Kenyon and Bromley 2020; Benavidez et al. 2022). The lack of knowledge of small KBOs is a major impediment for modelers; the population of small objects determines the frequency of disruptive impacts among larger bodies, yet they themselves are also susceptible to disruption (e.g., O'Brien and Greenberg 2003; Bottke et al. 2015).

3. **Unknown nature of the initial SFD of the PKB.** A critical component of collisional modeling work is an accurate starting SFD for the PKB (e.g., Johansen et al. 2015; Nesvorný and Vokrouhlický 2016; Nesvorný et al. 2018). Dynamical evolution scenarios of the giant planet instability using such a SFD must be able to recreate the observed Kuiper belt from Neptune's outward migration as well as generate the right number of captured objects in the outer main belt, Hildas, Trojans, irregular satellite, scattered disk, and Oort cloud populations. They must also be consistent with the observed Kuiper belt SFD for $D < 100$ km bodies, the shapes of the SFDs found in various daughter populations, and the crater SFDs found on outer solar system worlds.

4. **Unknown disruption scaling law.** The scaling law controlling KBO disruption events is not well determined. Many disruption laws for KBO-like ice-rock targets have been proposed in the literature (e.g., Benz and Asphaug 1999; Leinhardt and Stewart 2009; Jutzi et al. 2010; Kenyon and Bromley 2020; Benavidez et al. 2022) but the only direct constraints that exist are those derived from high velocity shot experiments striking icy targets in the laboratory (e.g., Leinhardt and Stewart 2009). A potential problem with determining a disruption scaling law for comets and KBOs is that they likely have highly porous internal structures, a property that is challenging to simulate with existing numerical hydrocode simulations (e.g., Jutzi et al. 2015).

As mentioned above, there have been two attempts to model the collisional evolution of the PKB after the New Horizons flybys of Pluto-Charon and Arrokoth: Kenyon and Bromley (2020) and Benavidez et al. (2022). While both models have intriguing attributes, their published model SFDs can only match portions of the predicted shape of the Kuiper belt SFD as determined from various observational constraints as well as the crater SFDs found on Charon and Arrokoth (Morbidelli et al. 2021). This suggests their models may be missing something; perhaps one or more of the items from #1-#4 above.

Even if these groups had identified a good match to the predictions of Morbidelli et al. (2021), however, it would not necessarily indicate uniqueness for their parameter choices. As discussed by Bottke et al. (2005a,b), who attempted to model the collisional evolution of the main belt SFD, a single SFD used as a constraint for a collisional model will lead to an envelope of model parameter possibilities, each capable of producing a comparable fit. In such a circumstance, additional constraints are needed to rule out model possibilities. For example, Bottke et al. (2005a,b) used the number and frequency of family-forming events in the main belt to overcome degeneracies in the problem.

This takes us to the issue of how to best take on our ultimate goal of modeling the collisional evolution of the PKB and its primary daughter populations, which include the Kuiper belt, the destabilized population, and the Jupiter Trojans. It is a complex problem with many unknowns, even more than represented by #1-#4 above. After some preliminary tests, we decided we had to find a way to make the problem more manageable.

Our solution was to set aside the Kuiper belt portion of the problem (for now) and instead focus on the PKB, destabilized population, and Jupiter Trojans. For the last two, we have "easy to use" constraints that cover a wide range of object sizes: crater SFDs from impacts of the destabilized population onto the giant planet satellites and the Jupiter Trojan SFD. Together, they yield enough information for us to solve for a plausible KBO disruption law (#4).

For the former, we determined the shape of destabilized population's SFD at early times using crater constraints from two satellites with ancient surfaces, Iapetus and Phoebe. Together, they provide constraints for projectile diameters between a few meters and nearly 100 km. For the latter, we synthesize a reasonable estimate of the present-day Jupiter Trojan SFD (L4 and L5 combined) using the best available ground- and space-based observations. These objects range from a few km to nearly 200 km. Our derivation of these constraints will be discussed in Sec. 3.

It is important to emphasize that this does not mean our model ignores key constraints from the current Kuiper belt. Instead, our work builds on existing collisional and dynamical models, especially those where the results have already been tested against many types of Kuiper belt constraints. For example:

- Our model of Neptune's migration through the PKB, and its interaction with a large population of Pluto-sized objects, can explain the sizes of the resonant and non-resonant populations in the Kuiper belt (Nesvorný and Vokrouhlický 2016). In turn, this work sets the initial number of $D > 100$ and $D > 2000$ km objects needed in the PKB to explain the current Kuiper belt populations.
- Our PKB population also contains enough $D > 100$ km objects to explain the number captured as Jupiter Trojans, Hildas, and outer main belt asteroids (Nesvorný et al. 2013; 2018; Vokrouhlický et al. 2016).
- Previous work has used the populations of well-separated binaries in the Jupiter Trojans and dynamically hot portion of the Kuiper belt to set the maximum amount of collisional evolution that could take place in the PKB and Kuiper belt populations (Nesvorný et al. 2018; Nesvorný and Vokrouhlický 2019). They tell us that Neptune's migration through the PKB must start <

100 Myr after the dissipation of the solar nebula. As we will show, our model results are well below this value.

The structure of our paper is as follows. Our work will use the results of dynamical simulations of the giant planet instability, PKB evolution, and Jupiter Trojan capture capable of reproducing the available observational constraints. A discussion of these models and their dynamical predictions will be given in Sec. 2, with their use in calculating collision probabilities and impact velocities for the PKB, destabilized population, and Jupiter Trojans discussed in Sec. 4.3

Our estimate of the shape of the PKB's initial SFD will be informed by a collection of sources, namely new observations, dynamical modeling work, collisional modeling work of the main asteroid belt, planetesimal formation models, and the size distributions of PKB daughter populations. We will also discuss potential PKB SFDs used in previous work. This will be discussed in Sec. 4.2

To deal with the unknown disruption scale law for KBOs, our collisional evolution model will test ~$10^4$ possibilities against constraints. The disruption laws that produce the best matches will be compared to those in the literature. A discussion of our methodology can be found in Sec. 4.4.

All of these components will be used within the Boulder collisional evolution code (Morbidelli et al. 2009). Our goal is to track the collisional evolution of the PKB, destabilized population, and Jupiter Trojans, with the latter two SFDs tested against constraints. Our model results will be discussed in Sec. 5.

From there, we will further verify our model results by testing them against constraints from a variety of sources (e.g., crater SFDs on the giant planet satellites, superbolide impacts on Jupiter, small impacts on Saturn's rings, the debiased SFDs of Jupiter family and long period comets). We will also discuss how our results relate to the predictions of Morbidelli et al. (2021). Comparisons between our best fit model and these data will be given in Sec. 6.

Finally, in Sec. 7, we will briefly discuss some of the more provocative implications of our results for the nature of comets, interplanetary dust particles, and interstellar objects. A longer discussion of these issues is reserved for Appendix A.

## 2. The giant planet instability and Neptune's migration across the primordial Kuiper belt

In order to explore the collisional evolution of the PKB, we first need to characterize how it experienced dynamical evolution. Here we use a model of outer solar system evolution that has been developed over the last decade by co-author D. Nesvorný and several colleagues (Nesvorný and Morbidelli 2012; Nesvorný et al. 2013; Nesvorný 2015a; 2015b; Nesvorný and Vokrouhlický 2016; Nesvorný et al. 2017; Nesvorný et al. 2019a; see also Nesvorný 2018 for a review). It describes how the giant planets dynamically evolved to new orbits as a consequence of a post-nebula giant planet instability (e.g., Tsiganis et al. 2005). It also allows us to quantify how the PKB was transformed and dynamically depleted by the outward migration of Neptune through the PKB.

To make things easier to understand, we defined the time periods around the giant planet instability as Stages 1, 2, and 3, respectively (Fig. 1).

**Stage 1**. Stage 1 is defined as the time interval $\Delta t_0$ between the end of the solar nebula and when Neptune enters the PKB. In Stage 1, the PKB has become modestly excited by numerous Pluto-sized objects. Here we assume the initial mass of the PKB is several tens of Earth masses, as suggested by models of Neptune's migration and the giant planet instability (e.g., see Nesvorný 2018 and references therein). This large size, combined with modest dynamical excitation from embedded Pluto-sized or larger objects, implies that considerable collisional evolution should take place within Stage 1.

**Stage 2**. Stage 2 is defined as the time interval $\Delta t_1$ between Neptune's migration across the PKB and the giant planet instability, which starts when giant planet-giant planet encounters begin. Dynamical models show that such events are typically achieved when Neptune reaches ~28 au (e.g., Nesvorný 2018 and references therein). The precise timing of $\Delta t_1$ depends on the spatial density of objects in the PKB. In the two dynamical models we will use in this project, $\Delta t_1$ lasts 10.5 or 32.5 Myr. The PKB becomes highly excited in this stage, with nearly all KBOs pushed onto giant planet-crossing orbits. This action creates the destabilized population, which continues to undergo considerable collisional evolution.

**Stage 3**. Stage 3 are all the actions that take place after the giant planet instability. We define it as the time interval between $\Delta t_0 + \Delta t_1$ and the present day ~4.5 Gyr later. Stage 3 includes the capture of the Jupiter Trojans by giant planet interactions, the origin of the scattered disk, and much of the bombardment of the giant planet satellites.

We now discuss the dynamical evolution of each stage in more detail.

## 2.1 Stage 1. Before the Giant Planet Instability

Stage 1 starts shortly after the giant planets form on nearly circular and coplanar orbits within the solar nebula. The likely time of solar nebular dispersion is 2-10 Myr after CAI formation (Kruijer et al. 2017; Weiss and Bottke 2021). Gas driven migration likely caused the giant planets to evolve into a resonant chain (i.e., the planets were trapped in mutual mean motion resonances with one another), but the stability of chains may have been limited after the gas disk went away (see review by Nesvorný 2018).

Many different resonant chain models were considered in Nesvorný and Morbidelli (2012; hereafter NM12). Over ~$10^4$ numerical simulations, they followed what happened to outer planet systems that contained four, five, and six giant planets (i.e., Jupiter, Saturn, and 2-4 Neptune-sized bodies), with the bodies usually stretched between ~5 and ~17 au. Model success in NM12 was defined by four criteria: there had to be four planets at the end, the planets had to have plausible orbits, Jupiter's eccentricity had to be close to its current value, and the planets had to migrate fast enough satisfy terrestrial planet and asteroid belt constraints (see also Nesvorný 2018).

The putative planetesimal population in the giant planet zone that existed between ~5 and ~17 au is poorly constrained at present. Here we assume it did not play a meaningful role in any of the collisional or dynamical set pieces relevant to our modeling work.

The starting location for the main portion of the PKB is located just beyond the giant planet zone between ~24 and 30 au, with the latter value set by the current location of Neptune (e.g., Tsiganis et al. 2005). Most of the KBO-like objects in various comet and small body reservoirs captured in Stage 3 come from this part of the PKB population (e.g., Dones et al. 2015). In NM12, the net mass of this part of the PKB was set to ~15-20 Earth masses, though this value does not account for losses by collisional evolution within the PKB. As we will show below, by including collisions, we can achieve reasonable results by starting with ~30 Earth masses of material.

Numerical simulations show a less massive portion of the PKB may extend beyond 30 au and could continue to ~47 au, the location of the outer edge of the cold classical KB (Nesvorný et al. 2019a). To prevent Neptune from migrating beyond ~30 au by planetesimal-driven migration (Gomes et al. 2004), however, the spatial density of KBOs in the PKB must decrease as a function of distance from the Sun (Fig. 1). Modeling work shows that Neptune stops its migration when the spatial density of bodies drops below 1-1.5 Earth masses per au (Nesvorný 2018).

Here we assume the extended disk contained about 1/400th of the population that existed between 24-30 au (Nesvorný et al. 2020). So, if the initial PKB population was ~30 Earth masses between 24-30 au, the extended disk would have 0.075 Earth masses. This kind of distribution implies that the cold classical Kuiper belt, located between the 3:2 and 2:1 mean motion resonances with Neptune (i.e., between the semimajor axis range of 42 to 47 au), could form in situ with a more limited population (e.g., Fraser et al. 2014). It also provides a rationale for why the cold classicals have a large fraction of well-separated binaries; they never encountered Neptune during its migration (Nesvorný et al. 2019b; 2020).

The PKB in Stage 1 becomes dynamically excited in two ways. First, Neptune's perturbations steadily erode its inner edge, allowing some KBOs to escape into the giant planet zone. Most of these bodies are passed down to Jupiter, where they are thrown out of the solar system. The net change in angular momentum causes the planets to undergo slow but steady migration (Tsiganis et al. 2005). Second, the majority of KBOs become excited by gravitational perturbations from the largest objects in the population, namely ~1000-4000 Pluto-sized objects (Nesvorný and Vokrouhlický 2016). This self-stirring mechanism increases impact speeds between KBOs and produces collisional evolution. In fact, if $\Delta t_0$ lasts long enough, most KBO collisions could occur in this interval.

A limit on the number of collisions taking place in the PKB is set by the observed population of well-separated KBO binaries with components that are $D > 100$ km (Petit and Mousis 2004; Nesvorný et al. 2018; Nesvorný and Vokrouhlický 2019). These binaries were presumably created by the planetesimal formation process called the streaming instability (Youdin and Goodman 2005; Nesvorný et al. 2019b; 2020; see review of the mechanism in Johansen et al. 2015;). Collisions can disrupt KBO binaries, so too many collisions in Stage 1 will lead to too few binaries compared to observations. Modeling work shows that $\Delta t_0$ must be less than 100 Myr to explain the capture of the Patroclus−Menoetius binary in the Jupiter Trojans in the aftermath of the giant planet instability (Nesvorný et al. 2018). As we will show, our best fits for $\Delta t_0$ are considerably lower than this threshold.

## 2.2 Stage 2. Neptune's Migration Across the PKB

Stage 2 begins when Neptune enters the PKB and begins to migrate outward by ejecting KBOs into the giant planet zone. We refer to those objects that reach planet-crossing orbits as the destabilized population, and they play a major role in deciphering the nature of collisional evolution in the PKB.

As Neptune migrates, a roughly 50-50 mix of objects are thrown inward and outward. The inward ejected objects are eventually handed down to Jupiter, which commonly throws them out of the solar system, while the outward ejected bodies come back for more Neptune encounters and the opportunity to be thrown inward. This continues until Neptune stops migrating at 30 au, with objects in the newly formed scattered disk continuing to have Neptune encounters over the following 4.5 Gyr. Many of these bodies will be lost to inward scattering events over that time (see bottom of Fig. 1). The long-lived survivors of the destabilized population now make up the scattered disk of Neptune, or scattered disk for short (Duncan and Levison 1997).

During migration, gravitational interactions between KBOs and Neptune produce dynamical friction, damping Neptune's eccentricity enough for it to reach its current value of ~0.01 (Nesvorný 2020). Concurrently, many KBOs and Pluto itself become trapped in mean motion resonances with Neptune as it moves outward. Neptune encounters with Pluto-sized objects in the PKB, however, produce gravitational jolts that can shake some KBOs out of resonance. The net effect of erratic/grainy outward migration for Neptune and dynamical friction plausibly explains the fraction of resonant and non-resonant objects observed in the Kuiper belt (Nesvorný and Vokrouhlický 2016).

The depletion of the PKB over time generally follows exponential decay, with an e-folding time comparable to the timescale of Neptune's crossing the disk. So, if Neptune takes 10 Myr to get across the disk, the e-folding time of the PKB's depletion is 10 Myr.

As mentioned above, numerical simulations of the survival of KBO binaries indicate that Stages 1 and 2 lasted ≲100 Myr (Nesvorný et al. 2018), but the exact time of the giant planet instability (i.e., $\Delta t_0 + \Delta t_1$) is unknown. Shorter transit timescales for Neptune (i.e., shorter $\Delta t_1$) mean less collisional evolution takes place among objects that will eventually become part of the destabilized population. The results in Nesvorný et al. (2018) indicate that shorter Stage 1 and 2 intervals also yield much better odds that the Patroclus−Menoetius binary would be captured intact within the Trojans. That favors faster passages and lower $\Delta t_1$ times.

In our work below, we will test a range of $\Delta t_0 + \Delta t_1$ values to determine which ones lead to successful outcomes.

## 2.3 Stage 3. The Giant Planet Instability and its Aftermath

In our dynamical models of the giant planet instability, the giant planets break out of their resonant chain as Neptune approaches 28 au or so. The reason is that the planets have been forced to interact with ejected mass from the PKB, which in turn has caused them to migrate (i.e., Jupiter migrates inward, the other planets migrate outward). The breakout allows the giant planets to have encounters with one another while surrounded by a plethora of objects from the destabilized population.

Numerical results from NM12 indicate that five planet systems most often produce four giant planets with orbits that resemble the present ones. The missing ice giant is ejected after multiple encounters with Jupiter (Nesvorný 2011; Batygin et al. 2012), but not before getting it to migrate inward rapidly over a timescale of 1 Myr. The dynamical runs used in this paper come from the most successful of the $\sim 10^4$ runs, where success is defined by the ability to reproduce additional small body constraints when tested in detail (e.g., see also Nesvorný 2018 and references therein).

As the PKB was eviscerated by Neptune's outward migration, KBOs were driven onto unstable orbits with the giant planets. Most of this destabilized population was ejected from the Solar System after being passed down to a close encounter with Jupiter, but some hit the planets, and others were captured in small body reservoirs such as the main belt (Levison et al. 2009, Vokrouhlický et al. 2016), Jupiter Trojans (Morbidelli et al. 2005, Nesvorný et al. 2013), irregular satellites (Nesvorný et al. 2007, 2014), Kuiper belt (Malhotra 1993, Gomes 2003, Hahn and Malhotra 2005; Levison et al. 2008a; Dawson and Murray-Clay 2012; Nesvorný 2015a,b; Nesvorný and Vokrouhlický 2016), scattered disk (Brasser and Morbidelli 2013, Kaib and Sheppard 2016, Nesvorný et al. 2016), and Oort cloud (Brasser et al. 2006, 2007, 2008; Levison et al. 2010; Kaib et al. 2011; Brasser and Morbidelli 2013; Vokrouhlický et al. 2019). A key focus of this paper will be to examine happens collisionally to the unstable ejected bodies and to those captured as Jupiter's Trojans.

Overall, the PKB dynamically lost a factor of $\sim 1000$ in population as Neptune migrated across it. These objects were driven onto Neptune-crossing orbits, where they become the destabilized population. Accordingly, by assuming a starting population with a few thousand Pluto-sized bodies, it is possible to end up with two Pluto-sized bodies (i.e., Pluto, Eris) in the present-day Kuiper belt and scattered disk.

The two numerical simulations of the giant planet instability used in this paper (Nesvorný et al. 2013; 2017) show that ~1% of the destabilized population will eventually hit Jupiter, while about a third as many will end their existence by striking Saturn, Uranus, and Neptune. A much smaller net fraction will hit the icy satellites; the work of Zahnle et al. (2003) suggests $\sim 10^{-4}$ hit Io, a large satellite orbiting deep within Jupiter's gravitational well, $\sim 10^{-6}$ for Iapetus, a satellite about half the Moon's size that is far from Saturn, and $\sim 10^{-8}$ for Phoebe, a ~200 km diameter irregular satellite of Saturn. The cratering history of the latter two objects will play a key role in defining our constraints for the destabilized population's SFD.

## 3. Model Constraints

### 3.1 The Size Distribution of the Destabilized Population Derived from Craters

#### 3.1.1. Craters on Pluto, Charon, and Arrokoth

One of the key clues telling us about the unusual nature of the PKB SFD comes from the crater histories of Pluto/Charon (resonant objects in the Kuiper belt) and Arrokoth (located in the cold classical Kuiper belt). All of these objects were once residents of the PKB. Crater counts of Charon show that its cumulative SFD follows a modestly steep power law slope of $q = -2.8 \pm 0.6$ for $D_{crat} > 10-15$ km, while $\sim 1 < D_{crat} < 10-15$ km craters follow a slope of $q = -0.7 \pm 0.2$ (Singer et al. 2019; Robbins and Singer 2021). The latter value is surprising, in that it predicts a severe paucity of craters, and the projectiles that made them, over a large size range.

When Arrokoth craters are folded in, the mystery extends to even smaller sizes. Arrokoth has five 0.55–1.15 km craters and one ~7 km crater (Spencer et al. 2020). Morbidelli et al. (2021) examined the likelihood that different crater production populations could produce this unusual distribution. They found that their most probable production population had $q \sim -1.25$, and that shallow slopes of $q = -0.7 \pm 0.2$ were statistically unlikely. By combining crater constraints from both Charon and Arrokoth together, and solving for the likely projectile SFD, they showed that the slope of small impactors on both worlds was probably $q = -1.1 \pm 0.1$, and that this slope extended from $0.03 < D < 1$ km. For $D > 1$ km projectiles, the steeper slope seen on Pluto-Charon for $D_{crat} > 10-15$ km comes into play.

One more intriguing issue raised by Morbidelli et al. (2021) is the abundance of Kuiper belt dust estimated by the New Horizons dust counter, presumably produced by collisional processes (Poppe et al. 2019). Calculations show that the mass of particles between 0.5 and 500 μm derived from the Kuiper belt is $3.5 \times 10^{18}$ kg. In order to achieve such a high value, the power slope of the Kuiper belt SFD must become relatively steep at some size below $D < 0.03$ km. Morbidelli et al. (2021) argued that a change slope of $q \sim -3$ starting near 0.02 km would explain the dust constraint (see their Fig. 5). We concur with this prediction, for reasons we will discuss below.

These works provide a fascinating snapshot into the history of the PKB. They tell us that Pluto-Charon-Arrokoth were rarely hit by projectiles between $D \sim 30$ m and 1 km. For the moment, we will ignore the modest slope differences above between $q = -0.7 \pm 0.2$ and $q = -1.1 \pm 0.1$. It has been postulated that these shallow slopes might mean that few primordial bodies ever formed in this size range, or that PKB impacts produce very few fragments in this size range (e.g., Singer et al. 2019). In this work, however, we favor a different interpretation in agreement with Morbidelli et al. (2021).

The impactor SFD identified above is reminiscent of the main belt SFD, which has a fairly steep slope of $q = -2.7$ for $D > 2-3$ km bodies and a shallow slope of $q = -1.2$ for $\sim 0.2 < D < 2-3$ km (e.g., Bottke et al. 2005a,b; 2015; 2020). Numerical modeling work indicates these slopes are a byproduct of collisional evolution, with the main belt SFD taking on a wavy shape that is controlled by the asteroid disruption scaling law (see review in Bottke et al.

2015). Given that the PKB was initially massive and became excited, it was inevitable that it and its daughter populations would experience extensive collisional evolution. The difference is that the PKB SFD is governed by a disruption scaling law applicable to KBOs, which are likely to be porous ice-rock bodies (Morbidelli et al. 2021).

In order to model the collisional evolution of the PKB SFD, we need to identify the most useful constraints. While we considered using craters on Pluto, Charon, and Arrokoth for this purpose ala Morbidelli et al. (2021), we ultimately decided it may not the best place to start. One issue is that the definitive crater counts for Pluto have yet to be reported in the literature (S. Robbins, personal communication). A second issue is that Pluto's craters might have experienced relaxation and erosion in its topography, given that they do not precisely match the crater SFD on Charon (Singer et al. 2019). A third issue is the one mentioned above; there is difference in inferred slope for $D < 10$ km craters inferred for Charon alone ($q = -0.7 \pm 0.2$; Robbins and Singer 2021) and those derived from Arrokoth ($q = -1.1 \pm 0.1$; Spencer et al. 2020; Morbidelli et al. 2021). Morbidelli et al. (2021) gave reasons why the inferred slope for both should be $q \sim -1.2$, but the difference in slopes is still an open question. A fourth issue is that while Morbidelli et al. (2021) inferred a change in slope for projectile sizes of 20 m, the crater SFDs on Pluto, Charon, and Arrokoth craters do not go to small enough sizes to prove it is true. All these issues warrant careful study, which we intend to pursue in a follow-up paper.

### 3.1.2. Craters on Iapetus and Phoebe

As an alternative, we decided to consider the bombardment histories of the giant planet satellites, which were hit by the destabilized population. Some of these worlds have a larger range of observed crater sizes than the combination of Pluto, Charon, and Arrokoth, leading to valuable constraints. This topic of cratering on the giant planet satellites is extensive, though, so much of our analysis of certain issues, like their relative and absolute surface ages, will be reserved for a follow-up paper.

After examining the available crater SFDs for Jupiter's satellites (e.g., Zahnle et al. 2003; Schenk et al. 2004), Saturn's satellites (Porco et al. 2005; Kirchoff and Schenk 2010; Bierhaus et al. 2013; Thomas et al. 2007; 2013), and Uranus's satellites (Kirchoff et al. 2022), we have decided to focus on the crater histories of two worlds: Iapetus and Phoebe. Each one will be discussed below.

**Iapetus.** Iapetus is the most distant large satellite of Saturn. It has a mean diameter of 1469.0 ± 6 km, roughly 1.5 times the size of Ceres, and a bulk density of 1.088 ± 0.013 g cm$^{-3}$. It has a semimajor axis of about 59 Saturn radii, considerably more distant than that of Titan and Hyperion, which are at 20 and 24 Saturn radii, respectively. This distance is important because it effectively rules out the possibility that Iapetus was meaningfully affected by a putative satellite instability/disruption event occurring among the inner Saturn satellites (e.g., Ćuk et al. 2016; see also Ida 2019).

Our main interest in Iapetus for this paper come from its most ancient surfaces. Despite being far from Saturn, which minimizes the effects of gravitational focusing, Iapetus has twenty basins that are nearly 100 km in diameter and three that are larger than 400 km (Kirchoff and Schenk 2010) (see Fig. 14). Accordingly, Iapetus's largest craters give us a sense of the impactor SFD's shape for $D > 10$ km projectiles hitting the Saturn system from the destabilized population.

**Phoebe.** Phoebe is the largest remaining irregular satellite of Saturn. It has a shape that is close to an oblate spheroid, $(218.8 \pm 2.8) \times (217.0 \pm 1.2) \times (203.6 \pm 0.6$ km), with a mean diameter of $213 \pm 1.4$ km (Castillo-Rogez et al. 2012). It has been classified as a C-type body, giving it similar spectroscopic signatures to carbonaceous chondrite meteorites (Porco et al. 2005). Its bulk density of 1.634 g cm$^{-3}$ is modestly higher than many icy satellites in the Saturn system (Jacobson et al. 2006). Phoebe orbits Saturn with a semimajor axis of 215 Saturn radii, an eccentricity of 0.164, and an inclination relative to the ecliptic of 173.04°. This places it nearly 3.6 times as far from Saturn as Iapetus.

We focus on Phoebe because Cassini images provide us with an incredibly wide range of crater sizes. The only comparable crater set among outer solar system worlds to date comes from Hyperion (Thomas et al. 2007), but we find its SFD more complex to interpret at smaller sizes. At the larger end of Phoebe's crater SFD, there are at least seven craters that are 50–100 km in diameter. At the smaller end, there are many thousands of $D_{crat} < 0.2$ km craters have been recorded on a collective surface area of 3000 km$^2$ (Porco et al. 2005; Kirchoff and Schenk 2010). This means Phoebe's net crater SFD has a large enough dynamical range that it can potentially test the Morbidelli et al. (2021) prediction that the projectile SFD becomes steeper for D < 20 m projectiles. If true, Phoebe is a Rosetta Stone for setting the nature of the KBO disruption law.

A potential issue with using Phoebe to understand the destabilized population's impact history is its unusual history. Phoebe is likely an escaped PKB object that was captured onto a retrograde orbit around Saturn during the giant planet instability (Nesvorný et al. 2003a; 2007; 2014). That means the dominant bombardment populations striking Phoebe have changed with time.

For example, Bottke et al. (2010) showed that the initial irregular satellite population at Saturn was devastated by intense collisional evolution. Their work predicts that Phoebe's surface was shattered and reset multiple times by large early impacts. The timescale for irregular satellite collisional evolution is rapid; most occurs on the order of tens of Myr after capture. This interval is short enough that any craters formed after that time were likely dominated by heliocentric projectiles from the destabilized population. This would explain why the shape of Phoebe's crater SFD for $D_{crat} > 10$-15 km craters is comparable to that of other Saturnian satellites (Kirchoff and Schenk 2010).

An additional factor to consider is that irregular satellites, being captured KBOs, should follow the same disruption laws as KBOs. Collisional algorithms suggest they should grind themselves into an evolved SFD whose shape at small sizes that is similar to the destabilized population's SFD (e.g., O'Brien and Greenberg 2003; Bottke et al. 2015). This assumes, of course, that the ejecta produced by collisions between prograde and retrograde satellites remain within the stable irregular satellite zone rather than take on a small enough orbital velocity that they would fall onto the planet; e.g., Levison et al. 2008b).

If collisional evolution of the irregular satellites works as we suggest, it may be challenging to distinguish the source of Phoebe's craters; they could be from a combination of irregular satellites, which dominate early impacts, and heliocentric projectiles, which dominate late impacts, with both producing a similarly shaped crater SFD at small sizes. For the purposes of this paper, however, we only need the shape of Phoebe's crater SFD at small sizes, which constrains what happens to a KBO-like SFD undergoing collisional evolution.

The Cassini spacecraft obtained high resolution images of Phoebe's surface when it entered the Saturn system (Kirchoff and Schenk 2010). Both Porco et al. (2005) (counts by P. Thomas) and Kirchoff and Schenk (2010) have measured the spatial densities of craters on Phoebe terrains, and both groups have graciously provided us with spreadsheets of their data (M. Kirchoff, P. Thomas, personal communication). Each group found comparable SFDs for small craters, but we focus here on the craters provided by P. Thomas, who examined three regions observed at different resolutions. His crater sizes go from 0.003 km < $D_{crat}$ < ~100 km, though incompleteness at small sizes means the range where a useful crater SFD can be identified is ~0.01 km < $D_{crat}$ < ~100 km. The data are shown as cumulative SFDs in Fig. 14, with the data placed into root-2 size bins.

By removing crater counts where incompleteness may be playing a role, we find that the crater SFD between 0.5 km < $D_{crat}$ < 10 km has a power law slope of $q$ ~ -1.2. This slope is similar to the that calculated on Charon and Arrokoth over a comparable size range by Morbidelli et al. (2021). Similarly shallow slopes over comparable crater size ranges have also been identified on several major satellites (e.g., Europa, Enceladus, Hyperion, Miranda) and many minor ones (Prometheus, Pandora, Epimetheus, Janus, Telesto, Calypso) (Zahnle et al. 2003; Thomas et al. 2007; 2013; Bierhaus et al. 2013; Singer et al. 2019; Kirchoff et al. 2022). This shallow slope is also the basis for the CASE A crater production SFD suggested by Zahnle et al. (2003). Accordingly, we argue that an impactor SFD with this shape has bombarded all of the giant planet satellites.

The most intriguing feature in the small crater SFD for Phoebe is that the slope substantially steepens for 0.1 < $D_{crat}$ < 0.5 km craters. As discussed in Sec. 3.1.1., this slope change was not observed on Arrokoth, possibly because of limited resolution, but it was deduced by Morbidelli et al. (2021) on the premise that an increase in slope was needed to explain the abundance of Kuiper belt dust detected by New Horizons's dust counter. We cannot rule out the possibility that this change was produced by secondary or sesquinary cratering, but both bodies are small enough that returning ejecta should hit at low impact velocities (i.e., < 70 and < 100 m/s for Phoebe and Hyperion, respectively). It seems unlikely this debris would produce craters with shapes identical to those formed at much higher velocities. We would also point out that secondary or sesquinary craters have yet been identified on asteroids comparable in size to Phoebe (Bottke et al. 2020). Our interpretation is that this SFD represents the crater production population, and we will treat it as such below.

### 3.1.3. Synthesis

Put together, Iapetus and Phoebe can be used to assemble a crater SFD stretching from 0.1 km to ~1000 km. While these surfaces represent an integrated history of past bombardment, the inferred shapes of their SFDs are likely dominated by that of the destabilized population at early times. As such, they are a treasure trove of information on the destabilized population's early collisional history as well as the shape of the disruption law controlling KBO behavior.

To use this information in our model, we created a synthesis projectile SFD from these data. We started by first obtaining a synthesis crater SFD, using $0.1 < D_{\text{crat}} < 100$ km craters from Phoebe and $D_{\text{crat}} > 100$ km craters from Iapetus. This involved making certain assumptions for the P. Thomas datasets, who identified crater SFDs for three terrains observed at different resolutions. When two different crater SFDs from Phoebe were found to overlap, we adopted the SFD with the higher spatial density of craters that could still be considered part of the continuous SFD. We also excluded craters at sizes that were clearly suffering from incompleteness (i.e., those trending toward horizontal lines on a cumulative plot). From here, we grafted the shapes of the Iapetus and Phoebe crater SFDs together, assuming they were continuous.

Next, we converted the synthesis crater SFD into a projectile SFD using a crater scaling law, specifically the Holsapple and Housen (2007) formulation of the Pi-group scaling law (e.g., used by Marchi et al. 2015 and Bottke et al. 2020; see also Tatsumi and Sugita 2018):

$$D_t = kd \left[ \frac{gd}{2V_P^2} \left(\frac{\rho}{\delta}\right)^{2\nu/\mu} + \left(\frac{Y}{\rho V_P^2}\right)^{(2+\mu)/2} \left(\frac{\rho}{\delta}\right)^{\nu(2+\mu)/\mu} \right]^{-\mu/(2+\mu)} \quad (1)$$

Here the transient crater diameter, defined by $D_t$, can be found using the impactor properties (impactor diameter $d$, velocity perpendicular to the surface $V_p$, bulk density δ) together with the target properties (density of target material ρ, strength of target material $Y$, surface gravity $g$). Additional parameters ($k$, $\nu$, $\mu$) account for the nature of the target terrain (i.e., whether it is hard rock, cohesive soil, or porous material), while the yield strength $Y$ corresponds to the nature of the target materials. Finally, we account for the collapse of the transient crater, such that the final crater size is $D_{\text{crat}} = 1.2\, D_t$ (e.g., as used for asteroids in Bottke et al. 2020).

A key issue for our work is determining reasonable values for these scaling law parameters. In the literature, many different choices have been made to model crater formation on the giant planet satellites and Charon/Arrokoth (e.g., see Zahnle et al. 2003; Dones et al. 2009; Singer et al. 2019, Morbidelli et al. 2021; etc.), but testing these results is challenging. For example, we have yet to observe an actual impact event onto an icy satellite, there are no large-scale impact experiments that have been performed into low temperature ice, and basic information on probable target properties for the giant planet satellites is minimal (e.g., we do not know the strength and porosity of near-surface ice). It is easy to show with Eq. (1) that different assumptions can lead to an enormous range of values for a parameter we call $f$, defined as the ratio of crater-to-projectile sizes.

Our path though this thicket is to select parameters for Eq. (1) based on what has been learned about cratering from potentially analogous bodies, such as the carbonaceous chondrite-like asteroid (253) Mathilde, which is 53 km in diameter, and the dwarf planet (1) Ceres, which is 940 km in diameter. These worlds may share many commonalities with the giant planet satellites. For example:
- Ceres is comparable in diameter, bulk density, and gravitational acceleration to the mid-sized satellites of Saturn and Uranus. They are all spheroidal and have diameters between 400 and 1500 km in diameter. Similarly, Mathilde and Ceres bracket the sizes of Phoebe and other smaller giant planet satellites.
- The depth and diameter of craters on Ceres are comparable to those found on the mid-sized satellites, suggesting their surfaces undergo similar rheologic behavior when hit by projectiles (Schenk et al. 2021).
- Ceres may have ice as a dominant crustal component as well as an ocean world-like internal structure (e.g., Fu et al. 2017; De Sanctis et al. 2020; Park et al. 2020; Raymond et al. 2020; Schmidt et al. 2020).

With this basis, we lean on the results of Bottke et al. (2020), who reproduced the crater SFDs on Mathilde and Ceres using the following parameters for Eq. (1): strength of cohesive soil parameters $k = 1.03$, $\nu = 0.4$, $\mu = 0.41$, yield strength $Y = 2 \times 10^7$ dynes cm$^{-2}$, and identical projectile and near surface bulk density values (i.e., both with 1.3 g cm$^{-3}$). The advantage of their work is the impactors hitting Mathilde and Ceres come from the main belt SFD, whose shape has been substantiated down to small sizes by observational data. If one has excellent knowledge of both the projectile and crater SFDs, the nature of the crater scaling law can be derived empirically. Bottke et al. (2020) used this method to test their Eq. (1) parameters for both worlds (and on many other asteroids observed by spacecraft).

In this paper, we will use the Eq. (1) values above for all of the giant planet satellites, except we will assume the projectile and near surface bulk densities are 1 g cm$^{-3}$. This approximation is arguably basic, but as we will show below, it does allow us to reasonably reproduce the crater SFDs on most giant planet satellites. For Iapetus and Phoebe, we will use impact velocities of ~8 km/s, values derived from our dynamical model using the formalism discussed in Zahnle et al. (2003). Other values, such as surface gravity $g$, were taken from Table 1 of Zahnle et al. (2003).

One last issue concerns the statistical errors on a power law slope derived from twenty Iapetus basins. A projectile SFD derived from these data could potentially have a wide range of values, especially given uncertainties in Eq. (1). After considerable testing, we settled on a shallow SFD for projectiles that are $D > 10$ km (i.e., cumulative $q = -1.1$). As we will discuss below, this value matches the expected shape coming from the destabilized population's SFD.

The conversion of our target crater function into a target projectile function is shown in Fig. 2a. It was normalized to the predicted number of $D = 50$ km bodies in our starting SFD, which we will discuss in Sec. 4.2. We call this shape hereafter the *Iapetus-Phoebe SFD*, or IP SFD.

PLACE FIGURE 2 HERE

## 3.2 The Size Distribution of Jupiter's Trojan Asteroids

A key constraint for our collisional model is the combined SFD of Jupiter's Trojans. Jupiter's Trojans are small bodies that orbit around the L4 and L5 Lagrange points of Jupiter, which are located 60° in front of and behind Jupiter, respectively. They are contained within Jupiter's 1:1 mean motion resonance, with Jupiter having a semimajor axis of 5.2 au. Reviews on the Trojan populations and their properties can be found Marzari et al. (2002) and Emery et al. (2015).

Numerical simulations show that Jupiter's Trojans are likely to be a daughter population of the PKB (e.g., Morbidelli et al. 2005; Nesvorný et al. 2013). Their origin is linked to giant planet encounters taking place during the giant planet instability. At these times, giant planet encounters frequently occur when the giant planets themselves are surrounded by objects from the destabilized population. In an encounter between Jupiter and a Neptune-sized body, Jupiter "jumps" to a new orbit, which means its L4/L5 zones also jump on top of wandering KBOs that happen to be at the right place at the right time. Numerical simulations show that the capture probability of PKB objects into these stable zones is $(5 \pm 3) \times 10^{-7}$ (Nesvorný et al. 2013).

This Trojan provenance model can reproduce the observed orbital distribution of Trojans, including their high inclinations, and the observed number of very large Trojans. It also explains why the Trojans are dominated by D- and P-type bodies, a taxonomic class that is consistent with KBOs. The stochastic nature of Trojan capture events can even produce a modest asymmetry in the populations within L4 and L5, though it is not clear these populations are statistically different from one another when it comes to the largest bodies (e.g., see Fig. 1 of Marschall et al. 2022).

The capture process described above is size independent, so it is often assumed that the SFD of the combined L4 and L5 Trojans should be a scaled version of the Kuiper belt SFD, the scattered disk SFD, and so on (e.g., Morbidelli et al. 2021). The accuracy of this statement depends on the degree of collision evolution taking place among the bodies trapped in various PKB sub-populations. Any assessment must not only consider that the Trojans undergo collisions en route to their final destination, but also experience billions of years of collisional evolution after capture. This raises the possibility of some divergence between the various SFDs. As an example, consider the irregular satellites, which are a captured subpopulation of the destabilized population (Nesvorný et al. 2007; 2014). Simulations show they were decimated by collisional evolution to such a degree that their current SFDs have no obvious connection to their source SFD, the PKB SFD (Bottke et al. 2010).

Accordingly, in order to test our collisional evolution model, we need the best available estimate of the combined L4/L5 Trojan SFD as a constraint. Our investigation into this issue shows there is no one-size-fits-all solution, partly because there are differences in the literature between estimates but also because the Trojans are observationally incomplete for smaller sizes.

For the largest Trojans, complications arise from the unusual shapes of some bodies. Consider the Trojan (1437) Diomedes. A combination of occultations, lightcurve modeling work, and photometry suggest that this Trojan has a three-dimensional shape of $(284 \pm 61 \text{ km}) \times (126 \pm 35 \text{ km}) \times (65 \pm 24 \text{ km})$ and a mean diameter of 132.5 km (Sato et al. 2000). Infrared observations of the same body from the IRAS, WISE, and AKARI space-based telescopes, however, suggest mean diameters of 164.31 ± 4.1 km, 172.60 ± 3.42 km, and 117.79 ± 1.18 km (Tedesco et al. 2004; Usui et al. 2011; Grav et al. 2012). Given the unusual shape of Diomedes, it could be argued that all of these values have some validity. A look through the literature indicates several other large Trojans have modestly divergent diameters when the same sources of observational data are considered.

This led us to move toward simplicity, so we adopt a Trojan SFD for larger objects based on the formulation made by Nesvorný et al. (2018). They used IRAS diameters for $D > 80$ km Trojans and then pivoted to another method for the remaining objects. They converted Trojan absolute magnitude $H$ values from the Minor Planet Center to diameters using the following relationship (Fowler and Chilemi 1992; Appendix in Pravec and Harris 2007):

$$D_{ast}(km) = 1329 \times 10^{-H/5} \, p_v^{-1/2} \qquad (2)$$

A representative value of 0.07 was chosen for the visual geometric albedo $p_v$ of the bodies based on the results of WISE data from Grav et al. (2011). Our SFD for these bodies is shown in Fig. 2b. It was assumed that the Trojan population for $D > 10$ km in the Minor Planet Database was observationally complete.

Our resultant SFD in Fig. 2b has a small dip near 60 km. We suspect this feature is an artifact of the Trojan shape effects discussed above. Estimates of the shape of the Trojan magnitude distribution from Grav et al. (2011), Wong and Brown (2015), Yoshida and Terai (2017), and Uehata et al. (2022) show no evidence for such a feature. Whether artifact or real, however, this feature does not meaningfully affect our fits between model and data, as we will discuss below.

Our resultant SFD also left off a few of the largest Trojans whose reported sizes had considerable variation in the literature. These bodies may or may not be far from the continuum SFD for $D > 100$ km bodies, and we wanted to avoid having our fitting procedure chase objects with unusual sizes. Our motivation for this strategy comes from our Monte Carlo code experience, where the largest bodies drawn randomly from a SFD can occasionally be much bigger or smaller than expected. If the Trojans were indeed captured as suggested, their largest bodies could easily be anomalously big or small for the same reason.

Finally, for $D < 10$ km Trojans, we looked to results from observational surveys of small Trojans obtained from the powerful Suburu telescope teamed with the Suprime-Cam/Hyper Suprime-Cam instrument (Yoshida and Nakamura 2005; Wong and Brown 2015; Yoshida and Terai 2017, Uehata et al. 2022). We caution that these results can be interpreted in different ways. It cannot be ruled out that small Trojans follow a single power law from larger to smaller sizes. Our favored interpretation, however, that the small Trojans follow a broken power law, with power law slopes for the cumulative Trojan SFD changing from $q = -2.25$ for $D > 5$ km bodies to $q = -1.8$ for $D < 5$ km. Note that the aforementioned knee in the SFD was observed to occur at $H \sim 15$, with the diameter calculated from Eq. 2 with $p_v = 0.07$.

We find that the Trojan SFD in Nesvorný et al. (2018) for $D > 10$ km is consistent with these results, with the power law slope between $10 < D < 40$ km found to be $q = -2.26$. This led us to extrapolate that slope down to objects that were between $5 < D < 10$ km. From here, we grafted on the cumulative SFD for $D < 5$ km bodies discussed above. This task was made easier by F. Yoshida (personal communication), who graciously provided us with a list of Trojan diameters from her paper. Our synthesis is shown in Fig. 2b.

In Figs. 2a and 2b, we also make direct comparisons of the IP SFD with the Trojan SFD, with each shown as dashed lines. They show that the two shapes are dissimilar to one another. The origin of these shapes will be explored in Sec. 5.

## 4. Modeling Collisional Evolution in the Primordial Kuiper Belt and Trojan Populations

### 4.1 Boulder Collision Evolution Code

Our collisional modeling runs use the code Boulder, which is described in Morbidelli et al. (2009). Note there are two versions of Boulder, one that computes the evolution of the relative velocity of the population of bodies, and one where the velocity is input from a dynamical excitation model (e.g., Nesvorný et al. 2018). We use the latter here, but in Morbidelli et al. (2009), only the former was described.

Boulder is capable of simulating the collisional fragmentation of multiple planetesimal populations using a statistical particle in-the-box approach. It is a well-tested code that has been used to model the collisional evolution of planetesimals in the terrestrial planet region, the early asteroid belt, asteroid families, Hildas, Trojans and Trojans families, irregular satellites, and the survival of binary objects in the primordial Kuiper belt (e.g., Morbidelli et al. 2009; Levison et al. 2010; Bottke et al. 2010; Brož et al. 2013; Cibulková et al. 2014; Rozehnal et al. 2016; Nesvorný et al. 2018; Zain et al. 2020; Nesvorný et al. 2018; Nesvorný and Vokrouhlický 2019; Marshall et al. 2022).

In Boulder, the results of smoothed particle hydrodynamics (SPH)/N-body impact experiments were used to estimate the fragment SFD produced in disruption events (Durda et al. 2004; 2007). For a given impact between a projectile and a target object, the code computes the impact energy $Q$, defined as the kinetic energy of the projectile per unit mass of the target. The kinetic energy per unit mass needed to disrupt the target and eject 50% of the material to escape velocity is defined as $Q_D^*(D)$ (see Sec. 4.4). The value of $Q_D^*(D)$, hereafter $Q_D^*$, is a function input into the code. The fragment SFD ejected from the collision is chosen from a look-up table of fragment SFDs previously computed according to $Q / Q_D^*$. This allows it to simulate cratering and supercatastrophic disruption events in a relatively realistic manner.

With that said, the reader should be aware of analytical results that suggest the shape of the fragment SFD for smaller breakup events is not an important factor in generating the shape of a larger population's SFD via collisional evolution (O'Brien and Greenberg 2003; see also Bottke et al. 2015). This means our choice for Boulder's fragment SFD for small bodies does not have to perfectly reflect reality to obtain reasonable model SFDs for the PKB, destabilized population, and Trojans.

For each of our PKB runs, Boulder requires the following input components, each which will be discussed in some detail below.

1. The initial SFD of the PKB (Sec. 4.2).

2. The collision probabilities and impact velocities for (i) the destabilized population striking one another and the stable remnants of the PKB, and (ii) the population that will become Jupiter Trojans (Sec. 4.3).

3. A $Q_D^*$ disruption scaling law suitable for KBOs, which are likely porous ice-rock mixtures (Sec. 4.4).

**4.2 Choosing the Initial Size Distribution of the Primordial Kuiper Belt**

One of the most sought-after components of the PKB is its initial SFD. With it in hand, modelers can constrain the kinds of planetesimal formation mechanisms that were active early in the most distant reaches of the outer solar system. It is also needed as a starting point for our collisional evolution modeling work. The problem is that it is difficult to know what to choose; as discussed above, observations of smaller KBOs are limited, and the PKB/Kuiper belt experienced an unknown degree of collisional evolution.

**4.2.1. Insights from the main asteroid belt**

A limiting factor on the initial SFD chosen for use in our model is that it must have the kind of shape that allows it to fulfill existing dynamical modeling constraints (e.g., it must be able to deliver reasonable numbers of KBOs to various sub-populations during the giant planet instability and reproduce the nature of the observed Kuiper belt during Neptune's migration; Nesvorný and Vokrouhlický 2016). As a result, we do not attempt to test certain SFDs suggested in the literature (e.g., the initial SFD suggested by Schlichting et al. 2013)

The challenge in choosing an initial SFD is similar to the one faced by Bottke et al. (2005a,b), who modeled the collisional evolution of the primordial asteroid belt. To make progress, one needs to deduce the initial SFD of the primordial main belt and be confident that choice works in concert with dynamical evolution models. Given the uncertainties involved with such efforts, Bottke et al. (2005a) decided to consider what has been learned from numerical SPH simulations of asteroid disruption events (e.g., Benz and Asphaug 1999). They showed that asteroids in the gravity regime, defined as $D > 0.2$ km bodies, are increasingly difficult to disrupt as they become larger. For enormous asteroids like Vesta (530 km) or Ceres (930 km), the self-gravity is so high that they cannot be disrupted short of being hit by a projectile that rivals the size of the target body itself. This implies that the shape of the main belt SFD is very difficult to change for objects that are several hundreds of km in diameter; in effect, it can be considered a primordial shape, though dynamical mechanisms have decreased the population at all sizes over time (Bottke et al. 2005a,b). This same argument was adopted for the PKB by Nesvorný et al. (2018), among others, and we will use it here.

The main belt, KBO, and Trojans SFDs all have a "bump" in their cumulative size frequency distribution (SFD) for $D \sim 100$ km objects (i.e., the SFD is substantially steeper for $D > 100$ km bodies and is substantially shallower for $D < 100$ km objects). That means most of the mass contained within the connected slopes of these SFDs are in ~100 km bodies. Accordingly, the bump is telling us something important about the nature of the initial SFD and planetesimal formation processes.

Aspects of the origin of the main belt bump were explored in Bottke et al. (2005a). Using a collisional model, they examined what happened with a variety of initial SFDs. They focused on those where the observed cumulative power law slope of $q = -4.5$, found between objects that were 100, 110, and 120 km and those that were many hundreds of km, was extended to smaller sizes. They showed that collisions could not reproduce the observed shape of the main belt SFD, which has a fairly shallow slope for $D < 100$ km bodies. The reason is that $D \sim 50$ to ~100 km objects are exceedingly difficult to disrupt, so if you start with too many of them, you cannot get rid of them without producing noticeable collisional damage elsewhere in the main belt SFD. Bottke et al. (2005a) concluded from this that the bump near $D \sim 100$ km had to be a "fossil" from planetesimal formation processes, and that relatively few $D < 100$ km bodies were produced by the same mechanism.

These results are consistent with modeling work on planetesimal formation processes that suggest that asteroids and KBOs were "born big" (Morbidelli et al. 2009). Models and analytical work show that turbulent gas motion in protoplanetary disks can create regions where small particles, or pebbles, can become aerodynamically concentrated. When their spatial densities become high enough, they undergo collapse by their mutual gravity and form a planetesimal. The process is likely to produce 100-km-class bodies (e.g., Youdin and Goodman 2005; Cuzzi et al. 2008; 2010; Nesvorný et al. 2019b; 2021a; Klahr and Schreiber 2020; 2021). It has also been demonstrated that this scenario can reproduce the sizes, orbits, and inclination distributions of well separated binaries in the Kuiper belt region.

**4.2.2. Candidate SFDs for the Primordial Kuiper Belt**

The next question is what to expect from planetesimal formation processes for $D < 100$ km bodies. Given that constraints are limited, the literature offers different options. For example, numerical results from Nesvorný et al. (2020) indicate the same gravitational collapse processes that can make a 100-km-class well-separated binary can also, at the same time, eject numerous smaller clumps. These refugees might explain the origin of Arrokoth, the unusual two-lobed object observed by the New Horizons spacecraft. Alternatively, perhaps Arrokoth was formed directly by the streaming instability (McKinnon et al. 2020) or some other turbulent concentration mechanism (e.g., Cuzzi et al. 2008; 2010).

Here we consider two candidate shapes for the initial SFD for $D < 100$ km bodies.

**Candidate A. The initial PKB had a shallow SFD for $D < 100$ km objects.** Collisional evolution models of the main belt indicate that the best fits come from initial SFDs where the $D < 100$ km objects follow a shallow slope to small sizes (i.e., Bottke et al. 2005a,b used a cumulative slope of $q = -1.2$). This starting shape implies that most objects in the main belt smaller than a few tens of km are likely to be fragments from the disruption of larger bodies (Bottke et al. 2005a,b).  Given that most main belt asteroids are akin to carbonaceous chondrites (e.g., DeMeo and Carry 2014), and such bodies are most likely planetesimals formed in the giant planet zone (e.g., Kleine et al. 2020), we can infer that the initial shape of the main belt SFD should be close to that of giant planet zone planetesimals.  In other words, in the region immediately adjacent to the PKB, the initial SFD was shallow for $D < 100$ km planetesimals. A reasonable inference would be that the PKB should start with a similar shape.

Additional justification for a shallow shaped SFD comes from recent observational studies of the cold classical Kuiper belt (CCKB) by the DECam Ecliptic Exploration Project (DEEP) (Napier et al. 2023). Using the 4-meter Blanco telescope at Cerro Tololo Inter-American Observatory in Chile with the Dark Energy Camera (DECam), the goal of DEEP was to probe the faint end of the Kuiper belt by discovering thousands of objects down to magnitude $r \sim 26.5$ (roughly 30 to 40 km) and measure the SFD. The CCKB is considered the least collisionally evolved component of the PKB (e.g., Morbidelli et al. 2021), so its shape is arguably a good proxy for the initial PKB SFD.

DEEP's analysis of 20 half-nights of data, covering 60 sq. deg. of sky, allowed them to go deeper than previous CCKB studies, while their methodology allowed them to carefully characterize the efficiency and false-positive rates for their moving-object detection pipeline. They determined that the CCKB had a faint end slope of $q \sim -1.35 \pm 0.2$. We note that their slope predictions yield results that agree with estimates of the small KBO population faintward of magnitude $r \sim 24.5$ from Bernstein et al. (2004), who used Hubble Space Telescope observations to make their assessment. Their shallow slope does disagree with predictions from the literature (e.g., Adams et al. 2014; Fraser et al. 2014; Kavelaars et al. 2021), who argued for $q \sim -2$ by assessing the shape of the SFD for relatively large KBOs and then extrapolating that slope to smaller KBOs. Given that direct measurements are preferable to extrapolations, and that the Napier et al. (2023) results are consistent with expectations from Bottke et al. (2005a,b), we argue it is reasonable to use Candidate A SFD in our modeling work.

**Candidate B.  The initial PKB had a Trojan-like SFD for smaller objects.** A second possibility is that PKB had an initial SFD for $D < 100$ km bodies that was similar to that of the Trojan asteroids. This school of thought was proposed by Fraser et al. (2014) and was recently summarized in Morbidelli et al. (2021) (see also Bernstein et al. 2004; Fuentes et al. 2009; Adams et al. 2014; Emery et al. 2015). It argues that the initial SFD for $D > 100$ km was steep, with $q = -4.5$ to $-7.5$ for the hot and cold sub-populations, respectively (Fraser et al. 2014), and was Trojan-like for $D < 100$ km bodies, with $q \sim -2$ for bodies that were many tens of km. Given that dynamical models show the Jupiter Trojans were captured from the PKB (e.g., Morbidelli et al. 2005; Nesvorný et al. 2013), and that small Trojans are more easily observable than small KBOs, it arguably makes sense to connect the observed KB and Trojan SFDs together to deduce the shape of the PKB SFD.  Accordingly, the Candidate B SFD is also plausible for our modeling work.

### 4.2.3. Choosing a Candidate SFD

A key aspect of our work is that our model SFDs have to be able to reproduce the shapes of the two SFDs discussed in Sec. 3, namely the ancient crater SFDs found on Iapetus/Phoebe, presumably created by the destabilized PKB population shortly after the giant planet instability, and the present-day Trojan SFD. Assuming the differences are a byproduct of collisional evolution, our task in this section is to identify the most likely starting SFD from the available information.

For Candidate A, our constraints provide us with a straightforward time order to collisional evolution. The initially shallow SFD in the PKB evolves to a wavy SFD in the destabilized population that can reproduce the IP SFD (Fig. 2a). The captured Trojan population has to go even further, evolving from a wavy SFD to a $q \sim -2$ power law over 4.5 Gyr (Fig. 2b).

For Candidate B, we have a more complicated situation, in that the initial PKB SFD and the modern-day Trojan SFD are assumed to have the same shape. This suggests that the Trojans preserve the original SFD (i.e., $q \sim -2$ for $D < 100$ km bodies) and that evolution toward the IP SFD in the destabilized population occurs after the giant planet instability (i.e., it would explain the differences between the two SFDs in Fig. 2a,b). The problem is that the destabilized population undergoes rapid dynamical decay after the instability, leaving limited time for collisional evolution.

In order to quantify whether the latter scenario was possible, we examined runs in Nesvorný et al. (2018). In their Supplementary Fig. 3a, an initial Trojan-like SFD was found to evolve to the IP SFD-like shape, but it required the order of 400 Myr of collisional evolution within a massive PKB. That high degree of comminution is implausible for the destabilized population. Our additional test runs showed the Candidate B SFD does not change very much if collision evolution is limited to < 100 Myr in a massive PKB (e.g., see Supplementary Fig. 3b in Nesvorný et al. 2018).

The other possibility for Candidate B is that the Trojans, Kuiper belt, and destabilized population all have SFDs with similar shapes, at least in a statistical sense. We argue the existing data does not favor this scenario, but that is not the same as saying it cannot be true. For example:

- As discussed above, Napier et al. (2023) finds that the faint end slope of the CCKB is $q \sim -1.35 \pm 0.2$, not -2.0 as observed in the Trojans. This result potentially rules out the Candidate B SFD, but the DEEP observations have yet to be verified by comparable observations. More KBO observations, perhaps with the Vera Rubin telescope, may be needed to settle this issue.

- Inspection of the crater/basin SFDs found on the giant planet satellites in Schenk et al. (2004) (Jupiter satellites) and Kirchoff et al. (2013) (Saturnian satellites) indicates their power law slopes become fairly shallow for $D_{crat} > 200$ and 100 km craters, respectively, reaching values near $q \sim -1.2$. This change is consistent with the Candidate A SFD and the inferred shallow shape of the IP SFD for $D > 10$ km projectiles (Fig. 2a). The complicating issue here is that there are relatively few craters/basins that are made by such projectiles. When error bars are included for these crater/basin SFDs, it can be shown that both the Candidate A and B SFDs fit the data, though chi-squared testing methods tell us the fits are better with Candidate A.

- The Candidate B scenario predicts that the Trojans of Jupiter and Neptune should have the same SFDs, given that they were captured from the destabilized population at approximately the same time. Given the greater distance of Neptune's Trojans from the Sun, they should have lower intrinsic collision probabilities than Jupiter's Trojans and thus should be less collisionally evolved. Observations by Sheppard and Trujillo (2010), however, indicate that Neptune's Trojans are missing intermediate-sized planetesimals relative to expectation from Jupiter's Trojans (i.e., $D < 100$ km planetesimals). This apparent difference can be most easily explained by the Candidate A scenario, where Neptune's Trojans started with shallow SFD for $D < 100$ km planetesimals. At this time, though, it is unclear whether the existing observational evidence for the Neptune Trojans is strong enough to rule out Candidate B.

- The modeling work by Bottke et al. (2005a,b) indicates that planetesimals in the giant planet zone started with a shallow SFD for $D < 100$ km bodies. We assert that it makes sense that bodies in the PKB, which is adjacent to the giant planet zone, would start with a similarly-shaped SFD. This argument, however, is not proof, and nature may have a mind of its own.

The preponderance of evidence given above favors the Candidate A SFD over the Candidate B SFD. Accordingly, we will focus on it in our work. Nevertheless, we cannot rule out the Candidate B SFD at this time. Further testing of Candidate B will be left for a future paper, depending on whether the evidence supporting the Candidate B SFD becomes stronger with time.

Our initial SFD for the PKB is shown in Fig. 3, with some of its characteristics described in Table 1. It is identical to predictions made by Nesvorný et al. (2018) for $D > 100$ km bodies, who found it by combining observational constraints, model results of hydrodynamical simulations of the streaming instability (e.g., Simon et al. 2017; Nesvorný et al. 2019b; 2020; see their Supplementary Fig. 4), and dynamical simulations of the capture of KBOs in mean motion resonances by an outward migrating Neptune (Nesvorný and Vokrouhlický 2016; see their Fig. 15).

PLACE FIGURE 3 HERE
PLACE TABLE 1 HERE

The characteristic size of planetesimals formed by the streaming instability in their simulations was $D \sim 100$ km, and the initial disk mass ($M_{\rm disk}$) was set to 30 Earth masses. Here we assume there were ~2000 Pluto-sized objects in the PKB (Nesvorný and Vokrouhlický 2016). It is possible even larger objects once existed in the PKB, but additional work is needed to confirm this possibility. This SFD follows $q = -2.5$ between $D \sim 300$ km and Pluto-sized bodies ($D = 2370$ km), as suggested by observations of large KBOs (Brown 2008).

Our power law slope for $D < 70$ km objects in Fig. 3 does not follow that of Nesvorný and Vokrouhlický (2016), who assumed it followed a Trojan-like SFD power law slope of q = -2.1. It was instead given a value of $q = -1.1$, similar to that assumed for the primordial main belt by Bottke et al. (2005a,b).

Using Boulder, we track collisions over the full range of objects found in Fig. 3, which goes from nearly meter-sized bodies to objects larger than Pluto. The largest bodies, however, are almost impossible to disrupt, so they do not play a meaningful role in the story presented here.

## 4.3 Selecting Collision Probabilities and Impact Velocities for Collisional Evolution

Our next task is to choose the appropriate collision probabilities and impact velocities for objects striking one another as they move from the PKB to the destabilized population and to Jupiter's Trojans. As discussed in Sec. 3, these values are time dependent, with the PKB undergoing excitation and dynamical depletion, so characterizing how they change is critical to obtaining an accurate collisional evolution solution.

### 4.3.1. Example collision probabilities for test bodies between 0.1 and 50 au

In order to try to make our results easier to understand, especially given the wide-ranging orbits of our bodies across the outer solar system, we performed the following test calculation. We created a random population of 10,000 test bodies spread in semimajor axis $a$ between 0.1 and 50 au. The eccentricity ($e$) and inclination ($i$) values of the bodies were chosen to follow a Rayleigh distribution with mean $e$ = 0.2 and mean $i$ = 0.1 rad. From this, we calculated the intrinsic collision probabilities $P_i$ and mean impact velocities $V_{imp}$ for each test body against all other test bodies that crossed its orbit using the formalism of Bottke et al. (1994). Our results are shown in Fig. 4.

PLACE FIGURE 4 HERE

We find that the intrinsic collision probabilities fall off with $P_i \propto a^{-3.5}$, while impact velocities decrease with a dependance $V_{imp} \propto a^{-0.5}$. These values can also be deduced from the collision probability equation $P = n\, \sigma\, V$, with the density of bodies per volume of space decaying as $a^{-3}$, $\sigma$ being the mutual cross section, and $V$ decaying as $a^{-0.5}$. Accordingly, in our test population, an object at 8 au is 90 times more likely to be struck than an object at 30 au (i.e., $(30/8)^{3.4}$ = 90), with the impact velocities at 8 au twice as high as 30 au (i.e., $(30/8)^{0.5}$ = 1.9). The reason for this change comes from two factors: (i) the volume of space available for orbits increases away from the Sun, and (ii) the orbital velocities of bodies orbiting the Sun decreases away from the Sun.

Our test case is just an example, but it allows us to draw some inferences about collisional evolution in the outer solar system. Consider that while the PKB is enormous, its distant location, with most of the mass between ~24-30 au, means it has relatively small intrinsic collision probabilities, generally ranging from ~$10^{-21}$ to ~$10^{-22}$ km$^{-2}$ yr$^{-1}$, and low impact speeds, generally ranging from ~0.5-3 km s$^{-1}$. Conversely, Jupiter's Trojans, located at 5 au, have values that are between ~$10^{-17}$ to ~$10^{-18}$ km$^{-2}$ yr$^{-1}$, nearly four orders of magnitude higher than the PKB, while their mutual impact velocities are also higher (~5 km s$^{-1}$) (e.g., Marzari et al. 1996; Dahlgren 1998). In addition, while the ratio of the Jupiter Trojan population to the PKB is < $10^{-6}$ (e.g., Vokrouhlický et al. 2019), the ratio of their possible residence times is 4.5 Gyr compared to a few tens of Myr for objects in the destabilized population, yielding the order of ~$10^2$ (Nesvorný et al. 2018). Putting these components together, the Jupiter Trojans, despite their relatively small population, may experience as much collision evolution over 4.5 Gyr as objects residing in the PKB for tens of Myr. These calculations also tell us that the Jupiter Trojans, despite their smaller population size, are probably more collisionally evolved than the scattered disk, which are the long-lived survivors of the destabilized population. This outcome is more consistent with the Candidate A SFD than the Candidate B SFD.

Inferences can also be made about collisions among bodies in the destabilized population. For example, after the giant planet instability, the vast majority of destabilized objects eventually enter into deep giant planet-crossing orbits with Jupiter, Saturn, etc. Their $P_i$ values will increase while they reside in that zone, with individual objects lasting a few Myr to tens of Myr on average (Nesvorný et al. 2018). When put together, it means that that substantial collisional evolution can occur for KBOs just prior to them being trapped within stable zones in the main belt, Hilda, Jupiter Trojans, and Jupiter irregular satellite populations.

Conversely, long-lived members of the destabilized population, many which end up in the scattered disk between 30-50 au and beyond, generally have $P_i$ values that are lower than those for objects residing in the PKB. This limits how much collisional evolution scattered disk objects experience from other scattered disk objects. With that said, some scattered disk objects reside for billions of years on orbits that cross the stable Kuiper belt population, which is ~0.1% the size of the PKB. Accordingly, we can expect objects in the Kuiper belt and destabilized population/scattered disk to show slow but steady collisional evolution over billions of years. These changes might be revealed in the crater SFDs found on younger terrains on outer solar system worlds (e.g., Europa, Ganymede, Enceladus, Ariel).

**4.3.2. Collisions among the destabilized population and Jupiter Trojans**

With those concepts in place, we are ready to calculate the collision probabilities and impact velocities of PKB objects in the destabilized population/scattered disk and Jupiter Trojans. Here we followed the strategies used in Bottke et al. (2005b) and Nesvorný et al. (2018) to find the parameters needed to model a population undergoing dynamical upheaval.

For Stage 1, our initial PKB was made of $10^6$ particles spread between 23.5 and 30 au. The surface density of the particles followed $1/r$ from the Sun, with $r$ being heliocentric distance. This value means the same number of objects would exist in any chosen $\Delta r$ bin.

In previous work, co-author D. Nesvorný tested how rapidly Neptune would migrate through dynamically cold and hot disks of different masses. In general, he found faster migration came from colder and/or more massive disks and slower migration from hotter and/or less massive disks. A complicating issue is that ~$10^3$ Pluto-sized bodies within the disk take varying times to get different masses excited, so the relationship is not one to one.

As an approximation, we assumed here that the PKB disk was quickly excited by embedded Plutos, with the steady state orbits of the bodies following a Rayleigh distribution with mean $e$ = 0.05 and mean $i$ = 0.025 rad (Nesvorný et al. 2018). This behavior is reasonable when our main purpose is to calculate collision probabilities and impact velocities between PKB objects. We also assumed the PKB remained relatively unchanged for a time $\Delta t_0$, the time between the dissipation of the gas disk and Neptune entering the PKB. In our model, we tested values where $\Delta t_0$ = 0, 10, 20, and 30 Myr.

For Stage 2, we consider Neptune's interaction with the PKB. We defined $\Delta t_1$ as the time from when Neptune enters the PKB (and began to migrate across it) to the start of the giant planet instability. Numerical simulations of this behavior capable of matching solar system constraints require careful work, so here use two pre-existing giant planet instability runs from the literature.
1. Nesvorný et al. (2017) tracked the evolution of the scattered disk with a giant planet instability taking place at $\Delta t_1$ = 10.5 Myr.
2. Nesvorný et al. (2013) studied the capture of the Jupiter Trojans with a giant planet instability taking place at $\Delta t_1$ = 32.5 Myr.

The $\Delta t_1$ values differ between the runs because each simulation made contrasting assumptions about the mass in the PKB's outer disk, with the former having more mass (a net mass of 20 Earth masses) than the latter (a net mass of about 15 Earth masses). This means that our dynamical runs are not self-consistent with the PKB's mass when Neptune enters the disk or when the instability occurs. At best, they are approximate; our PKB starts with 30 Earth masses but steadily loses debris from collisional evolution, with longer $\Delta t_1$ values corresponding to more mass loss via comminution. A fully self-consistent model with collisions and dynamics that also matches giant planet instability constraints will have to wait for future efforts with faster codes.

Finally, for Stage 3, we tracked what happened to test bodies that were ejected onto planet-crossing orbits by Neptune's migration. This population defines the destabilized population.

All told, the combination of four $\Delta t_0$ values and two $\Delta t_1$ values means we have eight different model histories to examine. A summary of these values is provided in Table 2.

PLACE TABLE 2 HERE

Our next step was to choose a representative sample of 5000 test bodies from Stages 1-2 that were destined to reach the destabilized population in Stage 3 and become long-lived enough to become part of the scattered disk. These test bodies were used to calculate collision probabilities ($P_i$) and impact velocities ($V_{imp}$) against the background test body population at every timestep in our simulations.

Before they can be included in the Boulder code, though, we must also account for dynamical depletion within the PKB and destabilized population. In Bottke et al. (2005b), we tracked $P_i$ and dynamical depletion separately and accounted for the latter in the model SFD at every timestep using the code CoDDEM. This method works but it is cumbersome to implement. For our Boulder runs, we instead normalized our collision probability by the starting population at every timestep:

$$P_{pop} = P_i \left(\frac{N_{surv}}{N_{pop}}\right) \tag{3}$$

with $N_{\text{surv}}$ being the number of test bodies left in the simulation at time $t$ and $N_{\text{pop}}$ being the starting population. Our derived values $P_{\text{pop}}$ and $V_{\text{imp}}$ for the destabilized population are shown in Fig. 5.

PLACE FIGURE 5 HERE

The situation for the Jupiter Trojans is more challenging. Numerical simulations show that the capture rate of Jupiter Trojans from the PKB is ~5 × $10^{-7}$ (Nesvorný et al. 2013). This means that in our standard giant planet instability simulations with $10^6$ test bodies, almost none will be captured. To overcome this problem, we took advantage of a residence time map of orbital locations from Nesvorný et al. (2013). It shows the orbital parameters most likely to yield Jupiter Trojans (i.e., objects with low to modest eccentricities residing for some interval near 5 au).

Using our giant planet instability runs, we identified destabilized population test bodies that reached those orbits at the same time Jupiter was undergoing encounters with a Neptune-sized body. Jupiter's orbit is jolted by these encounters, allowing it to serendipitously capture KBOs into L4 and L5 that are at the right place and time. The dynamical tracks of these specific test bodies were used to calculate $P_{\text{pop}}$ and $V_{\text{imp}}$ values for our model Trojans. This method allows us to account for collisions on model Trojans from PKB bodies, the destabilized population before and after capture, and "background" Trojans after capture. Our results are shown in Fig. 6.

PLACE FIGURE 6 HERE

These collision probability and impact velocity values were entered into the Boulder collision evolution code as look-up tables. The code then interpolates to get the appropriate parameters for every collision timestep.

## 4.4. Selecting a Disruption Law for Kuiper Belt Objects

A major challenge in modeling the collisional disruption of KBOs is that we do not know the disruption scaling law applicable to such objects. Disruption scaling laws are commonly defined by the parameter $Q_D^*$, defined as the critical impact specific energy or the energy per unit target mass needed to disrupt the target and send 50% of its mass away at escape velocity. The projectile capable of disrupting an object $D_{\text{target}}$ is defined as $d_{\text{disrupt}}$:

$$d_{disrupt} = \left(\frac{2Q_D^*}{V_{imp}^2}\right)^{1/3} D_{target} \tag{4}$$

There are several potential formulations in the literature for KBO disruptions that are based on numerical hydrocode modeling work (e.g., Benz and Asphaug 1999; Leinhardt and Stewart 2009; Jutzi et al. 2010). The issue is that there is no easy way to test which one is best. KBOs also have poorly understood physical properties and substantial porosities, a combination which is already challenging to treat for carbonaceous chondrite asteroids, let alone porous ice-rich bodies (e.g., Jutzi et al. 2015).

The alternative way to assess $Q_D^*$ functions is to use collisional evolution models like Boulder, with the model SFDs and disruption predictions compared against observational constraints. Several groups have done this for the main belt (e.g., see review by Bottke et al. 2015, with a different perspective provided by Holsapple et al. 2019). For example, spacecraft imaging of main belt asteroids and their crater SFDs provide us with information on the projectile SFD making those craters (e.g., Bottke et al. 2020).

For the Kuiper belt SFD, which is more difficult to observe, modelers to date have mainly tried to constrain their simulations using the known KBOs larger than several tens of km in diameter (e.g., Davis and Farinella 1997; Kenyon and Bromley 2004; Pan and Sari 2005; Schlichting et al. 2013; Nesvorný et al. 2018; see Kenyon et al. 2008 for a review). Only recently has information on small KBOs become available via the craters found on Pluto, Charon, and Arrokoth (Singer et al. 2019; Spencer et al. 2020; Morbidelli et al. 2021). This has allowed recent studies to probe the nature of the impactor population making those craters (e.g., Kenyon and Bromley 2020; Benavidez et al. 2022).

As discussed in the introduction, the components that allow a collisional model to test different $Q_D^*$ functions for KBOs are as follows:

a) A dynamical evolution model of the PKB that matches constraints (which allows accurate computation of $P_{\text{pop}}$ and $V_{\text{imp}}$), and,

b) An accurate estimate of the initial SFD of the PKB,

c) A set of robust constraints for the shape of the model SFDs over all sizes.

Assuming our choices for (a)-(c) are reasonable, we can use the Boulder collisional evolution model test a variety of $Q_D^*$ functions. We defined them using the following equation from Benz and Asphaug (1999) that was rewritten by Bottke et al. (2020):

$$Q_D^*(R) = aR^\alpha + bR^\beta \tag{5}$$

$$a = \frac{Q_{D_{LAB}}^*}{R_{LAB}^\alpha} \frac{1}{1 - \frac{\alpha}{\beta}\left(\frac{R_{LAB}}{R_{min}}\right)^{\beta-\alpha}} \tag{6}$$

$$b = -\frac{\alpha}{\beta} a R_{min}^{\alpha-\beta} \tag{7}$$

Here the $Q_D^*$ function has the shape of a hyperbola that passes through a normalization point $(Q_{D_{LAB}}^*, D_{LAB})$ determined using laboratory impact experiments, with $R = D / 2$ and $R_{LAB} = D_{LAB}/2$. The values α and β help define the left and right slopes of the hyperbola, which is shaped like a "V". The minimum $Q_D^*$ value in the hyperbola $(Q_{D_{min}}^*)$ is defined to be at the location $R_{min} = D_{min} / 2$.

In order to make our choice of $(Q_{D_{LAB}}^*, D_{LAB})$, we turned to the laboratory ice shot experiments described in Leinhardt and Stewart (2009) (see their Fig.

11). They show the outcomes of numerous vertical shot experiments into ice, with $D_{LAB}$ ranging from 3 to 27 cm and $Q^*_{D\,LAB}$ ranging from ~4 × 10$^4$ to 9 × 10$^6$ erg g$^{-1}$. Rather than choose a single point for our runs, we decided it was better to sample across this cluster of experimental results. Accordingly, we assumed $D_{LAB}$ = 10 cm and that $Q^*_{D\,LAB}$ = 6.20 × 10$^4$, 2.65 × 10$^5$, 1.13 × 10$^6$, and 4.84 × 10$^6$ erg g$^{-1}$. These values, as well as those provided below, are also given in Table 2.

For α, which helps control the slope of the $Q^*_D$ function in the strength regime (i.e., where $D < D_{\min}$), we chose values between 0 and -1 incremented by -0.1. For β, which helps control the slope in the gravity regime (i.e., where $D > D_{\min}$), we chose values between 0.1 and 2 incremented by 0.1. Finally, we varied $D_{\min}$ between 10 and 100 m incremented by 10 m. In preliminary work, we tried a larger $D_{\min}$ range, but the runs with $D_{\min}$ > 100 m were so unsuccessful that we discarded them for our production runs. The reason these runs failed will become clear in our discussion of our modeling results.

Taken together, this gives us 4 × 11 × 20 × 10 = 8800 $Q^*_D$ functions to test for each choice of ($\Delta t_0$, $\Delta t_1$). This is too many to put onto a single plot, or even a series of plots, so we only display a small sample of our $Q^*_D$ functions in Fig. 7. It shows the three sets of $Q^*_D$ functions. The highest set of curves has ($Q^*_{D\,LAB}$, α, β, $D_{\min}$) = (4.84 × 10$^6$ erg g$^{-1}$, -0.5, [0.1, 2], and 20 m), the middle set has (2.65 × 10$^5$ erg g$^{-1}$, -0.8, [0.1, 2], and 50 m), and the lowest set has (4.84 × 10$^6$ erg g$^{-1}$, -1.0, [0.1, 2], and 70 m).

For reference, we have also plotted the asteroid disruption law from Benz and Asphaug (1999) in Fig. 7. Using their Eq. 6, it assumes the target's bulk density is 2.7 g cm$^{-3}$, $Q_0$ = 9 × 10$^7$ erg g$^{-1}$, a = -0.36, b = 1.36, B = 0.5 erg cm$^3$ g$^{-2}$, and impact velocity of 5 km s$^{-1}$.

PLACE FIGURE 7 HERE

Given all of these possible test cases, we have no choice but to employ automated testing metrics to evaluate the differences between our model results and constraints.

## 5. Model Runs and Results

Using the information from Sec. 4.1-4.4, we have everything we need for our Boulder model runs. Here we are running two distinct sets of trial runs: one for the destabilized population and one for the population that goes on to make Jupiter's Trojans.

Each trial run includes as input the initial SFD of the PKB, a $Q^*_D$ function (defined by the parameters in Table 2), and a look up table of $P_{pop}$ and $V_{imp}$ values for collisions that are defined for each set of $\Delta t_0$ and $\Delta t_1$ (Figs. 5 and 6). A random seed is also entered to account for stochastic variation in the timing/nature of breakup events. We assigned a single bulk density of ρ = 1.0 g cm$^{-3}$ for all KBOs and their fragments. This value splits the difference between smaller objects, which are likely to have ρ < 1.0 g cm$^{-3}$ (A'Hearn 2011; Blum et al. 2006; Sierks et al. 2015) and larger KBOs, many which have ρ > 1.0 g cm$^{-3}$ (Bierson and Nimmo 2019). The Boulder code does not have the means to assign distinct bulk densities to its objects; they are instead treated statistically within size bins.

For each set of trial runs, we are combining 8800 potential $Q_D^*$ functions with our four possible values of $\Delta t_0$ and our two choices for $\Delta t_1$. This means we need to sift through 70,400 trials to find our best fit cases. Recall that each set of trial runs has its own set of $P_{\text{pop}}$ and $V_{\text{imp}}$ values, so in the end, we are evaluating 70,400 × 2 = 140,800 trials runs. This quantity of model data cannot be examined by eye, so we developed an automated way to score the quality of the fits between model and observational constraints.

**5.1 Scoring the Results**

For our runs tracking the destabilized population, our constraints on the SFD come from the projectile population derived from the IP SFD (Fig. 2a). Recall that the crater populations used here do not provide the bombardment SFD at a single moment in time but are instead integrated history; they record impacts from the moment a surface can retain craters to the present day. Accordingly, our model SFD can only be tested by constructing a model crater SFD in the same manner.

We did this as follows. First, we tabulated the impact rate of PKB bodies on Jupiter over 4.5 Gyr from the numerical simulations of Nesvorný et al. (2013, 2017, 2019a). It is shown in Fig. 8. The values have been normalized over the total number of Jupiter impacts, so they can be readily applied to our PKB population. They show that the impact rate drops by four orders of magnitude over the course of the simulation. Accordingly, we would expect the highest impact flux (and moat of the largest impacts) to occur on Jupiter relatively soon after Neptune enters the PKB. As the impact flux decreases, the odds become increasing small that a large impact will occur, though stochastic events can and do occur from time to time.

PLACE FIGURE 8 HERE

We found that 1.1% of the PKB population strikes Jupiter over time. Accordingly, if we start with 2000 Pluto-sized objects and ~$10^8$ $D$ > 100 km bodies (Fig. 6), and that collisional evolution is minimal for these large bodies, we predict that ~20 and ~$10^6$ should have struck Jupiter, respectively.

From here, we divided the Jupiter impact flux curve into time segments that matched our output from the Boulder code (i.e., we printed out results every Myr between the start of the simulations and 100 Myr of elapsed time, every 50 Myr between 100-1000 Myr, and every 100 Myr between 1000 and 4500 Myr). Starting from the present day, namely 4500 Myr in simulation time, we worked backwards in time and tabulated the average SFD for the destabilized population in a given time bin (e.g., the SFD between 4400 and 4500 Myr, 4300 and 4400 Myr, etc.). These were multiplied by the appropriate normalized Jupiter impact flux for that time. By adding the results together in a cumulative sense, they could be compared with the shape of the IP SFD at every time step.

In this paper, our main concern is how well our integrated model SFD can reproduce the shape of the IP SFD. The model age of the Iapetus and Phoebe surfaces, as well as those of many other satellites, will be discussed in Paper

II. To determine the quality of the fit between our model SFD at a given timestep and the IP SFD, we applied the chi-squared relationship (e.g., Bottke et al. 2020):

$$\chi^2 = \sum_{i=1}^{M} \frac{(N_{model}(>D_i) - N_{obs}(>D_i))^2}{N_{obs}(>D_i)} \tag{8}$$

Here $D_i = 1,..., M$, stands for the diameters of observed and model projectiles. To obtain reduced $\chi^2$ values, we divide them by the value $M$, yielding the value we define here as $\chi^2_{norm}$. These values were calculated for all output timesteps over 70,400 runs.

For the Trojan runs, we followed a similar approach. First, we took the Trojan SFD from Fig. 2b and created a uniform distribution of 30 points across the SFD. This was necessary because our observed SFD is a combination of the known bodies and estimates of the SFD from the Suburu telescope with Suprime-Cam/Hypersuprime-cam. Second, we scaled the results of the PKB SFD from each Trojan trial by the fraction of bodies captured into the Trojan population (5 × $10^{-7}$). Third, we used Eq. 8 to evaluate the $\chi^2$ values at 4500 Myr, the end of the simulation. This calculation corresponds to checking whether the model SFD matches the observed one in the present day.

The next question is how to interpret these values. We effectively have two independent sets of $\chi^2$ values: one set for the destabilized population, where $\chi^2$ was evaluated at every output timestep, and an individual value for the Trojans, where $\chi^2$ was only evaluated at the end of the simulation. This leads to our first confounding issue, namely for a given set of input parameters, the destabilized population might have low $\chi^2$ values at certain timesteps in its trial run, while its counterpart Trojan value might be high, or vice versa. Only a relatively small collection of input parameters yields low $\chi^2$ values for both sets.

A second confounding issue is trying to determine what kind of match between our trial runs and the two SFDs in Fig. 2 constitute a "minimal" good fit. Recall that our $\chi^2$ method tells us how the model SFD, with equidistant values in log diameter space, conforms to the full shape of the observed SFD. This means the $\chi^2$ value for the destabilized population and the Trojans has effectively been turned into a relative score that cannot be evaluated using statistical methods. Given this, it is not clear how to determine the relative scores that are barely good enough to meet a goodness of fit metric. One can define threshold values for these relative scores, but the success space would have an arbitrary size.

Our compromise solution was to multiply both $\chi^2$ values together, giving us a single value that could tell us the relatively quality of a given fit for a set of input parameters. The best fit trial runs were then evaluated by eye to see how well our automated fitting procedure had worked against the test functions in Fig. 2.

## 5.2 Results

The goal of our scoring system was to find those parameters in Table 2 that did a good job of (i) reproducing the IP SFD over a long period of time during the early stages of the run, when bombardment of the outer planet satellites would have been at its highest, and (ii) the Trojan SFD in the present day. In many cases, we found low $\chi^2$ scores for (i), representing an excellent fit, but more moderate $\chi^2$ scores for (ii), representing a modest fit to the Trojans (i.e., they were still too wavy, with the model SFD not quite reaching the $q \sim -2$ power law slope needed between $5 < D < 100$ km bodies). More infrequently we found the converse; a good fit with (ii) but a poor fit with (i).

By sorting the multiplied scores from our two sets of 70,400 trials, we generated a "Top 10" list of best fit cases. Their parameters and $\chi^2$ scores are reported in Table 3, and the $Q_D^*$ functions are plotted in Fig. 9. Before discussing the trends in this list, we will first examine in detail our best fit examples for both the destabilized population and the Jupiter Trojans.

PLACE TABLE 3 HERE

PLACE FIGURE 9 HERE

### 5.2.1. Best fit for the destabilized population of the PKB

Our best fit run for the destabilized population is shown in Fig. 10. It is identified as #2 in Table 3, and its disruption law is one of the two blue curves in Fig. 9. Here we show the evolution of the model SFD as a series of four snapshots. We find that a small fraction of the $D > 100$ to 300 km bodies in the PKB undergo disruption over 4.5 Gyr. The fragments from these collisional events undergo further comminution, eventually creating a SFD that takes on a wave-like shape. The SFD is seen to match constraints at the 40 Myr timestep in Fig. 10 (see also Table 1).

PLACE FIGURE 10 HERE

The explanation for this wavy shape has roots that go back to Dohnanyi (1969), who analytically explored how collisional evolution would proceed among bodies in a SFD if their disruption scaling law was independent of size. He showed that such SFDs would evolve to a steady state with a cumulative power law exponent $q = -2.5$. Using this as a starting point, O'Brien and Greenberg (2003) used analytical and numerical methods to demonstrate that if the disruptions have some dependance on size, namely the strength per unit mass decreases with size, as shown in Figs. 8 and 9, $q$ can become steeper than -2.5. Here $q$ depends on the slope of the $Q_D^*$ function for $D < D_{min}$, which means it is controlled by the α parameter in Table 3. The best fit SFD in Fig. 10, where $D_{min} = 20$ m, has a Dohnanyi slope for $D < D_{min}$ of $q \sim -2.7$.

The transition between the strength and gravity regimes occurs at $D_{min} = 20$ m in Fig. 9. Beyond this point, the self-gravity of objects plays an increasingly important role in making massive objects more difficult to disrupt.

Bodies just beyond the $D > 20$ m boundary are still quite weak, though, and they are being struck by numerous smaller objects that make up the Dohnanyi slope SFD. The outcome is that $D > 20$ m bodies are readily obliterated, and the SFD evolves to a shallow slope. Our model results indicate the slope is generally near $q \sim -1.2$. This goes on until we reach target sizes that are close to those that can be disrupted by $D_{\min} = 20$ m projectiles. For the $Q_D^*$ function controlling Fig. 10, that point is reached at $D \sim 1$ km. The deficit or "valley" created near $D \sim 20$ m means there are few projectiles that can disrupt $D \sim 1$ km bodies, so the latter look like a "bump". For $D > 1$ km, the slope becomes steep again because there are few objects between 20 m $< D <$ 1 km that can disrupt $D > 1$ km bodies.

We note that a similar collisional cascade occurs in the main belt, but asteroids have $D_{\min}$ ~200 m (Bottke et al. 2005a,b; 2015; 2020). This also leads to a wavy SFD, except the shallow slope of $q \sim -1.2$ is found between $0.2 < D < 2$ km, with a steeper slope starting at $D > 2-3$ km.

If our starting SFD was a power law $q = -2$ for $D < 100$ km bodies (i.e., Candidate B SFD), and sufficient collisional evolution were to take place, the relative excess near $D \sim 1$ km would create yet another valley at the sizes disrupted by $D \sim 1$ km bodies, namely those $10 < D < 30$ km in diameter. In our situation, however, the reaction of the input SFD is contingent on the shape of the starting SFD. As shown in Fig. 10, disruption events slowly reduce the initial bump between $50 < D < 300$ km, with some fragments repopulating the limited population of $10 < D < 30$ km bodies. This effect helps keep this size range fairly steady in Fig. 10.

The interesting consequence is that fragments from these larger disruption events slowly build up the population of objects between several km $< D <$ 10 km in Fig. 10, making it steeper. Over time, they even cause the bump near $D \sim 1$ km to shift over to $D \sim 2$ km. This occurs because there are relatively few objects between several tens of meters and 1 km that can disrupt these bodies. As we will discuss in a follow-up paper, this behavior may explain the crater SFDs found on the younger terrains of Ganymede (Schenk et al. 2004).

### 5.2.2. Best fit for Jupiter's Trojans

Our best fit run for Jupiter's Trojans, identified as entry #1 in Table 3, is shown in Fig 11 (see also Table 1). The first two snapshots, at 1 and 10 Myr, show the collisional evolution of the PKB SFD that will eventually be captured in Jupiter's L4 and L5 zones. The results have been multiplied by the capture efficiency of the Trojans from Nesvorný et al. (2013), namely $5 \times 10^{-7}$, so they can be compared to the observed population of the Trojans.

PLACE FIGURE 11 HERE

The initial shape of the PKB SFD is different from that of the Trojans. The timestep of 1 Myr in Fig. 11 shows the model SFD has too many objects that are $D > 30$ km and far too few that are $D < 30$ km, not a surprise because the objects have yet to be captured. Comminution in the PKB gradually makes the fit better by 10 Myr, but the shape is still more reminiscent of the IP SFD than the Trojans.

Major changes take place after the giant planet instability; in this run, $\Delta t_0$ = 10 Myr and $\Delta t_1$ = 10.5 Myr (Table 3). The pre-Trojan population is placed onto giant planet crossing orbits, where they are then passed down by giant planet encounters to nearly circular orbits at ~5 au, the orbits where they can be captured as Jupiter's Trojans. In this interval, the destabilized population is near its maximum size. The intrinsic collision probabilities between small bodies increases by orders of magnitude for those moving down to Jupiter's semimajor axis, but this is offset by the rapid depletion of the population (Fig. 5). Eventually, the Trojans are captured, but they continue to be hit for some period of time by remnants of the destabilized population. As shown at the 40 Myr timestep of Fig. 11, the combination is enough to remove many $D > 30$ km bodies and steepen the SFD of $D < 30$ km bodies.

For the remaining time, Trojan collisional evolution is dominated by mutual impacts with other Trojans within their particular cloud (i.e., within L4 or L5). While the population is much smaller than the PKB, there is billions of years to work with and the collision probabilities are relatively high (Fig. 4). This allows the Trojan SFD to take on a nearly power law shape of $q = -2$ between $5 < D < 100$ km at 4.5 Gyr (Fig. 11).

The reader may notice that our model SFD is slightly wavy between few km $< D < 70$ km, with an inflection point near $D \sim 20$ km. We find it intriguing that the Trojan SFD determined from WISE infrared observations shows a similar feature (see Fig. 12 and 14 from Grav et al. 2011). While this is a potentially positive sign that our model results are reasonable, we caution that the veracity of this feature must be considered suspect until more Trojans near these sizes have confirmed diameters.

Unfortunately, the deduced SFD from WISE observations could only be calculated down to 10 km in diameter. In order to find the true SFD between $5 < D < 100$ km, the Trojan population may need to be observed by a next generation instrument (e.g., ground-based observations from the Vera Rubin telescope; space-based infrared observations from the NEO Surveyor mission).

### 5.2.3. Parameter trends from the runs

Overall, our Top 10 runs yield several trends telling us what is needed for our model SFDs to match constraints (see Table 3 and Fig. 9). First, we find that there is a strong preference in our $Q_D^*$ functions for $D_{\min}$ to be near 20-40 m, with the top choices given to 20 m. This is the same size range predicted by Morbidelli et al. (2021). This value sets the nature of the Dohnanyi-type SFD in the strength-scaling regime, which in turn generates the smallest craters found in the IP SFD (Fig. 2a). We will discuss additional constraints related to this portion of the SFD in Sec. 6.1.

Second, we find we need just the right amount of collisional evolution in the destabilized population and the Jupiter Trojans to match constraints. This is more easily explained by placing our 70,400 × 2 trials into several broad groups:

- **Easy**. Easy collision evolution means the $Q_D^*$ function has relatively low values (e.g., corresponding to the lower curves on Fig. 7)

- **Difficult**. Difficult collision evolution is the opposite, with the $Q_D^*$ function having relatively high values (i.e., corresponding to the higher values on Fig. 7)
- **Short**. The model uses lower values of $\Delta t_0$ and $\Delta t_1$, which means the PKB only exists for a relatively short time prior to the giant planet instability.
- **Long**. Here the $\Delta t_0$ and/or $\Delta t_1$ values are on the higher side of the values given in Table 2, which means the giant planet instability occurs relatively late.

We found that runs corresponding to "Easy/Long" can occasionally reproduce the Trojan SFD, but at the cost of too much collisional evolution in the destabilized population's SFD. Specifically, these model SFDs often yield shapes similar to the last timestep shown in Fig. 10, but with the steeper slope for $1 < D < 10$ km occurring at earlier times. In effect, the destabilized population's SFD is trying to follow the same evolutionary path as the Trojan SFD (Fig. 11), but its collision probabilities and impact velocities in Fig. 4 are too low to achieve that level of grinding.

Conversely, all "Difficult" runs, regardless of "Short" or "Long", did not produce enough collisional evolution among the Trojans to get rid of their wavy shape; they still look too much like the IP SFD. As a thought experiment, consider scenarios where the giant planet instability occurs at $\Delta t_0 + \Delta t_1 = 100$ Myr or even 700 Myr. Those models could only match the IP SFD constraints if the PKB were to undergo very slow collisional evolution, which in turn means the use of $Q_D^*$ functions that prevent disruption events in a massive PKB. The problem is that there would be no way for that $Q_D^*$ function to reproduce the Trojan SFD. Once the Trojans were captured, a restrictive $Q_D^*$ function would prevent subsequent disruptions, and that would leave behind a prominent wavy shaped SFD that would be readily observable today.

The best fit results in Table 3 and Fig. 9 appear to provide just the right amount of collisional evolution, in that they prevent too much from occurring in the PKB at early times while allowing the Trojans to get just the right amount at later times to match constraints. These runs correspond to the "Easy/Short" tree of outcomes, though perhaps a better moniker would be "Easy/Moderate". All of the runs in our Top 10 had some combination of $\Delta t_0 + \Delta t_1$ that yielded ~20 or ~30 Myr. The only run that had $\Delta t_1 = 32.5$ Myr also had $\Delta t_0 = 0$ Myr. These results suggest that the PKB had to experience a moderate degree of collisional evolution before the giant planet instability took place.

The differences between our top two cases are an amalgam of the above discussion. Our top case nicely fits the Trojans SFD, but its match to the IP SFD is modestly too steep for $1 < D < 10$ km objects. Conversely, our second-best case has a slightly worse fit near 10 km for the Trojan SFD (i.e., it is slightly wavier than our Trojan SFD), but it matches the IP SFD nicely between $1 < D < 10$ km. While the second pick is slightly worse overall from a combined $\chi^2$ perspective, we favor it over the first case as a better overall choice.

The $Q_D^*$ functions in Fig. 9 are much lower than the asteroid disruption law from Benz and Asphaug (1999). They tend to favor β values between 1.0-1.2, though their $Q_{D_{LAB}}^*$ and α values are more distributed across a wide range (Table 2). The higher values of $Q_{D_{LAB}}^*$ appear to corollate with larger values of $D_{\min}$ between 30-40 m, while the lower $Q^*_{D\,LAB}$ values prefer $D_{\min}$ = 20 m. It seems likely that the power law slopes generated from these values for $D < D_{\min}$ are all similar enough that they only modestly affect the $\chi^2$ score. With that said, our top two best fit cases have the lowest tested $Q^*_{D\,LAB}$ values and $D_{\min}$ = 20 m.

For reference, we have also plotted the disruption law from Leinhardt and Stewart (2009), which was based on laboratory and numerical hydrocode impact experiments into weak ice, in Fig. 9. Using Eq. 6 from Benz and Asphaug (1999), this curve assumes that bulk density is 0.93 g cm$^{-3}$, $Q_0 = 2 \times 10^5$ erg g$^{-1}$, a = -0.4, b = 1.3, B = 35 erg cm$^3$ g$^{-2}$, and impact velocity is 1 km s$^{-1}$.

Their $Q^*_D$ function lies close to the range of $Q^*_D$ functions shown by our best fit functions in the gravity regime. The main difference is their value of $D_{\min}$, which is 200 m rather than 20-40 m. As discussed above, this value would produce a steep slope in the model SFD for $D < 200$ m, and thus would not be able to reproduce the IP SFD.

An extrapolation of their gravity regime curve to D ~ 20 m would yield results close to our best fit results. The implication is that small bodies have substantially lower $Q^*_D$ values than previously predicted, though they would still be consistent with the lower range of the ice shot targets provided by Leinhardt and Stewart (2009).

We postulate that the differences could be caused by the porous, possibly fluffy nature of small KBOs (e.g., Nesvorný et al. 2021). Impacts into such targets are challenging to simulate in a controlled laboratory setting or within numerical codes. Ultimately, ground truth may be needed to understand how small KBOs react to impacts.

## 6. Verification of Model Results

Our best fit modeling work makes a number of predictions that can be tested against observational data. For the current era, this includes recent impacts on Jupiter detected by ground-based observers, impacts on Saturn's rings found in Cassini images, and the debiased SFDs of Jupiter family comets, long period comets, and Centaurs. For more ancient times, our results can be checked against crater and basin SFDs on the giant planet satellites. Finally, the lifetime of the PKB must be consistent with the existence of Kuiper belt and Jupiter Trojan binaries. For example, we find that the PKB likely survives 20-30 Myr, well within the < 100 Myr limit derived by Nesvorný et al. (2018) and Nesvorný and Vokrouhlický (2019).

Note that some verification issues are left for future papers. For example, interpreting the bombardment history of giant planet satellites and computing their model surface ages from the spatial densities of their craters is deferred to Paper II. The predicted impact histories of the bodies in the Pluto-Charon system, and that of the cold classical Kuiper belt object Arrokoth, will be the subject of Paper III.

### 6.1 Testing the Small Body Impact Flux in the Outer Solar System

The most comprehensive model of the small body impact flux on outer solar system worlds is found in Zahnle et al. (2003) (hereafter Z03). Using the crater SFDs found on several younger terrains (e.g., Europa, those found on Ganymede's

Gilgamesh basin), and their preferred crater scaling laws, Z03 estimated the likely projectile SFD striking the Jupiter system. From there, they calibrated the impact rate for this population on Jupiter by examining the close encounter rate of comets near Jupiter over the last several hundreds of years. These close encounter estimates were updated in Dones et al. (2009), and we will use them later in this section. From there, Z03 used numerical results from comet evolution simulations to determine the ratio of impact rates on Saturn, Uranus, and Neptune to that of Jupiter. That allowed them to use an Opik-type model to scale their Jupiter impact rates to the satellites of the giant planets and calculate the impact velocities of the projectiles striking those worlds (Zahnle et al. 1998; 2001; see Table 1 in Z03).

The nominal Z03 model for impacts on outer solar system worlds, based on the projectile SFD derived from Europa and Ganymede craters, was called "Case A". A second model called "Case B" was also developed based on the crater SFD found on Triton. In a follow-up study, however, Schenk and Zahnle (2007) argued that most of Triton's craters were not primary craters and therefore could not be used as part of a production function.

The chronological system of Z03 was not constructed to include non-primary craters, so the findings of Schenk and Zahnle (2007), if true, would invalidate Case B. Regardless, Case B is still commonly used in the literature, partly because its impactor SFD provides a better match than Case A for $D_{crat}$ < 10-15 km craters on certain terrains but also because there is still an on-going debate about the nature of the impactor SFD hitting Triton and other outer solar system worlds. For example, some have argued that asteroids from the inner solar system find a way to become planetocentric impactors in the giant planet systems (e.g., Denk et al. 2010). This issue will be discussed in Sec. 6.3.

Overall, we find that our model SFD is a good match to the predictions of Case A for projectiles that are approximately $D$ > 50 m. This is no surprise because, like Z03, our model uses craters on outer solar system bodies as constraints (i.e., Z03 used craters from Europa/Ganymede, and we used them from Iapetus/Phoebe). Like Z03, we predict that a heliocentric impactor SFD from the destabilized population/scattered disk made most of the larger craters found across the outer solar system.

A difference between Z03 and our model, however, comes from our predictions for $D$ < 50 m projectiles. This size range was not investigated by Z03. We have shown that our model SFD develops an increasingly Dohnanyi-type SFD for $D$ < 20-50 m bodies, and this signature is needed to reproduce the IP SFD. This raises the question of whether this predicted signature is seen in any other data set. The rest of Sec. 6.1 will concentrate on that issue.

### 6.1.1 Calculating the Present-Day Impact Flux on Jupiter

In order to calculate the present-day impact flux on Jupiter from our model results, we need to multiply the following components:

1. The model SFD from the destabilized population predicted for the present day,

2. The total fraction of PKB test bodies that have struck Jupiter between the start of the simulation and the present day, and

3. The normalized fractional rate of PKB test bodies hitting Jupiter per year in the present day.

Two of these components have already been calculated. For component 1, we can apply the model SFD for the PKB at 4500 Myr from our best fit run #2 in Table 3 (see Fig. 9). For component 2, numerical models of the giant planet instability indicate that 1.1% of all test bodies from the PKB will eventually hit Jupiter (Nesvorný et al. 2013; 2017). For the last component, we took advantage of the increased impact statistics produced by Nesvorný et al. (2019a), where they created numerous clones of test bodies in the destabilized population and tracked them forward in time to better determine the population of Centaurs that would be discovered in the present-day. Taking the Jupiter impacts from these runs, we collected all test body impacts that occurred between 4400 and 4500 Myr and calculated the median fraction that strike Jupiter per year. This worked out to be $4.2 \times 10^{-13}$. Multiplying this value by 1.11%, we find that the fraction of PKB bodies that strike Jupiter in the present day is $\sim 4.7 \times 10^{-15}$ yr$^{-1}$.

This value is more intuitive if we multiply it by our destabilized SFD. Using the last frame shown in Fig. 10, we find there are $\sim 10^{12}$ $D > 1$ km bodies. That value would yield an impact interval on Jupiter for $D > 1$ km bodies of $1 / [(10^{12})(4.7 \times 10^{-15} \text{ yr}^{-1})]$ or $\sim 210$ years. Performing the same calculation for the entire model SFD, we get the Jupiter impact rates shown in Fig. 12.

PLACE FIGURE 12 HERE

As a check on our work, we have added two calibration points for larger impacts. The first comes from Z03, who argue there have been six encounters within 4 Jupiter radii over the last 350 years that were from $D > 1.0$ [+0.5, −0.3] km objects (their equation 12; see also Schenk and Zahnle 2007; Dones et al. 2009). They also assert that the distribution of perijove distances for JFCs making close encounters with Jupiter is uniform, such that Jupiter impact rate is at least $4 \times 10^{-3}$ yr$^{-1}$. This value yields an impact interval of $\sim 250$ years, close to the value provided above. We show this as the blue dot on Fig. 12.

The second comes from Dones et al. (2009), who argue from observations and orbital simulations that at least 9 comets have crossed Callisto's orbit between 1950 and 1999 (Callisto's semimajor axis is 26.3 Jupiter radii). The smallest of these objects is thought to be $D > 0.5$ km (Lamy et al. 2004). Assuming these values are complete, they yield an impact rate on Jupiter of $4 \times 10^{-3}$ yr$^{-1}$ for $D > 0.5$ km comets. This value is shown in Fig. 12 as the red dot, with error bars from Dones et al. (2009).

Both of these values are consistent with our calculated impact rate on Jupiter, shown as the solid line in Fig. 12, and that of Case A from Z03. In the next set of tests, however, we consider impactors that are $D < 50$ m.

### 6.1.2 The impact flux of superbolides striking Jupiter

Over the last decade, Jupiter has become a compelling target for amateur ground-based observers. Modern equipment has allowed them to detect sporadic bright flashes in Jupiter's upper atmosphere. The first detection of a flash was made in 2010, and new detections have been made roughly every two years since that time (Hueso et al. 2010; 2013; 2018). The interpretation is that Jupiter is being struck by small bodies, with the so-called superbolides disrupting within the atmosphere after penetrating to some depth. The energy released in these events is comparable to the Chelyabinsk airburst, which was produced by the disruption of a ~20 m projectile about 30 km above the city of Chelyabinsk, Russia in 2013 (Popova et al. 2013; Hueso et al. 2013; 2018). In effect, the events are smaller versions of what was seen when the tidally-disrupted fragments of comet Shoemaker-Levy 9 hit Jupiter in 1994 (e.g., Harrington et al. 2004).

This work inspired professional ground-based observers to watch Jupiter with dedicated telescopes, which led to the detection of a large flash in October 2021 (Arimatsu et al. 2022). The kinetic energy of the impactor in this event was the order of two megatons of TNT, an order of magnitude more than previously observed flashes on Jupiter.

Observers have used this information to estimate the impact frequency of superbolides on Jupiter, after accounting for various observational biases. Using strengths and bulk densities associated with Jupiter-family comets, the cumulative flux of these small flashes, thought to be created by objects between 5-20 m in diameter or larger, is between ~10-65 impacts per year (Hueso et al. 2018). This range is plotted on Fig. 12 as the green box. The cumulative frequency of megaton-class impact on Jupiter is estimated to be ~1.3 [+3.1, -1.1] impacts per year (Arimatsu et al. 2022). Depending on the bolide's bulk density, its size could be between ~16 m (0.2 g cm$^{-3}$) to ~24 m (0.6 g cm$^{-3}$) to ~32 m (2.0 g cm$^{-3}$). This data range is plotted as the magenta box in Fig. 12, with the black horizontal line representing the best estimate of Arimatsu et al. (2022).

Both boxes are consistent with the predicted flux from our model. We believe they provide persuasive evidence that the projectile SFD evolves to a Dohnanyi-type SFD for D < 20 m, as predicted by both Morbidelli et al. (2021) and our model results. Additional tests are also possible by considering impacts on another large witness plate in the outer solar system, namely Saturn's rings.

### 6.1.3 The impact flux of small impactors on Saturn's rings

Saturn's ring system is a sizable target for small bodies. Consider that Saturn's A, B, and C rings have a collective surface area of 4.1 × 10$^{10}$ km$^2$, comparable to that of Saturn itself (i.e., 4.26 × 10$^{10}$ km$^2$). During the Cassini mission to Saturn, at a few select times, the spacecraft was able to obtain particular viewing geometries of the rings that made it possible to detect dusty clouds of icy debris (i.e., at Saturn's equinox; at high phase angles to the rings) (Tiscareno et al. 2013). These clouds are believed to be ejecta produced by small impact events on icy bodies within the rings themselves.

The size of the projectiles striking the rings was calculated by Tiscareno et al. (2013) from the brightness of the dusty clouds, with assumptions made about the nature of the ejecta particle SFDs within each cloud. Their estimates of the cumulative impact flux of impactors onto the A, B, and C rings are shown in Fig. 4 of Tiscareno et al. (2013). Using different ejecta particle SFDs for the debris clouds, with cumulative power law slopes of q = -3 and -4, they estimated the range of impactor sizes striking the rings was between ~40 cm and potentially ~20 meters in diameter.

To plot these data on Fig. 12, we need to convert the flux on Saturn's rings to that on Jupiter. We did this as follows. The results of the numerical giant planet instability runs from Nesvorný et al. (2013; 2017) indicate that the velocity of bodies entering Saturn's Hill sphere, or what is often called $V_\infty$, is close to ~8 km/s, while the ratio of Saturn impacts to Jupiter impacts is φ = 0.33. The gravitational focusing factor for objects encountering Saturn at a given distance $d$ is given by:

$$F = \left(1 + \frac{V_{esc}^2(d)}{V_\infty^2}\right) \tag{6}$$

with the escape $V_{esc}(d)$ at a given distance from Saturn is given by:

$$V_{esc} = \sqrt{\frac{2GM_s}{d}} \tag{6}$$

Here $GM_s$ is the gravitational constant multiplied by Saturn's mass, or 37931206.234 km$^3$ s$^{-2}$.

These variables allow us to calculate the gravitational focusing factors $F$ for projectiles hitting the Saturn's A, B, and C rings (i.e., $F_A$, $F_B$, $F_C$, or $F_{ring}$ for short) and that for Saturn itself. ($F_{saturn}$). For the former, we calculated the mean escape velocity of objects at each ring distance from Saturn (i.e., the C ring ranges from 74,568-92,000 km, the B ring ranges from 92,000-117,580 km, and the A ring ranges from 122,170-136,775 km). For the latter, we assumed Saturn's radius is $R_s$ = 58,232 km, which yields Saturn's escape velocity of $V_{esc}$ = 35.5 km/s. Using these values, we converted the cumulative impact flux values in Fig. 4 of Tiscareno et al. (2013) to a Jupiter impact flux by multiplying them by $\pi R_s^2$ ($F_{saturn}$ / $F_{ring}$) / φ. Our results are shown in Fig. 12 as the blue points with error bars.

The converted impact flux values match up well with an extrapolation of our model SFD; we expect the model SFD to follow a Dohnanyi-type power law slope to very small sizes (O'Brien and Greenberg 2003; Bottke et al. 2015). It suggests that the impactor hypothesis offered in Tiscareno et al. (2013), namely that the impactors come from a stream of Saturn-orbiting material produced by the previous breakup of a meteoroid, may not be needed. Instead, the dusty ejecta clouds observed by Cassini are likely to be from heliocentric impactors from the scattered disk.

## 6.2 The Ancient Crater Size Distributions on Outer Solar System Worlds

Our best fit model SFD for the destabilized population gives us the ability to interpret the cratered surfaces of many different outer solar system worlds, as well as make predictions on the largest impactors that could have hit them early in their history. There is also the fact that the destabilized population's SFD changes with time (Fig. 10), such that younger craters made by impactors that are between a few km and 10 km have a different SFD than older craters (Fig. 10). As discussed above, these topics are extensive enough that they will be the subject of a separate paper. Still, we believe it is useful to show comparisons between model crater SFDs based on our work and observed crater SFDs found on different worlds.

### 6.2.1 Jupiter's Satellites

We will start with crater data sets from the Galilean satellites, namely Europa, Ganymede, and Callisto (Fig. 13). The crater SFDs for Europa and Ganymede are described in Z03 (see their Fig. 1), and the Callisto counts come from Schenk et al. (2004) (see their Fig. 18.13). We chose them so we could examine the shape of different portions of our model SFD. The crater SFDs have been scaled by arbitrary factors so they can all be viewed on the same plot without the loss of any information.

PLACE FIGURE 13 HERE

A standard way to estimate the age of a solar system surface is by counting the number of impact craters found upon it. The greater the crater spatial densities, the older the surface. If one also understands the crater production rate, it is possible to use crater spatial densities to estimate the absolute age of the surface.

The number of craters counted on Europa and Ganymede per square million kilometers from Z03 correspond to relatively low crater spatial densities. For Europa, the crater SFDs come from counts taken across its surface. When compared to the CASE A crater production model from Z03, Europa's surface is between 30 and 70 Myr old. For Ganymede, the crater SFD shown comes from craters superposed on the large basin Gilgamesh, and they have a CASE A age of 800 Myr old (Z03). The crater SFD for Callisto represent global counts. No age is listed in Schenk et al. (2004), but Z03 suggest Callisto's surface age is very old.

Prior to discussing our crater production model, we note that all three of these worlds are 3-5 times larger than Ceres, the largest body where our preferred scaling law in Eq. (1) has been verified against crater data (Bottke et al. 2020). They also have gravitational accelerations that are more analogous to the Moon. These differences could mean that different parameters should be selected for Eq. (1). For this exercise, however, we will continue to use our standard parameters, leaving an exploration of this issue to Paper II.

Our model curves in Fig. 13, shown as black lines, represent our crater production function. It was calculated as follows. First, we took the model SFDs from our best fit case in Fig. 10 and used Eq. (1) to turn them into crater SFDs. Next, we multiplied the SFDs by the Jupiter impact rates from Fig. 8 appropriate for their output time. Taking the crater SFD for from the youngest output timestep (i.e., the time nearest the present day) and the next youngest timestep, we determined the mean between the two. This represents the shape of the crater SFD produced on our surface in that time interval ($\Delta t$).  Next, we scaled the mean crater SFD by multiplying it by $\Delta t$.  Moving backward in time, we added each new scaled model SFDs to the previous ones until reaching the beginning of the simulation. These integrated model crater SFDs, once properly scaled, can be compared with the observed crater SFDs.

The missing component needed to calculate the model surface age is the probability factor that scales Jupiter impacts to specific satellite impacts (e.g., Zahnle et al. 2003). These values are dependent on the dynamical model used to create the impactors. We will re-calculate them in Paper II in order to derive model surface ages for the giant planet satellites.

Using chi-squared methods to find the best fit between our crater production model and the observed craters SFDs, we find that the shapes of our model crater SFD are similar to the observed crater SFDs in Fig. 13. The lightly cratered surfaces of Europa and Gilgamesh basin on Ganymede are a better match to our results from relatively recent times, while Callisto's more heavily cratered surface is consistent with our results from ancient times.

Note that our production SFD does veer away from largest basins on Callisto. This result is a consequence of Eq. (1)'s tendency to strongly increase projectile size for large impacts. This quality was also seen on Vesta and Ceres by Bottke et al. (2020), and it may be in need some future reassessment. Overall, however, the major components of the wavy shaped SFD predicted by our model are confirmed.

### 6.2.2 Saturn's Satellites

Our crater production model was also compared with ancient craters on Saturn's satellites in Fig. 14.  For the satellites Mimas, Tethys, Dione, Rhea, and Iapetus, we used the crater data from Kirchoff and Schenk (2010). The terrains chosen by those authors for assessment have high spatial densities of craters; they represent some of the oldest surfaces on these worlds.

PLACE FIGURE 14 HERE

Here we plot craters and basins from a defined surface as filled circles, while the observed global distribution of $D_{\text{crat}}$ > 100 km craters and basins are shown as filled stars. Note that the circles and stars occasionally include the same data but may be offset from one another because they were scaled by different surface areas. Fig. 14 also includes craters from Phoebe and Hyperion that were reported in Porco et al. (2005) and Thomas et al. (2007).  Their counts were placed into root-2 size bins, which explains the equal spacing between the filled circles. To make comparisons between these crater SFDs and our model SFD, we again used the scaling law from Eq. (1) with the parameters provided in Sec 3.1.3.

As in Fig. 13, we used chi-squared methods to find the best fit between our crater production model and the observed craters SFDs. Overall, our model SFDs show general alignment with the crater SFDs, but there are some mismatches found for $D_{crat}$ < 15 km craters. A possible reason for this is that many of these terrains may be dominated by secondary and/or sesquinary crater populations (e.g., Zahnle et al. 2003).

For reference, secondary craters are those formed from the impact of sub-orbital ejecta from a single impact, while sesquinary craters are those formed from the impact of ejecta that initially escaped the target body, orbited the central body in the circumplanetary system, and then re-impacted the target body or another body in the circumplanetary system.

Secondary crater fields on the Moon, Mars, and elsewhere often have steep power law slopes (e.g., $q$ = -3 or more). Given that our crater production model has a shallow slope of $q$ = -1.2 for sub-10 km craters, it is relatively easy for secondaries and sesquinaries to overwhelm primary crater populations locally or even regionally (McEwen and Bierhaus 2006; Bierhaus et al. 2018).

Support for the apparent dominance of secondaries and sesquinaries on many Saturn satellite terrains come from the elongated crater populations identified on Tethys and Dione by Ferguson et al. (2020; 2022a,b). Of the craters they measured, most which were $D_{crat}$ < 15 km, they found most had East-West orientations. Such direction distributions are much more easily produced by sesquinaries than heliocentric impactors.

### 6.2.3 Uranus's Satellites

Uranus also has a number of mid-sized satellites with extensively cratered surfaces, Miranda, Ariel, Umbriel, Titania, and Oberon. They have sizes ranging in diameter from 472 km for Miranda to 1578 km for Titania. Their crater histories were recently reanalyzed using modern methods by Kirchoff et al. (2022). In their paper, Miranda was divided into four study terrains, Ariel and Titania was divided into two study terrains, and Umbriel and Oberon only had one study terrain apiece. Here we plot the crater SFDs from each world that were defined either as "cratered" terrains or, for Miranda, "cratered dense (or cratered D)" terrains by Kirchoff et al. (2022). Using chronology models from Z03, Kirchoff et al. (2022) showed that the crater spatial densities of these surfaces of Miranda, Umbriel, Titania, and Oberon had ancient surfaces, while those for Ariel were younger. They are shown in Fig. 15.

PLACE FIGURE 15 HERE

Using chi-squared methods to find a best fit between model and observed crater SFDs, we find the two are largely consistent with one another. The largest observed mismatch is with Ariel craters for $D_{crat}$ < 15 km. As suggested with Saturn's mid-sized moons, this could indicate the advent of secondaries or sesquinary craters, as seen on many of Saturn's satellites (Fig. 14). Overall, the crater SFDs on the Uranian satellites provide additional evidence that our model SFD in Fig. 10 is plausible.

### 6.2.4 Pluto, Charon, and Arrokoth

One of the ongoing mysteries presented in Z03 was the unknown impactor SFD striking various giant planet satellites. Given the differences between their CASE A and CASE B crater production SFDs, the uncertainties about the contribution of secondaries and sesquinaries, etc., there was no easy way to rule out certain models and scenarios. Some clarity was brought to this issue by the New Horizons mission, which imaged the cratered surfaces of Pluto, Charon, the smaller moons Nix and Hydra, and the cold classical Kuiper belt object Arrokoth (Robbins et al. 2017; Singer et al. 2019; Spencer et al. 2020; Robbins and Singer 2021). These observations provided new constraints on the impactor SFD striking KBOs, which would be a combination of the PKB, the Kuiper belt, and the destabilized population/scattered disk (e.g., Morbidelli et al. 2021). From here, we can glean new insights into the general shape of the destabilized population's SFD striking the giant planets.

The crater SFD found on Charon by Singer et al. (2019) showed a very shallow power law slope for craters $D_{crat}$ < 10-15 km, and a steeper slope for craters larger than this size. Robbins et al. (2018) found similar results, and the different crater counts were reconciled in Robbins and Singer (2021) (see also Ali-Dib 2022). The new work identified that smaller craters follow a cumulative slope $q$ = -0.7 ± 0.2, while larger craters had -2.8 ± 0.6. The former is shallower than any of the crater SFDs found on the giant planet satellites so far. For example, the IP SFD, which has data in the size range, shows $q$ ~ -1.2 for $D_{crat}$ < 10 km craters. The authors also argue that this shallow shape was not produced by crater erasure processes.

A more complex picture emerges if we incorporate into the story the limited number of craters found on Arrokoth (Morbidelli et al. 2021). Given that Arrokoth is a cold classical Kuiper belt object, the populations that can strike it over time are more limited. They can come from the cold classicals themselves, portions of the excited PKB/Kuiper belt, and portions of the destabilized population/scattered disk. Arrokoth has several craters that are sub-km, no craters with 1 < $D_{crat}$ < 7 km, and one crater that is $D_{crat}$ > 7 km. Using a Monte Carlo code, Morbidelli et al. (2021) showed that the cumulative power law slope of the projectile population making these craters would likely have -1.5 < $q$ < -1.2 (see also Spencer et al. 2020; who found $q$ ~ -1.3), and that values of $q$ = -0.7 ± 0.2 were statistically improbable (i.e., see their Fig. 3). The former values are consistent with the results of the model SFD created in Fig. 10.

In our collisional model, we have only seen one example of a $q$ = -0.7 ± 0.2 slope in the relevant size range, and that was for a limited time. When Neptune migrates across the PKB, it dynamically excites many KBOs to higher eccentricities and larger impact velocities with one another, as seen in Figs. 1 and 5. For a short interval, while the PKB and destabilized populations are still massive, the higher impact velocities cause the Dohnanyi-type SFD for $D$ < $D_{min}$ to shift to larger values. This change allows impactors in the Dohnanyi-type SFD to disrupt modestly larger bodies, which in turn decreases the slope of the SFD between several tens of meters < $D$ < 1 km to $q$ ~ -0.9 for a limited time. The $q$ ~ - 0.9 slope does not last very long, with the system eventually re-equilibrating to $q$ ~ -1.2, but while it lasts, its value is within the error bars of $q$ = -0.7 ± 0.2 (Robbins and Singer 2021).

We find this effect to be interesting, but it seems unlikely to us that such early impacts, while intense, could overshadow the impact flux affecting Charon over the subsequent 4.5 Gyr of its history. Accordingly, we are presently unsure how to explain the slope differences between Robbins and Singer (2021) and Morbidelli et al. (2021).

Morbidelli et al. (2021) predicted that the Kuiper belt should have power law slopes of $q = -3$ for $D < 20$ m, $q = -1.0$ to $-1.2$ for $20$ m $< D < 2$ km, and $q = -3$ for $2 < D < 100$ km. Our preferred slopes for the destabilized population at early times are largely consistent with these predictions: $q = -2.7$ for $D < 20$ m, $q = -1.2$ for $30$ m $< D < 1$ km, $q \sim -2.5$ for $1 < D < 10$ km, and $q \sim -1.5$ for $10 < D < 100$ km. The main mismatch comes from their prediction of the Trojan-like SFD for $D > 2$ km; our model results suggest this size range has a wavy shape driven by collisional evolution.

The "odd constraint out" is Charon's observed slope, which is anomalous compared to the other outer solar system bodies studied to date. Until we visit additional KBOs, or large Centaurs whose crater records are still intact from their time in the Kuiper belt/scattered disk, we have no easy way to gauge the importance of this shape difference.

With that said, the Kuiper belt SFD has considerable observational uncertainties, particularly at smaller sizes, as shown in Parker (2021) (e.g., their Fig. 4). Perhaps a solution can be found by treating the collisional evolution of the PKB, destabilized population, Kuiper belt, and cold classical Kuiper belt as distinct populations in Boulder, each with their own SFDs. We will explore whether this model can explain the cratered records of Pluto, Charon, Arrokoth, Nix, and Hydra in Paper III.

**6.3 Comparison of Model SFDs for the Main Belt and Destabilized Population**

One of the most unusual scenarios in the literature is that numerous craters on Saturn's satellites were produced by ejected main belt asteroids (e.g., Denk et al. 2010). This idea is based on the observation that crater SFDs on Saturn's satellites have a wavy shape broadly similar to crater SFDs found on the Moon, and so both may have been made by the same impactors. The upside of this hypothesis is that if it were true, lunar crater chronology could be used to date surfaces throughout the Saturn system. The downside is that ejected main belt asteroids need to find dynamical pathways that would allow them to be captured into planetocentric orbit around Saturn with low semimajor axis, eccentricity, and inclination values. This would allow them to hit the inner Saturn satellites at low impact velocities and create a crater SFD similar in shape to observations.

A problem with this scenario is that nearly all asteroids reaching the outer solar system do so after being ejected out of the solar system via an encounter with Jupiter (e.g., Bottke et al. 2002; Granvik et al. 2016; 2018). The only way to be captured into the Saturn system would be for the asteroids to undergo large numbers of planetary encounters, such that they achieve nearly circular, low inclination orbits with Saturn. Such behavior has yet to be noted by any numerical simulations discussed in the literature. Next, some of these bodies would need to wander into the L1 or L2 regions of Saturn, which would allow them to achieve a temporary Saturn orbit capture. The new "minimoons" would reside within Saturn's Hill Sphere for a limited time until they struck a body or wandered back out of the L1 or L2 regions. At Jupiter, this mechanism explains how a small fraction of Jupiter family comets become temporarily captured within its Hill Sphere (e.g., Kary and Dones 1996). But it is a relatively rare dynamical pathway for objects, even those on nearly circular orbits. Finally, even if a few bodies managed to beat the odds and follow all of these steps, they would need to lower their planetocentric semimajor axes, eccentricities and inclinations values enough to achieve low impact velocities with the inner satellites. At best, satellites encounters could achieve this behavior for tiny fraction of the minimoon population. Taken together, we can readily reject the asteroid impactor hypothesis on dynamical grounds.

We argue that a more satisfying solution to this issue is that collisional evolution in the main belt and PKB/destabilized populations yield wavy SFDs that have modestly similar shapes. While the $Q_D^*$ functions for asteroids and KBOs in Fig. 9 differ from one another, both produce steep Dohnanyi-type SFDs for objects with $D < D_{min}$, and this leads to collision grinding among larger objects up to sizes disrupted by a projectile with $D_{min}$. In the main belt, this means a shallow SFD of $q \sim -1.2$ between 0.2 and 2-3 km (Bottke et al. 2020), while in the destabilized population, we get a shallow slope of $q = \sim -1.2$ between ~0.03 and 1 km (Fig. 10; see also Morbidelli et al. 2021).

A comparison between the main belt SFD from Bottke et al. (2020) and the destabilized population SFD from Fig. 10 (at the 40 Myr timestep) is shown in Fig. 16. Both SFDs are wavy, have Dohnanyi SFDs at small sizes, and have vaguely similar shapes. The local maxima, however, are different: the main belt has a bump at $D \sim 2-3$ km while destabilized population has one at $D \sim 1$ km. This variance is caused by asteroids and KBOs following different $Q_D^*$ functions, with their $D_{min}$ values being 200 m and 20 m, respectively. The implication is that there is no longer any basis to use lunar crater chronologies in the outer solar system. Instead, new chronology models will need to be developed using the model SFD produced by the destabilized population.

PLACE FIGURE 16 HERE

**6.4 Neptune Trojans**

A potential test of our model SFD comes from the Neptune Trojans that populate the L4 and L5 orbital regions of Neptune. In our model, these worlds were captured from the destabilized population early after the giant planet instability (Nesvorný and Vokrouhlický 2009; Parker 2015; Gomes and Nesvorný 2016). According to Lin et al. (2021), 13 of the 21 known Neptune Trojans are stable for more than 1 Gyr (see their Table 1). They suggest that the Neptune L4 population has 149 [+95, -81] objects with absolute magnitude $H < 8$, roughly equivalent to prediction of 162 ± 73, 250, and an upper limit of 300 by Lin et al. (2019), Sheppard and Trujilo (2010), and Gladman et al. (2012), respectively. For reference, the Lin et al. (2021) value is 2.4 times than that of the Jupiter Trojans over the same magnitude range.

At face value, the larger population size would suggest a similarly-shaped SFD to that seen in the Jupiter Trojans (Fig. 2b). Insights gleaned from our collision probability simulations in Fig. 4, however, suggest that the intrinsic collision probability of Neptune Trojans striking one another is likely to be orders of magnitude lower than that of the Jupiter Trojans. Unless impacts from the destabilized population can make up the difference, which we suspect is unlikely, our prediction is the Neptune Trojan's SFD should have a shape that is comparable to an early version of the destabilized population's SFD in Fig. 10.

Estimates of the Neptune Trojan SFD from Sheppard and Trujilo (2010) show that the largest observed objects with $H < 8.5$ ($D > 90 \pm 20$ km, depending on the choice of albedo; see Eq. (1)) follow a cumulative power law slope $q = -5 \pm 1$, similar to that of the Kuiper belt (e.g., Nesvorný and Vokrouhlický 2016). The SFD then bends to a slope that is shallower than $q \sim -1.5$ for objects with an absolute magnitude $H > 8.5$. This value is inconsistent with the Jupiter Trojan SFD, whose $5 < D < 100$ km bodies follow $q \sim -2$ (Fig. 2b), but it is a match with our estimate of the initial PKB (Fig. 3) and the shape of the destabilized population's SFD (Fig. 10). It also may explain why no Neptune Trojans have been detected between $8.1 < H_r < 8.6$ (Lin et al. 2021).

Overall, the estimated shape of the Neptune Trojan SFD from the twenty or so known objects supports the hypothesis that the Neptune Trojans are less collisionally evolved than the Jupiter Trojans. While the degree of collisional evolution they have experienced needs to be modeled, it is possible that the largest Neptune Trojans still retain craters from their time in the PKB. Accordingly, these bodies may make compelling targets for spacecraft missions that fly past 30 au or those that intend to enter into orbit with Neptune.

### 6.5 Comet SFDs

#### 6.5.1 Jupiter family comet size distribution

The SFD of the Jupiter family comets (JFCs) has been investigated by many different groups over the last decade (e.g., a compilation of several different observational sets is shown in Lisse et al. 2020). The cumulative power law slopes of these comet datasets are between $-1.6 < q < -2.5$ for $4 < D < 10$ km bodies. Infrared observations of 98 distant comets using Spitzer over the same size range determined a SFD with a slope of $q = -1.9$ (Fernandez et al. 2013). It should be noted that nearly all of the comets were observed at more than 3 au from the Sun, yet a substantial fraction still showed activity. In Fig. 17, we have plotted a comparison between the Fernandez et al. (2013) data and a scaled version of our model SFD, using our results from the last timestep in Fig. 10.

PLACE FIGURE 17 HERE

The match is good between the two sets for $5$ km $< D < 20$ km, but the SFDs diverge for smaller objects. Given that our model SFD has the same shape as fresh crater SFDs on Europa and Ganymede, we suspect this difference is observational incompleteness for $D < 5$ km bodies.

As a further check on this discrepancy, we have also placed on Fig. 17 an estimate of the debiased JFC SFD found using infrared observations from the Wide-field Infrared Survey Explorer (WISE) mission (Bauer et al. 2017). Using a survey simulator, they created suites of synthetic comets to determine the observational selection effects for these bodies being detected by WISE. This allowed them to debias their detected number of 108 short period comets. Here they found that the power law slope of the net SFD was steeper over a larger range, with $q = -2.3 \pm 0.2$ for $1 < D < 20$ km.

We plotted a scaled version of this SFD in Fig. 17 so we could compare the shape of their debiased SFD to our model SFD. Note that we do not attempt to match their predicted number of JFCs here because we would need to link our SFD with both test body dynamics and a comet activity model. We consider this work interesting but beyond the scope of this paper. Instead, we normalized our SFD to their population of $D > 5$ km bodies.

Overall, our model SFD and debiased SFD from Bauer et al. (2017) are very close to one another for most of the bins in the JFC dataset. The results are mutually supportive, in that they suggest both model and data are reasonable. For $D < 2$ km bodies, however, we find a mismatch between model and observations. The cause is either observational incompleteness or the onset of a physical loss mechanism (e.g., Jewitt 2022) that presumably is affecting JFCs that come close to the Sun.

**6.5.2 Long period comet size distribution**

There have also been attempts to debias the long period comet (LPC) population. For example, using a survey simulator, Boe et al. (2019) debiased 150 detected LPCs detected by the Pan-STARRS1 near-Earth object survey. Their resultant SFD was found to be consistent with another LPC survey from Meech et al. (2004), who used the Wide Field Camera on the Hubble Space Telescope and the Low Resolution Imaging Spectrograph on the Keck II telescope to detect 21 LPCs and JFCs comet nuclei. These comets were observed at heliocentric distances between 20-30 au, far enough away that nuclei identification was relatively easy compared to that of comets observed closer to the Sun with sizable comae. Both SFDs are plotted in Fig. 18.

PLACE FIGURE 18 HERE

Dynamical modeling work indicates the LPCs were ejected into the Oort cloud during the giant planet instability via giant planet encounters, with their perihelion values raised by passing stars and galactic tides (e.g., Vokrouhlický et al. 2019). Accordingly, our expectation is that these comets should have a SFD similar to the destabilized population shortly after the giant planet instability. Here we plotted our model SFD from the second timestep in Fig. 10 (10 Myr timestep), which is modestly shallower than the third timestep (40 Myr). As before, we normalized our SFD to the cumulative number of $D = 2.5$ km bodies in Boe et al. (2019) to compare the shape of the SFDs.

Overall, we found a good match between our model SFD, the Boe et al. (2019) results, and the Meech et al. (2004) results for $D > 2$ km. For smaller sizes, the debiased SFD quickly becomes shallow compared to the model SFD. As stated in Boe et al. (2019), we do not know whether the paucity of small LPCs means the objects are harder to detect or that they have physically disrupted. Given that the mismatch between model and debiased data approaches an order of magnitude for $D > 0.5$ km bodies, and that Boe et al. (2021) presumably understands the observational biases associated with Pan-STARRS1 reasonably well, we suspect that the likely culprit is comet splitting or disruption events among LPCs that approach the Sun (e.g., Jewitt 2022).

We also plotted the debiased estimate of LPCs detected by WISE in Bauer et al. (2017). WISE detected 56 LPCs, with their debiased SFD shown in Fig. 18. Here their slope is much shallower than that of the JFCs, with $q = -1.0 \pm 0.1$ for $1 < D < 20$ km. We have also normalized our model SFD so it intersects the WISE SFD at $D = 4.5$ km.

The match between our model and the debiased SFD of the WISE LPCs is modest. We have yet to see a model SFD in our runs that could match $q = -1.0 \pm 0.1$ over $1 < D < 20$ km, so we suspect several possibilities: the detection efficiency of LPCs calculated for WISE is inaccurate, LPC comae are not being subtracted out as well as they could, and/or that there is some physical process that is being missed. Note that many of the WISE LPCs were detected inside 3-5 au, where activity is more prominent. LPCs undergo relatively frequent splitting events as approach the Sun, and this can affect both their activity, with new regions available for sublimating ices, and the size of the nuclei.

### 6.5.3 Centaurs

Another way to test the dynamical model used here, as well as the destabilized population's SFD, is to examine the observed population of Centaurs. These bodies come from the destabilized population/scattered disk and reside on giant planet crossing orbits. Centaurs are formally defined as objects with a semimajor axis smaller than that of Neptune, a Tisserand parameter with Jupiter larger than 3.05, and a perihelion distance $q > 7.35$ au (Gladman et al. 2008). Objects with $q < 7.35$ au are defined as JFCs. Note that the dynamical boundaries separating Centaurs from JFCs are defined for convenience, in much the same way a river is sometimes named for its upper and lower courses.

Nesvorný et al. (2019a) tested their dynamical model for the giant planet instability and the evolution of the PKB/destabilized population by predicting the present-day Centaur population. They compared their results to those of the Outer Solar System Origins Survey (OSSOS), a wide-field imaging survey that has detected nearly a thousand outer Solar System objects. Using an OSSOS survey simulator, they computed how many Centaurs OSSOS should have detected with semimajor axis $a < 30$ au, perihelion distance $q > 7.5$ au and diameter $D > 10$ km, defined as absolute magnitude $H < 13.7$ for a 6% albedo. They found it should have picked out $11 \pm 4$ Centaurs with $H < 13.7$, within the error bars of the 15 Centaurs actually detected by OSSOS.

A possible problem with the Nesvorný et al. (2019a) estimate, however, is that it used a Trojan-like SFD for $D < 100$ km bodies in the PKB (Fig. 2). This means their model Centaurs with $D < 50$ km follow a cumulative power law slope of $q = -2.1$ down to small sizes. This value is more aggressive than our model results from Fig. 10; our model results indicate that they should instead adopt the SFD from the last time step in Fig. 10.

Examining both SFDs side by side, we find our model SFD is lower by a factor of 1.6 near $D > 10$ km than that of Nesvorný et al. (2019a), but then catches up to it again at $D > 4$ km. Accordingly, the $11 \pm 4$ detections above should instead probably be $7 \pm 3$, which would put the model off by a factor of 2. With that said, there is variability among Centaur albedos; many have values larger than 10% (Johnston 2018), while Spitzer observations indicate the observed Centaurs have mean albedos similar to comets and are a few percent (Lisse et al. 2020). The use of lower or higher mean albedos in this model would shrink this gap between model and observations but would not eliminate it.

Given our match with the other constraints (see previous sections), we postulate that this difference could be from small number statistics or stochastic processes (i.e., the Centaur population should fluctuate over short time intervals, though it is secularly decreasing over longer timescales). Regardless, the Nesvorný et al. (2019a) analysis should be repeated as additional Centaurs are discovered by OSSOS and other surveys.

## 7. Implications

Our collisional evolution model results for the PKB and destabilized population have implications for various solar system issues. Given the length of this paper, we decided it would be better to summarize some of our implications in this section and place a longer discussion of these topics into Appendix A.

**Hydrated Materials in Interplanetary Dust Particles and Comets**. The conventional wisdom is that most anhydrous IDPs reaching Earth are derived from comets or KBOs, while hydrous clays among IDPs are likely from collisions taking place among hydrated asteroids in the main belt. Using the cosmic ray exposure age of various IDPs collected in the stratosphere, Keller and Flynn (2022) report that the abundance of hydrous IDPs reaching Earth is somewhere between 20-35%. This result is arguably unexpected, in that Nesvorný et al. (2010) show that the abundance of IDPs reaching the Earth from the main belt is < 15%. This mismatch suggests the abundance of hydrous IDPs coming from Jupiter family comets and KBOs higher than expected.

One way to explain this difference is to extrapolate from our model results. We show that numerous 100-300 km bodies are disrupted in the PKB, destabilized population, and scattered disk, with many more potentially shattered and scrambled. We hypothesize that these events may be capable of dredging up and/or ejecting aqueously altered materials from the deep interiors of KBOs, provided they exist. The putative interior water creating the aqueously altered materials would come from the decay of radioactive nuclides like $^{26}$Al. Alternatively, it could be that impacts produce sufficient heating on the surface of KBOs that they create a layer of near-surface hydrous clays, some which get liberated as hydrous IDPs.

One way to test these ideas may be to examine objects like the Jupiter Trojan Eurybates, the largest remnant of the Eurybates family and a target of the Lucy mission. It may show evidence for exposed hydrated materials dredged up from its parent body's interior or possibly hydrated materials created by impact heating (Sec. A.1).

**Most Observed Comets are Fragments of Large KBOs**. The initial SFD of the PKB indicates that $D > 10$ km objects are largely primordial, while our collisional model shows that $D < 10$ km objects, as they become smaller, are increasingly fragments of large KBO collisions. Our predictions are in line with a considerable body of previous work. We postulate this may explain why so many comets have shapes reminiscent of near-Earth asteroids, which are themselves collisional fragments (Sec. A.2).

**The Sizes of Interstellar Comets.** The sizes of the interstellar comets 'Oumuamua and Borisov have diameters of 0.14-0.22 km and 0.8-1.0 km, respectively ('Oumuamua ISSI Team 2019; Jewitt and Luu 2019; Hui et al. 2020). While two objects by themselves are the definition of statistics of small numbers, this factor of five difference in their sizes is arguably unexpected. Assuming their detections were not strongly biased by observational selection effects, we used a Monte Carlo code to estimate what kind of synthesis SFD could produce this range. Our work shows that these two bodies were most likely derived from a shallow power law slope similar to the ones reported here for the destabilized population (i.e., $q \sim -1$ cumulative). This similarity may suggest that collisional processes strongly affect small comets formed in exoplanet systems prior to ejection.

## 8. Conclusions

We developed a collisional evolution model of the primordial Kuiper belt (PKB), its destabilized population, and the Jupiter Trojans asteroids. The destabilized population are KBOs pushed onto giant planet crossing orbits in the aftermath of the giant planet instability and Neptune's migration across the PKB. Some of these bodies go on to strike the giant planet satellites, and their long-lived remnants make up Neptune's scattered disk, the population that replenishes the Jupiter family comet and Centaur populations. A small portion of the destabilized population is captured in Jupiter's L4 and L5 zones during the giant planet instability, and these bodies make up the Jupiter Trojan population (Sec. 2).

**Model Constraints.** We tested our collisional evolution results by attempting to reproduce two key sets of constraints: the shapes of the basin/crater SFDs found on the oldest terrains of the giant planet satellites and the combined SFD of the Jupiter Trojans (Sec. 3). For the former, we created a synthesis projectile SFD using basins/craters from Iapetus and Phoebe (i.e., the IP SFD) together with crater scaling laws verified against the crater SFDs found on the largest main belt asteroids (e.g., Ceres, Vesta, Lutetia, Mathilde) (Sec. 3.1). For the latter, we used the latest estimates of the combined Jupiter Trojan SFD, including observations made from the Suburu telescope with Suprime-Cam and Hyper Suprime-Cam (Sec. 3.2).

**Initial SFD of the PKB.** Using insights from a range of sources (e.g., planetesimal formation models that use the streaming instability to reproduce the population of well separated binary KBOs; new observations of the cold classical Kuiper belt, collisional evolution models of the main asteroid belt; dynamical simulations of giant planet instability), we deduced a starting SFD for the PKB (Sec 4.2). It assumes there is a bump in the cumulative SFD near $D \sim 100$ km, as observed in current main belt and Kuiper belt. For smaller bodies, we assume the SFD follows a shallow power law slope with cumulative power law slope $q \sim -1.1$, like that inferred for the primordial asteroid belt (Bottke et al. 2005a,b).

**Collisional Evolution Model.** Our work takes advantage of numerical models of the giant planet instability that are capable of reproducing dynamical constraints across the solar system. These results allowed us to calculate the collision probabilities and impact velocities of PKB bodies, destabilized population bodies, and Jupiter Trojans. They were used as input into our collisional evolution model (Sec. 4.3).

Using the Boulder collisional evolution code, we tested a wide range of input parameters: the time Neptune takes to enter the PKB ($\Delta t_0$), the time it takes Neptune to cross the PKB prior to the giant planet instability ($\Delta t_1$), and the shape of the KBO disruption law, or what we call the $Q_D^*$ function, which is defined by four parameters (Table 2; Secs. 4.3-4.4). We ran a combined number of over 140,000 trial cases for the evolution of the destabilized population and the Jupiter Trojans, with the results of each set tested against the appropriate constraints using $\chi^2$ methods (Sec. 5).

**Model Results for Destabilized Population and Trojans**. We found that disruption events among $D$ ~100 to ~300 km bodies produced numerous $D < 10$ km fragments that continue to undergo collisional evolution. We also identified $Q_D^*$ functions that can reproduce the IP SFD and the Trojan SFDs (Table 3). From these functions, we find the easiest KBOs to disrupt from an impact energy per mass perspective are near $D_{\min}$ ~ 20 m. When put into a collisions model, we find these $Q_D^*$ functions lead to a steep Dohnanyi-like SFD for $D_{\min} < 20$ m (i.e., $q$ ~ -2.7). In contrast, the same value for main belt asteroids, which are dominated by carbonaceous chondrite-like bodies that presumably formed in the giant planet zone, is $D_{\min}$ ~ 200 m.

For the destabilized population, steep Dohnanyi-like SFDs with cumulative power law slopes of $q$ ~ -2.7 will disrupt numerous bodies with $D > 20$ m, which in turn creates a shallow SFD ($q$ ~ -1.2) for 30 m $< D <$ 1 km and a steeper SFD for $1 < D < 10$ km. The result is a wavy shaped SFD that is broadly similar to that of the main belt (Sec. 5.2.1).

Despite its smaller population, Jupiter Trojans experience more collisional evolution than typical members of the destabilized population. In part, this is because Trojan collision probabilities are orders of magnitude higher than the destabilized population, but also this is because Trojans are stable enough to maintain their current population size (within a factor of 2 or so) for ~4.5 Gyr. Collisions steepen the Trojan SFD over time until it takes on the observed $q$ ~ -2 slope for $5 < D < 100$ km (Sec. 5.2.2).

The remnants of the destabilized population in the scattered disk undergo a more limited degree of collisional evolution. These bodies can strike both themselves and the population of stable KBOs over many billions of years, leading to a SFD whose power law slope between $1 < D < 10$ km gradually steepens with time. This effect can be seen on large relatively young terrains on worlds like Ganymede (Sec. 5.2.1).

Our best fit runs indicate that the giant planet instability probably occurred ~20-30 Myr after the dissipation of the solar nebula, which in turn probably occurred a few Myr after CAIs (e.g., Weiss and Bottke 2021). This interval gives time for the PKB to experience some collisional evolution, but not so much to greatly steepen in the slope of $1 < D < 10$ km bodies. Accordingly, our work is supportive of the idea that the giant planet instability did not occur after hundreds of Myr (e.g., Nesvorný et al. 2018) (Sec. 5.2.3).

In order to further test our results, we compared our predictions to additional sets of data and found the following results.

**Superbolide Impacts on Jupiter.** We calculated the model impact flux of small comets striking Jupiter in the present day and found it was high enough to reproduce the frequency of "bright flashes" on Jupiter seen by amateur and professional observers. The size and frequency of superbolide impacts on Jupiter are consistent with our model's prediction that they come from the steep Dohnanyi-type portion of the scattered disk's SFD, which has a cumulative power law slope of $q = -2.7$. This value is consistent with the predicted shape of the projectile SFD suggested by Morbidelli et al. (2021), who argued the power law slope for small comets should be $q \sim -3$ (Sec. 6.1.2).

**Impacts on Saturn's Rings.** A similar test was performed to determine the flux of small bodies striking Saturn's rings. We found that the impact frequency of $0.01 < D < 1$ m bodies was consistent with constraints provided by specific ring images from the Cassini mission. The impactor sizes were deduced by the Cassini team from tiny clouds of debris, which were interpreted to be ejecta from small collision events on ring bodies (Sec. 6.1.3).

**Crater SFDs on Giant Planet Satellites.** Using the results of crater scaling laws with input parameters verified on asteroids like Mathilde and Ceres, we compared crater SFDs produced by our model to those found on the satellites of Jupiter, Saturn, and Uranus. We found general agreement between model and observations for our larger craters. The mismatches that did exist predominantly came from smaller craters on worlds where secondary and/or sesquinary craters may be important (Sec. 6.2.1-6.2.3).

**The Shallow SFD on Charon and Arrokoth.** We discussed the shallow SFD found on Charon ($q = -0.7 \pm 0.2$; Robbins and Singer 2021) for $D < 10$-15 km craters. While we did not model the collisional evolution of the Kuiper belt in this paper, we find that the shape of the destabilized population's SFD is larger than this value ($q = -1.2$). Conversely, our value was consistent with prediction made from an analysis of Arrokoth craters by Morbidelli et al. (2021). Additional modeling will be needed to see if Kuiper belt collisions can explain crater observations on Charon (Sec. 6.2.4).

**Neptune Trojans.** Observations of the Neptune Trojans indicate they have a shallower SFD for $D < 100$ km bodies than the Jupiter Trojans. This outcome is consistent with their capture from the early PKB/destabilized population. Collisional evolution is likely to be limited near 30 au, and thus the Neptune Trojan SFD may be relatively unchanged from early times (Sec. 6.4)

**Comparisons with Comet SFDs.** Our model SFD for the destabilized population was compared to debiased observations of the Jupiter-family comet (JFC) and long period comet (LPC) SFDs, which have modestly different shapes. We found the shape of the model SFD from the present-day was a good match to the debiased population of JFCs for $D > 2$ km bodies. Recall that these comets come from the scattered disk, a population that has been collisionally-coupled to the Kuiper belt for 4.5 Gyr. For $D < 2$ km objects, we found our results increasingly overestimate the observed population as we go to smaller sizes (Sec. 6.5.1). Given that our model can reproduce young crater SFDs on Europa and Ganymede, we suspect the difference is most easily explained by observational incompleteness and/or small comets disrupting as they approach the Sun

The shape of our model SFD was also able to reproduce that of the debiased $D > 2$ km LPC SFD found by Pan-STARRS1 (Boe et al. 2019), provided that we use collisional model results taken shortly after the giant planet instability. This implies that comets sent to the Oort cloud only experienced modest collisional evolution prior to their emplacement. For $D < 2$ km LPCs, we found our results overestimated observational constraints. We could mean that sub-km LPCs experienced substantial physical evolution and/or disruption en route to perihelion (Sec. 6.5.2).

**Centaurs**. Using a destabilized population/scattered disk that followed a Jupiter Trojan-like SFD for $D < 100$ km bodies, Nesvorný et al. (2019a) used the OSSOS survey simulator to predict how many Centaurs with $D > 10$ km should have been detected to date. They found 11 ± 4 object, whereas the observed number was 15. Here we substituted in the model SFD from our work, which has a "valley" near $D \sim 10$ km. This decreased the OSSOS detections to 7 ± 3. While still within a factor of 2 of the observed value, this does represent a mismatch. The origin of this discrepancy will require further investigation (Sec. 6.5.3).

**Acknowledgements**

We thank Michelle Kirchoff, Paul Schenk, and Peter Thomas for giving us access to their crater databases, which was extremely helpful for this project. We also thank our two anonymous referees for their useful and constructive comments that made this a better paper. The work in this paper was supported by NASA's Solar System Workings program through Grant 80NSSC18K0186, NASA's Lucy mission through contract NNM16AA08C, and NASA's NEO Surveyor mission through NASA Contract 80MSFC20C0045. The work of David Vokrouhlický was partially funded by grant 21-11058S from the Czech Science Foundation.

**Appendix**

**A.1 Hydrated Materials in Interplanetary Dust Particles and Comets**

Interplanetary dust particles (IDPs) are micron- to cm-sized objects that come from asteroids, comets, and possibly KBOs (e.g., Nesvorný et al. 2010). This makes them a potential source of information about the nature of objects that formed and evolved beyond the original orbit of Neptune.

Most IDPs larger than a few microns are driven inward toward the Sun by the effects of Poynting-Robertson (P-R) drag, which decreases both their semimajor axes and eccentricities. This creates a disk of IDPs, called the Zodiacal cloud, whose thermal emission can be observed using space-based telescopes (e.g., the Infrared Astronomical Satellite (IRAS), Spitzer). It also means that a small fraction of IDPs evolve far enough toward the Sun that they can impact Earth. Laboratory studies of IDPs suggest the majority have bulk compositions are similar to carbonaceous chondritic meteorites, specifically CIs and CMs (e.g., Bradley et al. 2003).

IDPs are often defined by their physical properties and are placed into two broad categories: chondritic porous (CP) and chondritic smooth (CS) IDPs (Bradley et al. 2003). CP IDPs look like fractal aggregates and have porosities as high as 70%. They are anhydrous materials that are either from anhydrous parent bodies or from regions of hydrous parent bodies where aqueous alteration was negligible. CS IDPs have hydrated silicates (clays) and occasionally carbonates. They probably come from parent bodies where aqueous alteration has occurred.

Traditionally, it has been assumed that CP IDPs, being predominantly anhydrous, come from comets, all which started their lives in the PKB and reached the inner solar system via the scattered disk or Oort cloud. The morphology and mineralogy of some CP IDPs also hint at connections to ice-rich parent bodies (Rietmeijer 2004; Zolensky et al. 2006). In this view, their parent bodies form with limited radiogenic nuclides like $^{26}$Al and are therefore too cold to produce flowing water and hydrated materials. Some CP IDPs could also come from the main asteroid belt, where comet-like bodies, in the form of D- and P-type asteroids, are found in substantial numbers. This connection also matches up with the spectral characteristics of these bodies (Bradley et al. 1996; Vernazza et al. 2015).

Following this logic, the hydrous materials in CS IDPs would need to come from an alternative source where aqueous alteration is possible, such as carbonaceous asteroids that now reside in the main belt. Several lines of evidence indicate that many CM or CI-like parent bodies in the main belt experienced aqueous alteration (e.g., studies of the links between IDPs and asteroid spectra; the presence of carbonate veins on Bennu; analysis of returned samples from Ryugu by Hayabusa2; Vernazza et al. 2015; Kaplan et al. 2020; Yokoyama et al. 2022).

The sources of IDP populations coming from the main belt and comets have been explored using dynamical models constrained by IRAS and Spitzer observations. Here we focus on modeling results and predictions from Nesvorný et al. (2010). They showed that IDPs from the main belt objects start with relatively low inclinations, like their parent objects. As they evolve inward by P-R drag, their ecliptic latitudes remain low. IDPs from Jupiter family comets (JFCs) have a broader distribution of ecliptic latitudes, and nearly-isotropic comets (NICs) have the broadest distribution of all. These distributions, when fit to IRAS observations of the Zodiacal cloud, indicate that 85% to 95% of the observed mid-infrared emission is produced by particles from JFCs. Less than 10% of the remaining contributions comes from NICs or the main belt, and the typical size of the particles that contribute to this emission are 100 µm in diameter. Further modeling reveals that > 85% of the total mass flux reaching Earth from IDPs is likely to come from disrupted Jupiter-family comets (though the relative importance of JFC and Kuiper Belt particles beyond Jupiter has yet to be quantified). The combined contribution from main belt bodies and NICs for these IDPs is therefore limited to < 15%. Accordingly, our expectation would be that > 90% of all IDPs should be anhydrous CP-type IDPs and < 10% would be hydrous CS-type IDPs.

The fraction of hydrous IDPs collected from the stratosphere, however, instead varies from about 20% to 35% (e.g., Genge et al. 2020; Keller and Flynn 2022). This value is substantially higher than the < 10% limit imposed by Nesvorný et al. (2010), and it could suggest a substantial fraction of IDPs from Jupiter-family comets and/or the Kuiper belt are hydrous.

Additional support for hydrous materials from comets and/or KBOs comes from the work of Keller and Flynn (2022), who investigated the exposure ages of IDPs to solar energetic particles. They found that many IDPs spent > 1 Myr in space prior to reaching Earth, and that makes it difficult for them to come directly from main belt asteroids or Jupiter-family comets; assuming their typical sizes are 20 µm, their P-R drift timescales are far too short. Instead, Keller and Flynn (2022) postulate these long exposure age IDPs come from the Kuiper belt. They assert that such materials comprise ~25% of all of the IDPs analyzed in NASA's stratospheric dust collections, and that nearly half of these were hydrous. While this value is not debiased, it nevertheless points in the direction that a major source of hydrous IDPs could be the Kuiper belt, either directly or indirectly through JFCs.

Phyllosilicates are relatively rare among the observed comets (Davidsson et al. 2016), but they have been noticed from time to time (e.g., spectral features on 9P/Tempel 1 and C/1995 O1 Hale-Bopp; Lisse et al. 2006, 2007). Hydration features have also been detected on a few KBOs (de Bergh et al. 2004). They seem to be more noticeable on irregular satellites, which are thought to be captured KBOs (Nesvorný et al. 2008). Himalia, the largest irregular of Jupiter, appears to be hydrated and is similar to C-type asteroids (Brown and Rhoden 2014; Takir and Emery 2012). Phoebe, the largest irregular of Saturn, has phyllosilicates exposed on its surface (Clark et al. 2005), with apparently all regions showing water bands (Fraser and Brown 2018). Sycorax, the largest irregular of Uranus, has a 0.7 µm spectral feature consistent with hydrated silicates (Sharkey 2023). Collectively, these bodies hint at a more complicated story for the source of hydrated IDPs.

Keller and Flynn (2022) argue that most aqueous altered IDPs are derived from low-velocity collisional processes taking place on the surfaces of comets and KBOs. Here impacts would produce transient heating events that melt ice and yield aqueously altered materials. For larger KBOs, these processes would presumably make them look like present-day Himalia, Phoebe, etc. A potential issue with this scenario is whether most IDPs are indeed from surface collisions. Consider that the majority of IDPs coming from the main belt are derived from the disruption of a few modest-sized asteroids (e.g., Nesvorný et al. 2003), while many IDPs observed in the Zodiacal cloud may be coming from the disruption of Jupiter family comets (Nesvorný et al. 2010). In both cases, limited ejected material would be residing near the surface. Clearly additional modeling is needed here to explore this issue.

We postulate that another way to explain these results would be to put them into the context of our collisional model results. We have argued that most of the mass of the initial PKB was in the form of ~100 km bodies (Sec. 4). Some bodies of this size or larger likely formed early enough to heat up from active radiogenic elements like $^{26}$Al, which could lead to flowing water and aqueous altered materials within their deep interiors. Their exteriors, however, would presumably remain primitive to a substantial depth, with low temperatures preventing water from percolating up to the near surface.

An example of such a body could be (87) Sylvia, a 280 km diameter P-type asteroid in the outer main belt that was likely captured from the PKB (Vokrouhlický et al. 2016). While Sylvia's exterior is spectrally similar to anhydrous IDPs, studies of its gravitational interactions with its satellites point to a differentiated interior (Carry et al. 2021). In fact, in thermal modeling work done by Carry et al. (2021), they suggest Sylvia has a three-layer structure: a central region dominated by a muddy ocean, surrounded by a porous layer free of water, and then a primordial outer layer that remains too cold for ice to melt.

Our model results show that disruptive collisions among 100-300 km diameter bodies create the fragments that produce the wavy shape of the destabilized population's SFD (and that of the Kuiper belt). If some of these bodies have Sylvia-like internal structures, disruptive collisions should mix interior and exterior materials in the largest remnant of the parent body and the ejected fragments. The degree of hydrous materials escaping in the form of fragments would depend on many factors: the accretion time and size of the parent body, the nature of the collision event dredging up aqueous altered materials, etc. Given that hydrated material is probably deep within large KBOs, our expectation is that most comets will be dominated by samples of the exterior, as is the case for asteroids in asteroid families (e.g., DellaGiustina et al. 2020).

An interesting test of our hypothesis may come from NASA's upcoming Lucy mission. It will be interesting to see whether Jupiter Trojan (3548) Eurybates, the largest remnant of the Eurybates family and a body with C-type spectra (Marshall et al. 2022), has exposed phyllosilicates, and if so, whether the putative source can be identified.

**A.2 Comments on Comets as the Byproducts of Collisions**

Our results may be useful to the ongoing debate about whether comets are primordial or fragments of larger bodies. For the former, primordial means that they were formed directly by planetesimal processes in the solar nebula, such that their shapes, compositions, etc. are mainly from that provenance. For the latter, a fragment means they were formed by a cratering or disruption event taking place on a larger body. Davidsson et al. (2016) refers to the latter comets as "collisional rubble-piles" produced by the gravitational re-accumulation of debris. That characterization may not be a good fit for all bodies made in collisions. For example, as defined by Richardson et al. (2002), certain objects within asteroid families, like (243) Ida or (951) Gaspra, might be better described as fractured or shattered objects rather than rubble piles, though all are gravitational aggregates. For the moment, we will assume that fragments of larger collisions fall somewhere on this spectrum of possibilities.

In the aftermath of studies of comet 67P/Churyumov-Gerasimenko by the Rosetta mission team, there have been several papers debating the nature of observed comets, most which are smaller than 10 km (Lamy et al. 2004; Belton 2014). Arguments favoring the pro-fragment case have been made by many groups (e.g., Michel and Richardson 2013; Morbidelli and Rickman 2015, Rickman et al. 2015, Jutzi et al. 2016, Schwartz et al. 2018, Campo Bagatin et al. 2020; Benavidez et al. 2022), while a pro-primordial case was made in an omnibus paper by Davidsson et al. (2016). Our results favor the pro-fragment case. For example, our results indicate many $D < 10$ km bodies in the destabilized population are fragments produced by collisions (Fig. 10). Moreover, our destabilized population's SFD is consistent with the shape of the debiased Jupiter family comet SFD in the present day (Fig. 17) and the long period comet SFD shortly after the giant planet instability (Fig. 18) (see Sec. 6.5).

Some pro-primordial arguments focus on the bilobed shape of comets, which could be produced by in a collapsing cloud of pebbles (e.g., the hamburger-like shapes of the two lobes from the 20 km object Arrokoth is probably primordial; McKinnon et al. 2020). Many comet shapes observed by spacecraft, however, also look analogous to contact binaries in the near-Earth asteroid population; numerical modeling work has shown that collisions can create such shapes (e.g., Rickman et al. 2015; Jutzi and Benz 2017; Schwartz et al. 2018; Campo Bagatin et al. 2020). We would also argue most NEOs are likely to be collisional aggregates produced by large main belt collisions (e.g., Bottke et al. 2005a,b).

Other pro-primordial arguments focus on the paucity of aqueous altered materials in comet 67P/Churyumov-Gerasimenko. As discussed in the last section, this could be explained by this comet coming from the primitive exterior of a KBO, or from a KBO that formed late enough that it experienced minimal aqueous alteration. Accordingly, the presence of exotic ices or supervolatiles on a comet does not necessarily mean that the object could not be a collisional fragment, especially since studies have shown that at smaller scales ($D < 10$ km), collisions do not produce substantial heating of the largest fragments (e.g. Jutzi and Benz 2017; Jutzi et al. 2017; Schwartz et al. 2018).

Our model results also counter additional pro-primordial arguments made by Davidsson et al. (2016). For example, our preferred starting conditions for the PKB indicates that smaller primordial comets are not a favored size of planetesimal formation mechanisms (Klahr and Schreiber 2020; 2021), though it is possible some might form this way. In addition, during planetesimal formation, as small particles gravitationally collapse to form predominately $D \sim 100$ km planetesimals, it can be shown that smaller gravitational aggregates form and are sometimes ejected from the system. Either scenario might explain ~20 km diameter objects like Arrokoth (Nesvorný et al. 2021).

Finally, our work indicates that collisional evolution can reproduce the observed SFD in the main belt, destabilized population, and in the Trojans within a single dynamical and collisional framework, provided most of the mass in the SFD is in the form of $D \sim 100$ km bodies. This removes the need for numerous $D < 10$ km comets created by primordial processes, and it sets up interesting predictions that can be tested by future missions.

**A.3 Comments on the Sizes of the Interstellar Comets 'Oumuamua and Borisov**

The discovery of the first interstellar comet on October 19, 2017 by the PanSTARRS survey sent shock waves through the astronomical community (Meech et al. 2017). Long anticipated but never seen, this object, named 1I/'Oumuamua, provides us with critical clues about the nature of planetesimal formation and small body evolution around other stars.

The origin of 'Oumuamua has been debated with vigor, with both simple and exotic mechanisms used to explain how its trajectory was influenced by non-gravitational forces during its close approach to the Sun (e.g., Bialy and Loeb 2018; 'Oumuamua ISSI Team 2019; Seligman and Laughlin 2020; Desch and Jackson 2021). In some ways, though, 'Oumuamua can be considered a standard comet, with colors consistent with Jupiter family comet nuclei and D-type asteroids (Jewitt and Luu 2019). Its most unusual physical property may be its axis ratio, which is thought to be 6:1 ('Oumuamua ISSI Team 2019). With that said, the object is small; its absolute magnitude ($H$) is 22.4 ± 0.04, which, for an albedo of 0.04 to 0.1, translates into a diameter of 140-220 m. We have little to no information on the nature of such small comets in our own solar system, so it is impossible to say whether 'Oumuamua's elongated or perhaps hamburger-like shape is unusual.

A second interstellar comet, named 2I/Borisov, was found on August 30, 2019 by amateur astronomer Gennadiy Borisov. Like 'Oumuamua, it has colors that are essentially identical to the mean color measured for the dust comae of long-period comets (Jewitt and Luu 2019). Borisov does not have appear to have a bizarre shape; estimates of its nucleus size suggest it is ~0.8 to ~1.0 km in diameter (Jewitt and Luu 2019; Hui et al. 2020). This value is substantially larger than 'Oumuamua.

It is surprising that the first two detected interstellar objects have such a wide range of sizes. One factor that might explain this would be their discovery circumstances: 'Oumuamua was discovered three days after its closest approach to Earth at 0.16 au, while Borisov was discovered near 3 au. Another factor would be the activity of the bodies at discovery; Borisov had a sizeable coma, while 'Oumuamua did not show any sizes of comet activity, such as coma or tail. Debiasing these two detections is beyond the scope of this paper, but as a thought experiment, we will assume here that their observation selection effects balance each other out. That allows us to explore the implications of their sizes using a Monte Carlo code.

We assumed that the SFD of interstellar comets reaching our system follow a generic power law with $P(D) = C D^n$, with $C$ being a constant, $D$ being diameter, and $n$ being the differential slope between -1.1 and -5.0. This makes the cumulative slope $q = n + 1$. In each trial, we used random deviates to extract two bodies from the power law SFD between 0.1 and 5 km and then determined whether they were between 0.14-0.22 km or 0.8-1.0 km. A successful run had one body in each size range. We tested each value of $n$ over $10^6$ trials and tallied the results. The relative probabilities of success are shown in Fig. A1.

PLACE FIGURE A1 HERE

Our results suggest that our two interstellar objects are far more likely to come from a shallow sloped SFD than a steep one. The relative probabilities of success for $q = -1$ are 3.8, 28, and 200 times more likely than $q = -2$, $-3$, or $-4$, respectively. The most successful values are similar to those found in model runs for this size range, where $q \sim -1.2$ (Fig. 10).

Accordingly, if the two interstellar objects are not enormously biased by observational selection effects relative to one another, and one can cope with statistics of small numbers, 'Oumuamua and Borisov provide us with insights into small body evolution in exoplanet systems. For example, our inferred shallow slope from Fig. 19 could be telling us that many interstellar comets are byproducts of collisional evolution, unless planetesimal formation processes naturally make a shallow SFD between 0.1 and 1 km. We argue this makes sense, since to be ejected from their home system, these comets probably had to have a giant planet encounter. Getting from their source region to the giant planet almost certainly involved some kind of dynamical excitation, which in turn implies collisional evolution within the source region.

Our results also hint at the possibility that collisional processes act in similar ways in various exoplanet systems, regardless of disk mass. This makes sense, in that whether we are dealing with asteroids or comets, their SFDs should develop a steep Dohnanyi-like SFD at small sizes, a shallow SFD for sub-km bodies, and a steep SFD for objects $\sim 1 < D < 10$ km bodies. Some evidence for this may be found in the inferred slope of exocomets within the Beta Pictoris system, whose comet diameters between $3 < D < 8$ km follow a cumulative power law slope $q = 2.6 \pm 0.8$ (Lecavelier des Etangs et al. 2022). This value is consistent with our results (Fig. 10) and those of the known comets (Figs. 17 and 18).

The other interesting issue concerns the highly elongated shape of 'Oumuamua. Its origin is a mystery. Some suggest it was made by tidal disruption of a comet-like object (Cuk 2018; Raymond et al. 2018; Zhang and Lin 2020), while others assert it could be made by "sandblasting" a modestly elongated progenitor with small particles over long time periods (Vavilov and Medvedev 2019). We offer no solution to this problem, but if small interstellar objects are derived from SFDs with a shallow power law slope, it could be that collisions played some role in 'Oumuamua's initial and perhaps final shape. Testing this possibility might lead to an interesting line of future research.

| Initial Size Frequency Distribution of the Primordial Kuiper Belt (Fig. 3) | | |
|---|---|---|
| $D$ (km) | $q$ | Notes |
| < 70 | -1.1 | Motivated in part by assumed shallow starting slope for main belt (Bottke et al. 2005a,b) and observations of cold classical KBOs (Napier et al. 2023). |
| 100 to 300 | -5.0 | Slopes identified in KBOs and Trojan SFD (Nesvorny and Vokrouhlicky 2016). |
| 300 to 2000 | -2.5 | Inferred slope needed for inferred SFD to reach ~2000 Pluto-sized objects (Nesvorny and Vokrouhlicky 2016). |
| Size Frequency Distribution of Destabilized Population at 40 Myr (Fig. 10) | | |
| < 0.02 | -2.7 | Evolution to Dohnanyi-type slope (O'Brien and Greenberg 2003). Slope observed in small body impact rate on Jupiter and Saturn's rings (Fig. 12). |
| ~0.1 to ~1 | -1.2 | Objects decimated by small bodies in steep slope (O'Brien and Greenberg 2003). Shallow slope similar to main belt SFD over comparable size range (Fig. 16). Slope observed in $D < 10$ km crater SFDs found on several worlds (e.g., Europa, Hyperion, Phoebe, Miranda) (Figs. 13-15). |
| ~1 to ~10 | -2.4 | Lack of $D < 1$ km projectiles leads to steeper slope (O'Brien and Greenberg 2003) Slope observed in $10 < D < 100$ km crater SFDs found on Callisto and many mid-sized satellites of Saturn and Uranus (Figs. 13-15). Slope observed on long period comet SFD with $D > 2$ km (Fig. 18). |
| Size Frequency Distribution of Destabilized Population at 4300 Myr (Fig. 10) | | |
| ~3 to ~10 | ~-2.5 | Slope becomes steeper; inflection point moves from ~1 km to ~2 km. Slope possibly found on Ganymede crater SFD (Schenk et al. 2004). Slope observed on Jupiter family comet SFD with $D > 2$ km (Fig. 17). |
| Size Frequency Distribution of Trojans at 4500 Myr (Fig. 11) | | |
| ~1 to ~5 | -1.8 | Objects decimated by smaller bodies that follow steeper slope. Shallow slope detected in L4 Trojans (see summary in Uehata et al. 2022) (Fig. 2). |
| ~10 to 50 | ~ -2 | Slight wave still remains from captured SFD. Feature may exist in Trojan SFD from WISE/NEOWISE observations (Grav et al. 2011). |

**Table 1. Discussion of some slopes for model destabilized population and Jupiter Trojans.**

The first column describes diameter range in question, while the second column as the cumulative power law slope found in that range. The third column describes the origin of the slope and possible connections to other data.

| Parameter | Tested Values |
|---|---|
| $Q^*_{D\,LAB}$ | $6.20 \times 10^4$, $2.65 \times 10^5$, $1.13 \times 10^6$, $4.84 \times 10^6$ erg g$^{-1}$ |
| α | [0, -1] incremented by -0.1 |
| β | [0.1, 2] incremented by 0.1 |
| $D_{min}$ | [10 m, 100 m], incremented by 10 m |
| $\Delta t_0$ | [0 Myr, 30 Myr], incremented by 10 Myr |
| $\Delta t_1$ | 10.5, 32.5 Myr |

**Table 2. Parameters for $Q^*_D$ disruption functions.**

See text for definitions. The time between the dispersal of the solar nebula and Neptune entering the PKB is defined by $\Delta t_0$, while the time between Neptune entering the PKB and the giant planet instability taking place is defined by $\Delta t_1$. Put together, this represents 70,400 trials to be tested against the IP SFD and another 70,400 trials to be tested against the Jupiter Trojan SFD (Fig. 2).

| Best Fit Rank | $\Delta t_0$ (Myr) | $\Delta t_1$ (Myr) | $Q^*_{D\,LAB}$ (erg g$^{-1}$) | $D_{min}$ (m) | $\alpha$ | $\beta$ | IP SFD $\chi^2$ | Trojan $\chi^2$ | Combined $\chi^2$ |
|---|---|---|---|---|---|---|---|---|---|
| 1 | 10 | 10.5 | $6.20 \times 10^4$ | 20 | -0.3 | 1.1 | 2.40 | 0.46 | 1.11 |
| 2 | 20 | 10.5 | $6.20 \times 10^4$ | 20 | -0.4 | 1.2 | 1.29 | 0.95 | 1.23 |
| 3 | 0 | 32.5 | $6.20 \times 10^4$ | 20 | -0.3 | 1.1 | 2.89 | 0.50 | 1.44 |
| 4 | 20 | 10.5 | $1.13 \times 10^6$ | 40 | -0.6 | 1.1 | 1.81 | 0.80 | 1.46 |
| 5 | 20 | 10.5 | $4.84 \times 10^6$ | 40 | -0.8 | 1.0 | 1.69 | 0.87 | 1.47 |
| 6 | 20 | 10.5 | $2.65 \times 10^5$ | 30 | -0.5 | 1.1 | 1.87 | 0.83 | 1.55 |
| 7 | 10 | 10.5 | $2.65 \times 10^5$ | 30 | -0.5 | 1.1 | 2.09 | 0.76 | 1.59 |
| 8 | 20 | 10.5 | $1.13 \times 10^6$ | 40 | -0.7 | 1.1 | 1.86 | 0.91 | 1.69 |
| 9 | 20 | 10.5 | $2.65 \times 10^5$ | 30 | -0.4 | 1.1 | 2.66 | 0.68 | 1.80 |
| 10 | 10 | 10.5 | $6.20 \times 10^4$ | 20 | -0.4 | 1.2 | 1.88 | 0.97 | 1.82 |

**Table 3. Top 10 Best Fit Cases for $Q^*_D$ disruption functions**

A compilation of our top ten best fit cases from the parameters in Table 2. The combined $\chi^2$ is obtained by multiplying the best fit $\chi^2$ values from our IP SFD fits from 70,400 trials with those of our Trojan SFD fits over 70,400 trials.

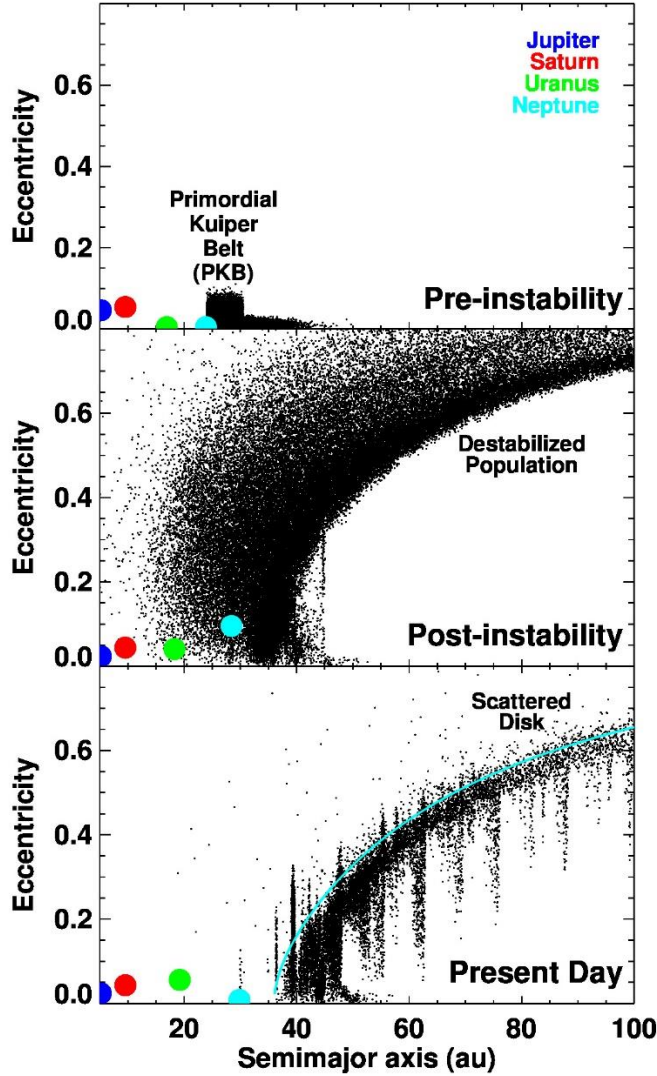

**Figure 1.** The evolution of the primordial Kuiper belt (PKB) in semimajor axis and eccentricity (*a, e*) over three snapshots in time (see Morbidelli et al. 2021). The orbits of Jupiter, Saturn, Uranus, and Neptune are shown as blue, red, green, and cyan dots, respectively. The top panel shows $t < \Delta t_0$, or the time just before Neptune enters the PKB. The PKB stretches from 24-50 au, but the spatial density of bodies decreases with distance from the Sun, with most of the mass located between 24-30 au. This means the cold classical Kuiper belt starts with its present-day mass. The middle panel shows $t > \Delta t_1$, or the time just after the giant planet instability. The bodies ejected onto Neptune-crossing orbits are defined as the destabilized population, and they can bombardment the giant planets and their satellites. A small fraction of them will eventually become Trojans, the irregular satellites, Oort cloud comets, and the scattered disk of Neptune. The bottom panel shows $t \gg \Delta t_1$. It is meant to show our current solar system, with Neptune and the Kuiper belt now in place. Only a remnant of the destabilized population is left behind in the scattered disk (i.e., above the cyan line). It feeds the Centaurs and the Jupiter family comets.

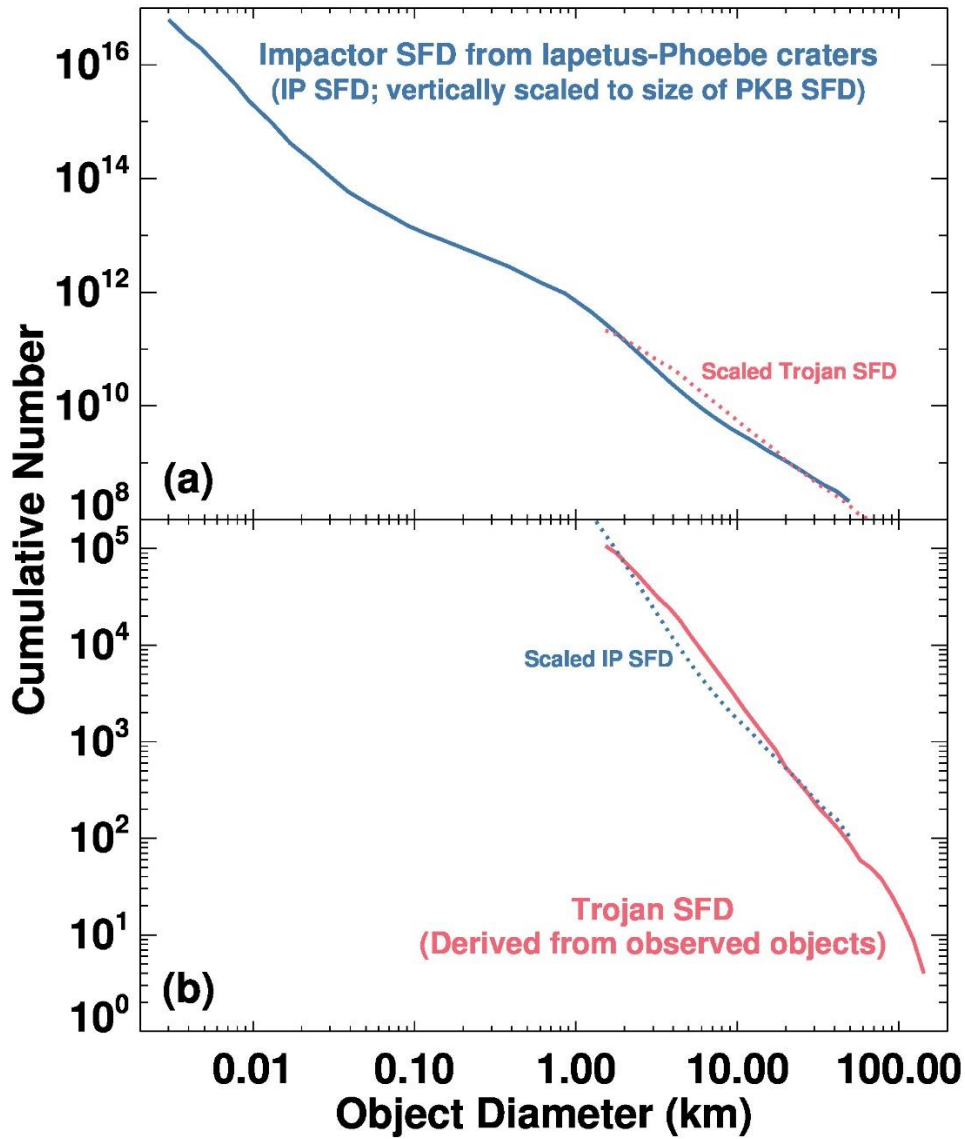

**Figure 2.** The size frequency distributions (SFDs) used to constrain our model destabilized population (blue) and Jupiter Trojans (red). (a) The projectile SFD derived from a combination of Iapetus and Phoebe craters (see Fig. 14 for crater data). Here we combine basins from Iapetus ($D_{\text{crat}} > 100$ km) and smaller craters from Phoebe ($D_{\text{crat}} < 100$ km). Projectiles were derived using the crater scaling law and impact velocities provided in Sec. 3.2.2. We call this the IP SFD in the text. A vertically scaled Trojan SFD is shown for comparison (red dashed line) (b) The Trojan SFD is derived from the known Trojans, assuming an albedo of 7% (Grav et al. 2011) and observations from Suprime-Cam/Hyper Suprime-Cam on the 9-meter Suburu telescope (Wong and Brown 2015; Yoshida and Terai 2017; Uehata et al. 2022). A vertically scaled IP SFD is shown for comparison (blue dashed line).

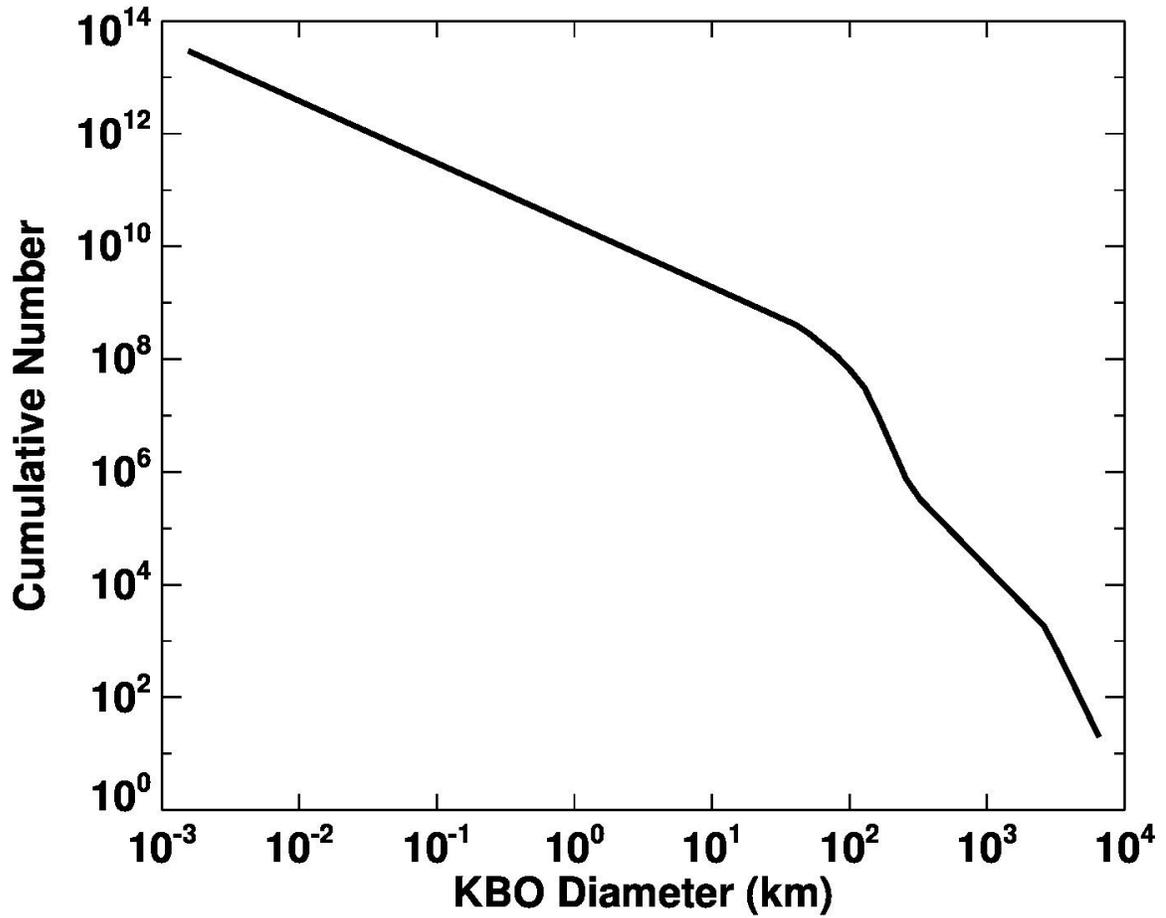

**Figure 3.** The starting size frequency distribution (SFD) of the PKB in our simulations. It has a shape similar to one provided by Nesvorný et al. (2018; see supplemental figures 2 and 4) (see also Nesvorný and Vokrouhlický 2016). We assume the population started with ~2000 Pluto-sized bodies and ~$10^8$ $D$ > 100 km bodies, values that allow the SFD to match dynamical constraints from many regions where KBOs have been captured (e.g., Kuiper belt, irregular satellites, Jupiter Trojans, outer asteroid belt, etc.). The shape for 100 < $D$ < 300 km was inferred from the Jupiter Trojans and Kuiper belt observations. The total mass of the PKB is 30 Earth masses. The cumulative power law slope for $D$ < 100 km bodies follows $q$ = -1.1, which means there are relatively few small bodies made by planetesimal formation processes.

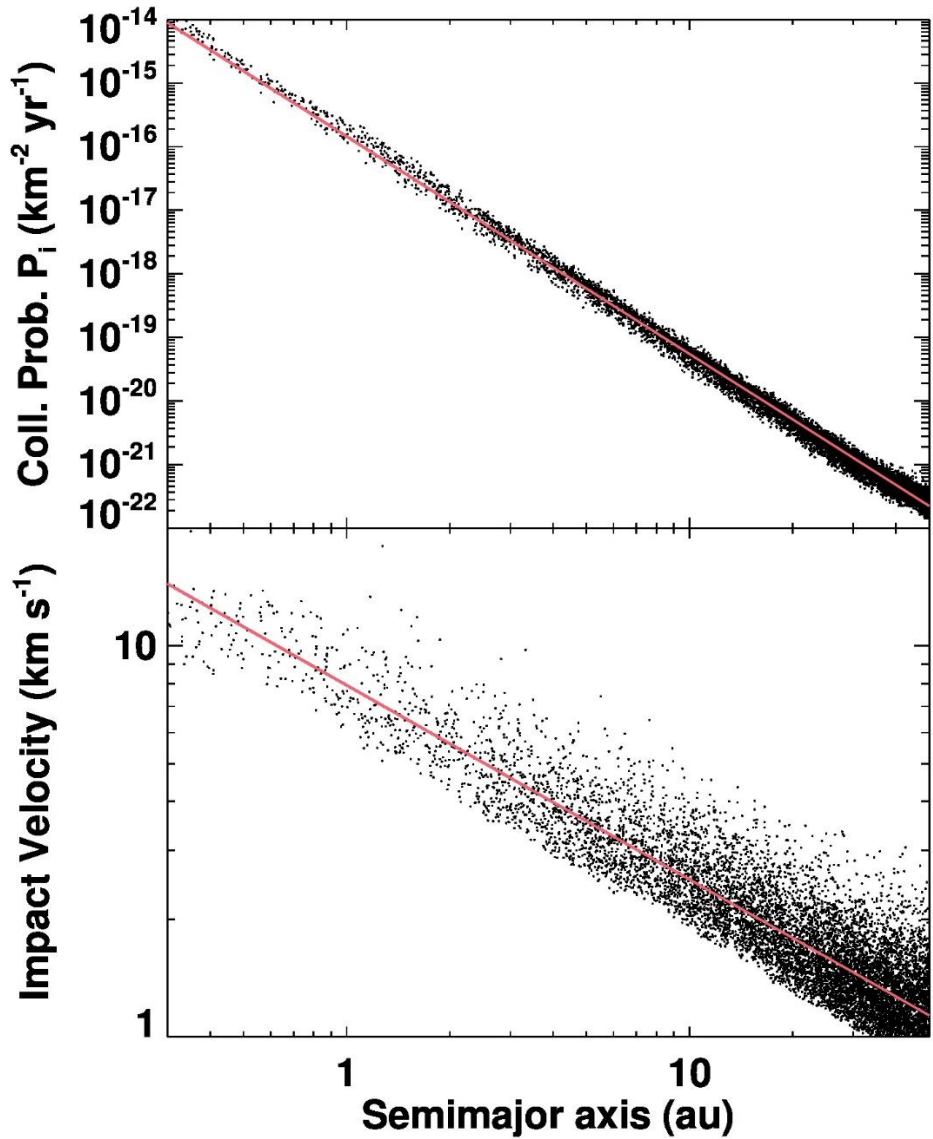

**Figure 4.** The intrinsic collision probabilities ($P_i$) (top) and impact velocities ($V_{imp}$) (bottom) of a Rayleigh distribution of 10,000 test bodies spread between 0.1 and 50 au in semimajor axis *a* with mean eccentricity *e* = 0.2 and mean inclination *i* = 0.1 rad. The values were calculated using the methodology found in Bottke et al. (1994). Here the intrinsic collision probabilities fall off with $P_i \propto a^{-3.5}$, while impact velocities decrease with a dependance $V_{imp} \propto a^{-0.5}$.

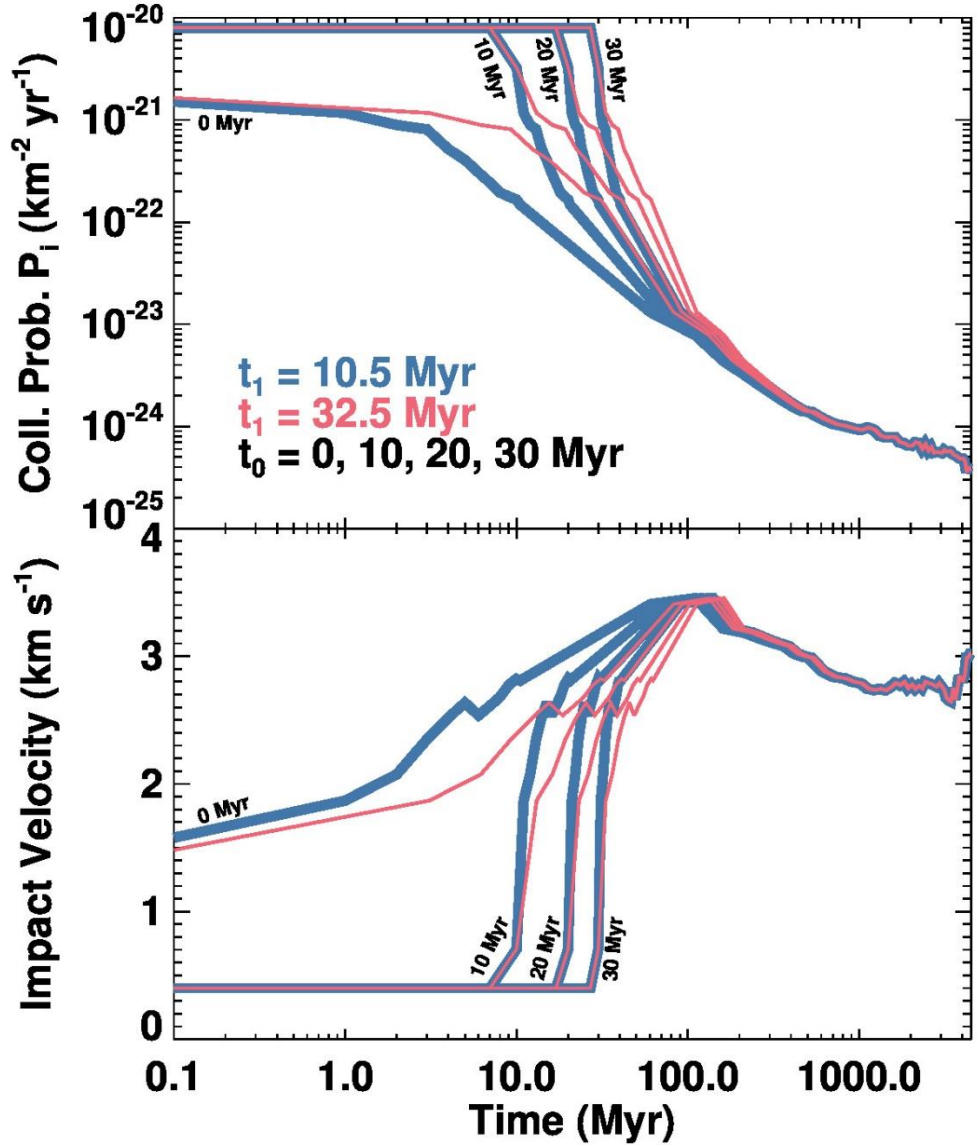

**Figure 5.** The collision probabilities ($P_{\mathrm{pop}}$) and impact velocities ($V_{\mathrm{imp}}$) for our chosen test bodies to strike the background population as they move from the PKB to the destabilized population over 4.5 Gyr (see Fig. 1). The intrinsic collision probabilities ($P_{\mathrm{i}}$) have been normalized by the starting test body population for use in the Boulder code. The different curves show the times $\Delta t_0$ when Neptune enters the PKB (i.e., 0, 10, 20, and 30 Myr). The blue and red colors correspond to the time $\Delta t_1$ for Neptune to undergo the giant planet instability (i.e., 10.5 and 32.5 Myr, respectively). Collision probabilities are highest at the beginning of the simulation before the PKB is dispersed. The impact velocities are highest when the destabilized population is most excited, but this is also when the population (and collision probabilities) are rapidly decreasing. At later times, collisional evolution slows down but does not cease because objects in the scattered disk can still be hit by each other and stable Kuiper belt objects.

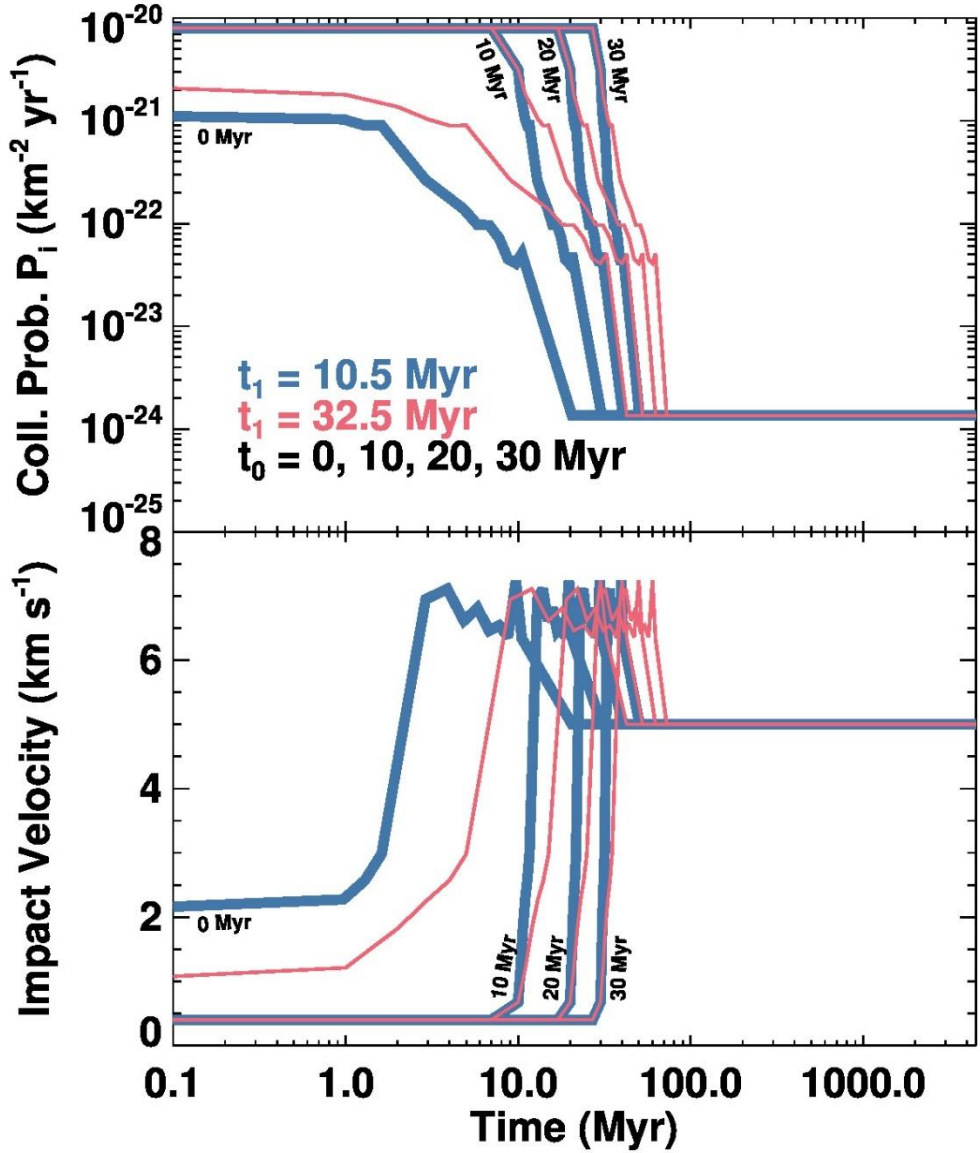

**Figure 6.** The collision probabilities ($P_{pop}$) and impact velocities ($V_{imp}$) for our chosen test bodies to become Jupiter Trojans. As in Fig. 5, the intrinsic collision probabilities ($P_i$) have been normalized by the starting test body population. The different curves show the times $\Delta t_0$ when Neptune enters the PKB (i.e., 0, 10, 20, 30 Myr). The blue and red colors correspond to the time $\Delta t_1$ for Neptune to undergo the giant planet instability (i.e., 10.5 and 32.5 Myr, respectively). Impact velocities reach their highest values when the test bodies get to the vicinity of Jupiter (e.g., Fig. 4). At late times, we assumed collisions are mainly with other Trojan asteroids.

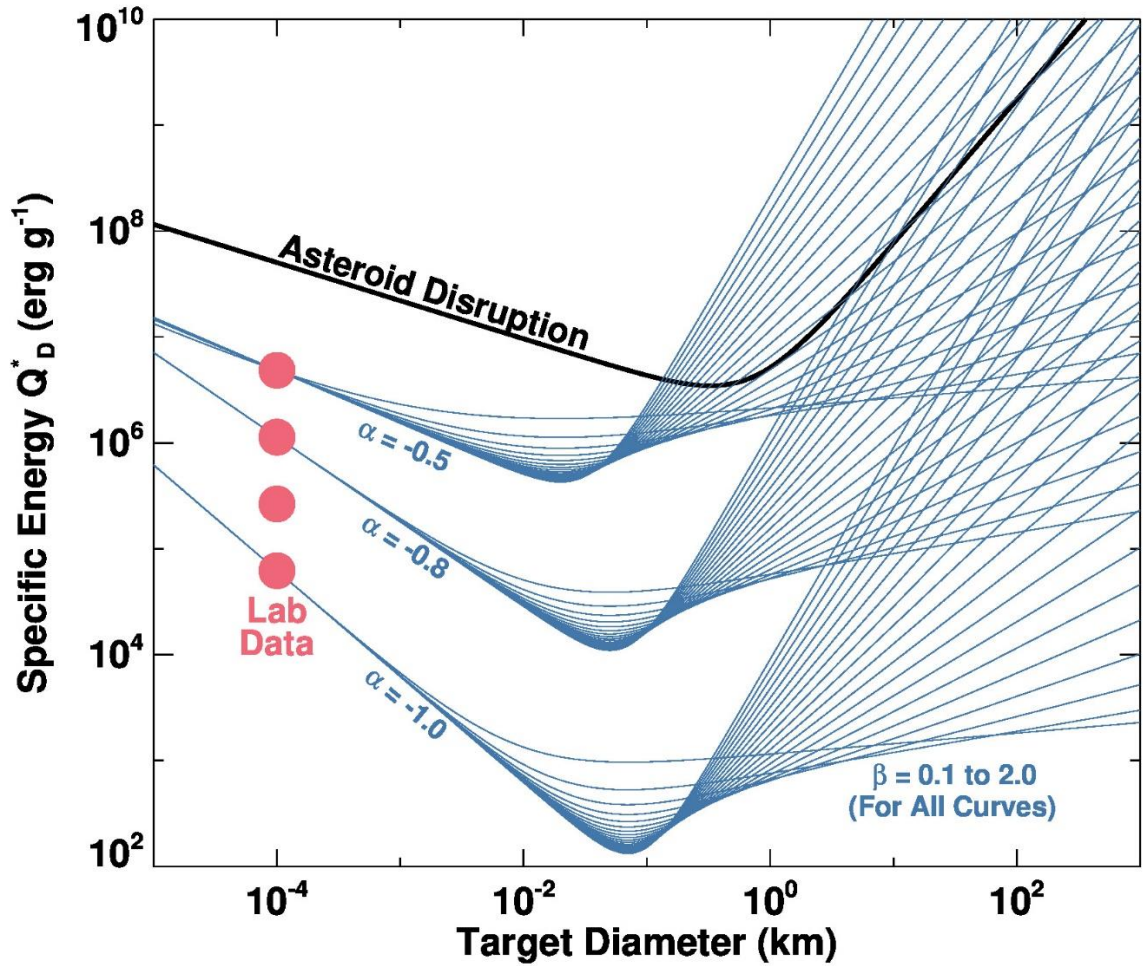

**Figure 7.** Examples of the KBO disruption laws ($Q_D^*$ functions) used in our collisional evolution model runs. Each function is defined by Eq. (5) using five parameters: $Q_{D_{LAB}}^*$, $D_{LAB}$, α, β, and $D_{\min}$. See Table 2 for the range of values explored in our runs. Here $D_{LAB}$ = 10 cm and $Q_{D_{LAB}}^*$ = 6.20 × 10$^4$, 2.65 × 10$^5$, 1.13 × 10$^6$, and 4.84 × 10$^6$ erg g$^{-1}$. The minimum of each curve, representing the easiest object to disrupt, is defined by $D_{\min}$. The values α and β provide the slopes of the left and right side of each $Q_D^*$ function, respectively. The asteroid disruption law for basaltic targets being hit at 5 km s$^{-1}$ from Benz and Asphaug (1999) is provided for reference.

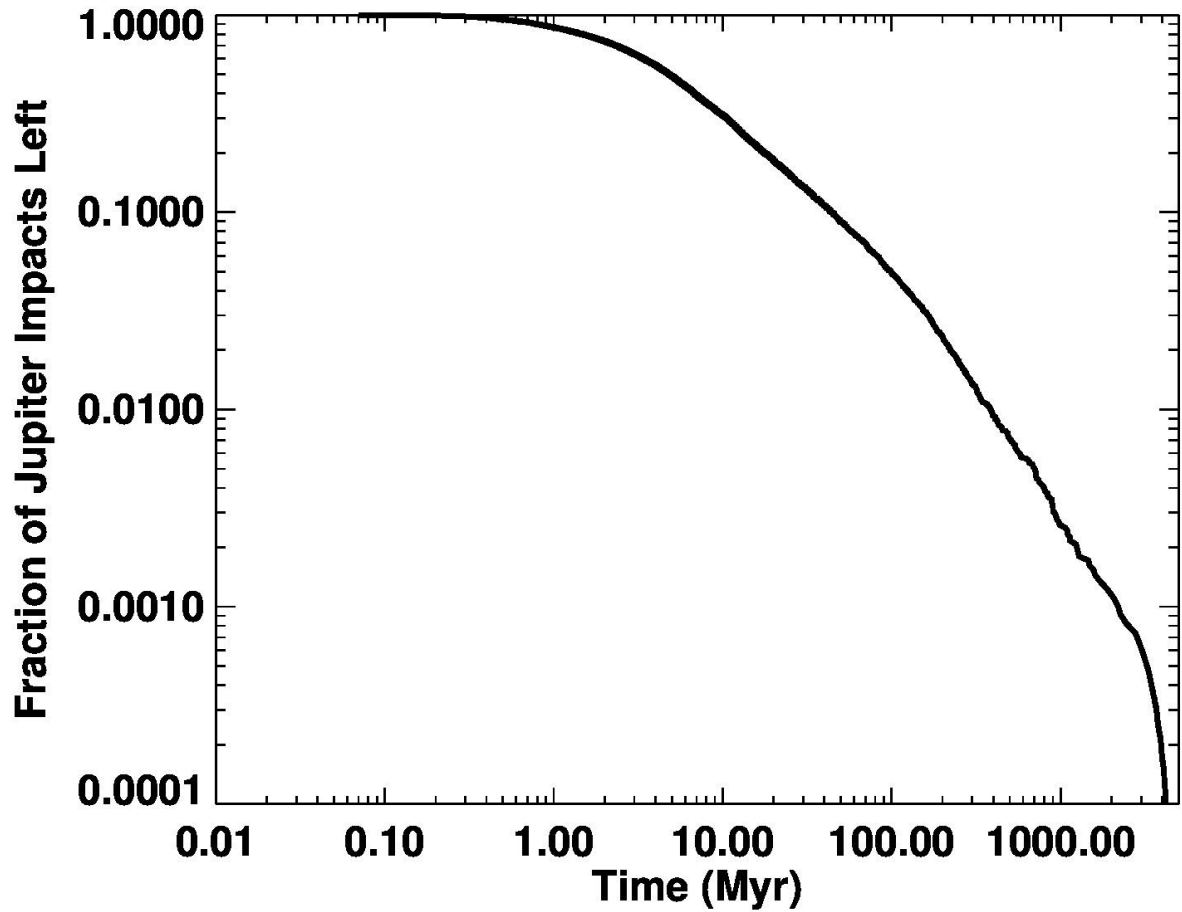

Figure 8. Fractional rate of impacts on Jupiter from the destabilized population and scattered disk (Fig. 1). Using the numerical simulations described in Nesvorný et al. (2019b), we tabulated the number and timing of test bodies striking Jupiter in the aftermath of Neptune's migration across the PKB and the giant planet instability. These values can be combined with our model SFDs of the PKB and destabilized population to get the impact rate on Jupiter over time. These values can then be scaled to estimate impact rates on the giant planet satellites using the methodology found in Zahnle et al. (2003).

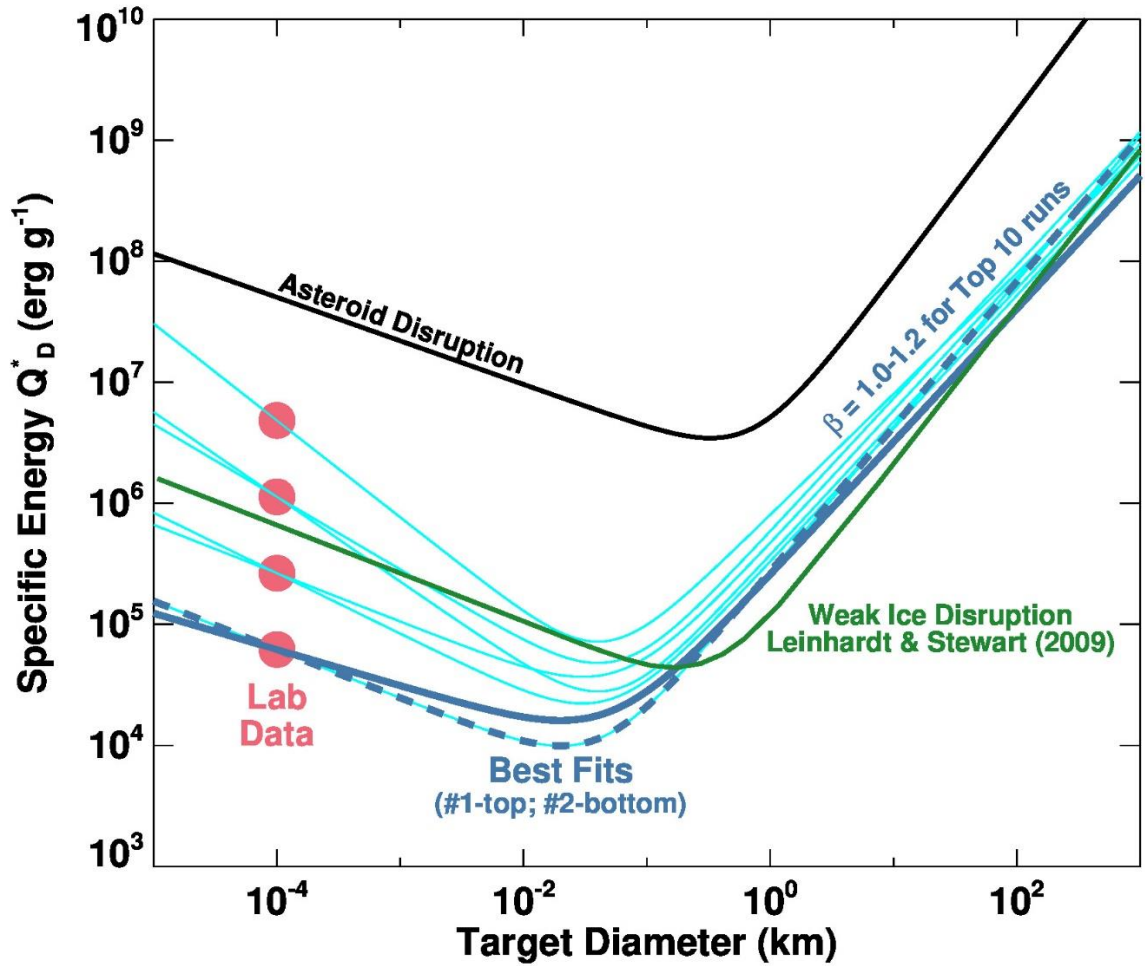

**Figure 9.** The top ten best fitting disruption laws ($Q_D^*$ functions) found for our collisional evolution model runs. See Fig. 7 for parameter definitions (including those for asteroid disruption law) and Table 3 for parameter choices. Our top two functions are the dark blue lines. They have $Q_{D_{LAB}}^* = 6.20 \times 10^4$ erg g$^{-1}$, $D_{\min}$ = 20 m, α = -0.3 to -0.4, and β = 1.1 to 1.2. The remainder are shown as cyan lines. They have a range of $Q_{D_{LAB}}^*$ and α values, but the other parameter values are more restricted (i.e., $D_{\min}$ = 20 to 40 m, β = 1.0 to 1.2). For reference, the green line shows the weak ice disruption law from Leinhardt and Stewart (2009). It is broadly consistent with our top ten $Q_D^*$ functions, but it assumes $D_{\min}$ = 200 m rather than values near 20 m.

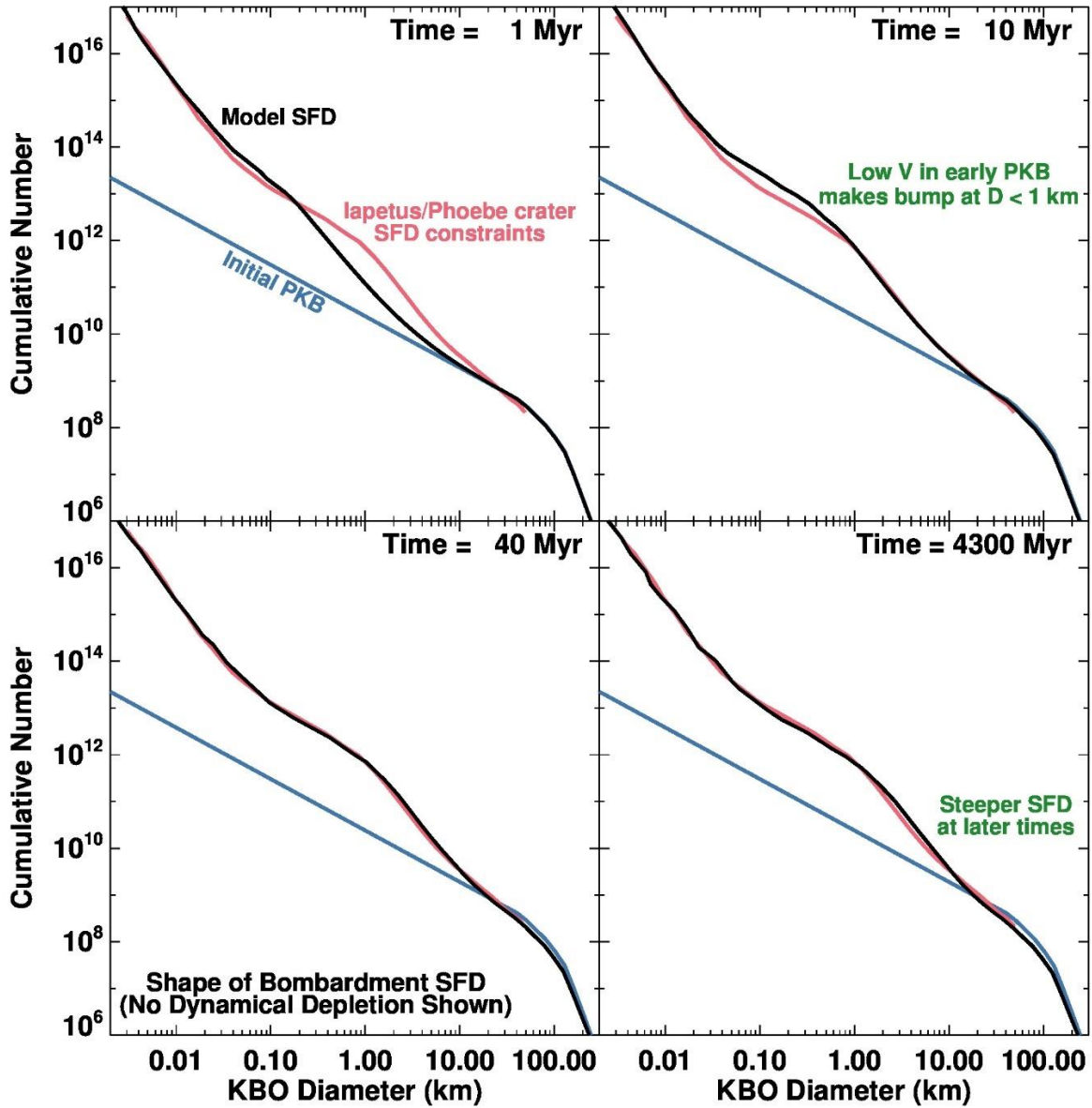

**Figure 10.** Four snapshots from the collisional evolution of the destabilized population, according to our best fit model SFD (Table 3, Run #2). This model reproduces the shape of the impactor SFD as determined from Iapetus and Phoebe craters (IP SFD, shown as the red line; see also Fig. 2). Here Neptune enters the PKB at $\Delta t_0 = 10$ Myr, while it takes $\Delta t_1 = 10.5$ Myr for Neptune's migration to trigger the giant planet instability. At 1 Myr, collisional evolution among the $D > 100$ km bodies creates numerous $D < 10$ km fragments that also undergo comminution. A sizable bump of fragments is produced at 10 Myr, but the low impact velocities in the PKB prevent a good fit to the IP SFD constraints. At 40 Myr, the match between the model SFD and the IP SFD is excellent. In the last timestep, collisional evolution over billions of years has caused the slope of the model SFD between $1 < D < 10$ km to become steeper than that observed in the IP SFD.

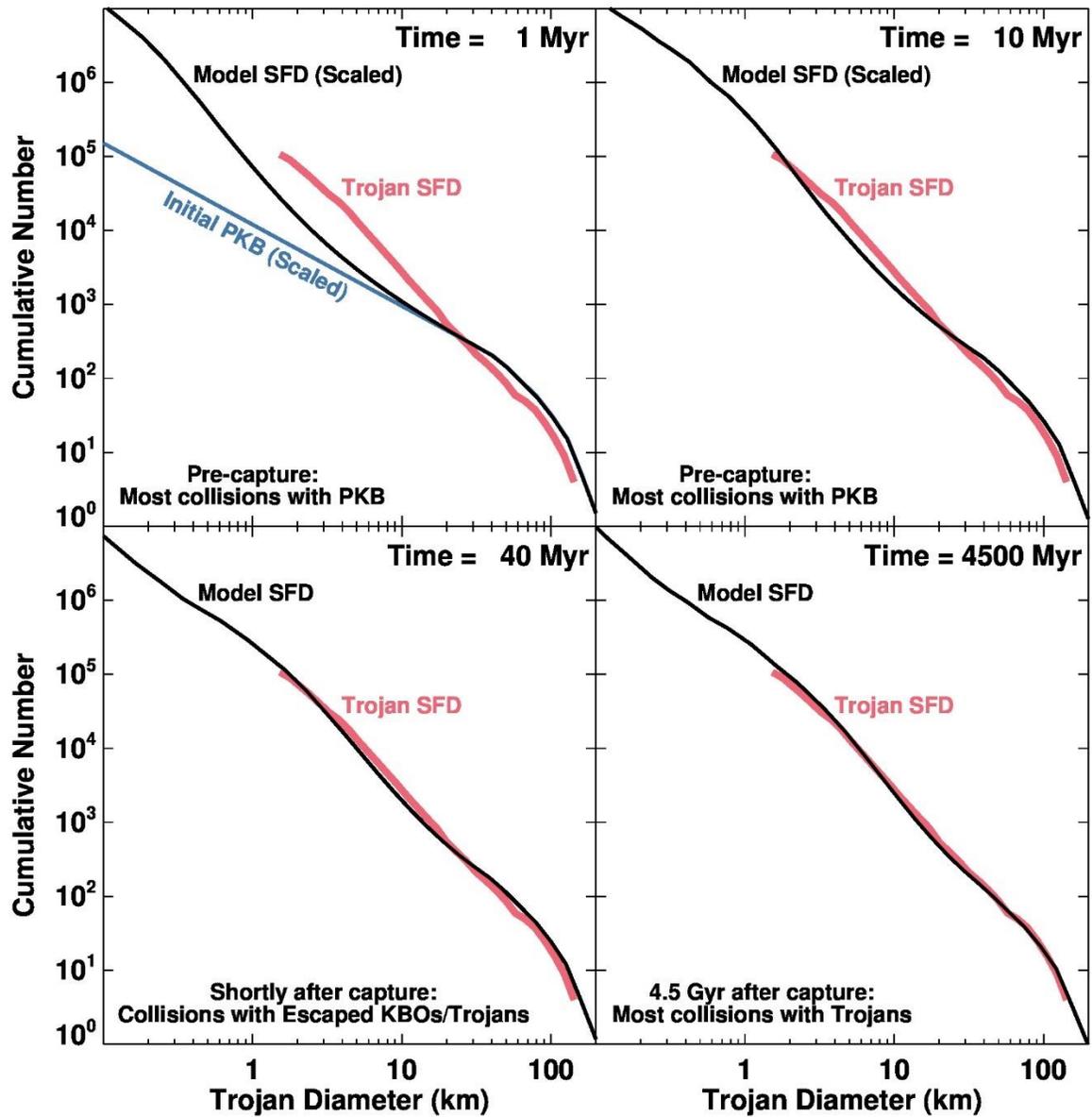

**Figure 11.** Four snapshots from the collisional evolution of the Jupiter Trojans, according to our best fit model SFD (Table 3, Run #1). Here Neptune enters the PKB at $\Delta t_0 = 20$ Myr, while it takes $\Delta t_1 = 10.5$ Myr for Neptune's migration to trigger the giant planet instability. The model SFD gradually goes from a wavy shape (timesteps 1, 10, 40 Myr) to a shape more akin to a power law between $5 < D < 100$ km at 4500 Myr. At early times, considerable collisional evolution takes place as the bodies move to 5 au, where they are destined to be captured in Jupiter's L4 and L5 locations. From there, they continue to be struck by the remnants of the destabilized population and other captured Trojans. The last timestep shows that the observed Trojan SFD (red line) is reasonably reproduced. Note that while our model shows the L4 and L5 populations together, it assumes they cannot strike one another.

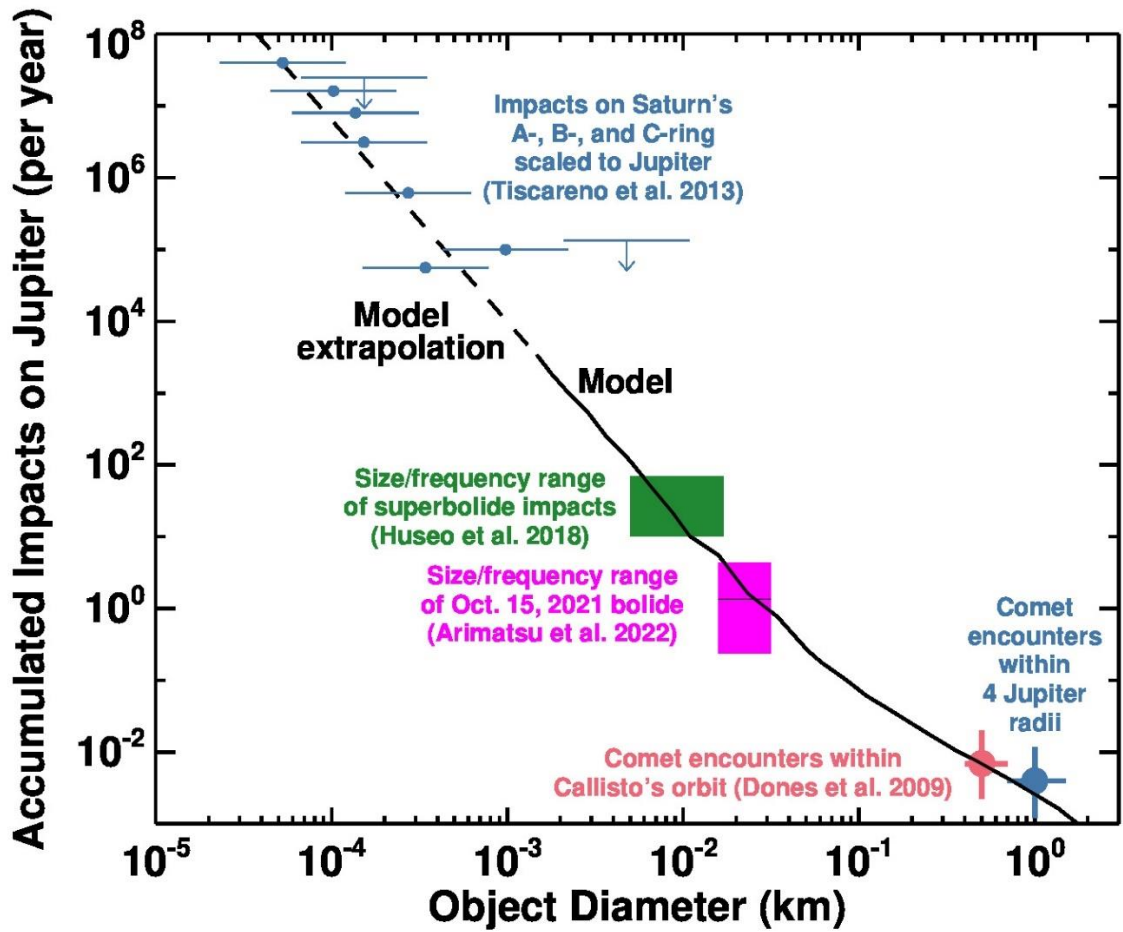

**Figure 12.** The model impact rate of comets on Jupiter. The model SFD from the last timestep in Fig. 10 (Table 3, Run #2) is combined with the fractional rate of impacts on Jupiter in the destabilized population and scattered disk (e.g., Fig. 8). The blue and red dots on the right shows the inferred impact rate deduced from comets making close encounters with Jupiter within 4 Jupiter radii and within Callisto's distance, respectively (Dones et al. 2009). The green and magenta squares show the impact flux of various superbolide sizes on Jupiter as determined by Huseo et al. (2018) and Arimatsu et al. (2022), respectively. The blue dots in the upper left show how the estimated impact flux on Saturn's rings from Tiscareno et al. (2013) should translate to Jupiter. The blue arrows represent upper limits. The projectiles smaller than a few tens of meters follow a Dohnanyi-like SFD with a cumulative power law slope of -2.7 (O'Brien and Greenberg 2003).

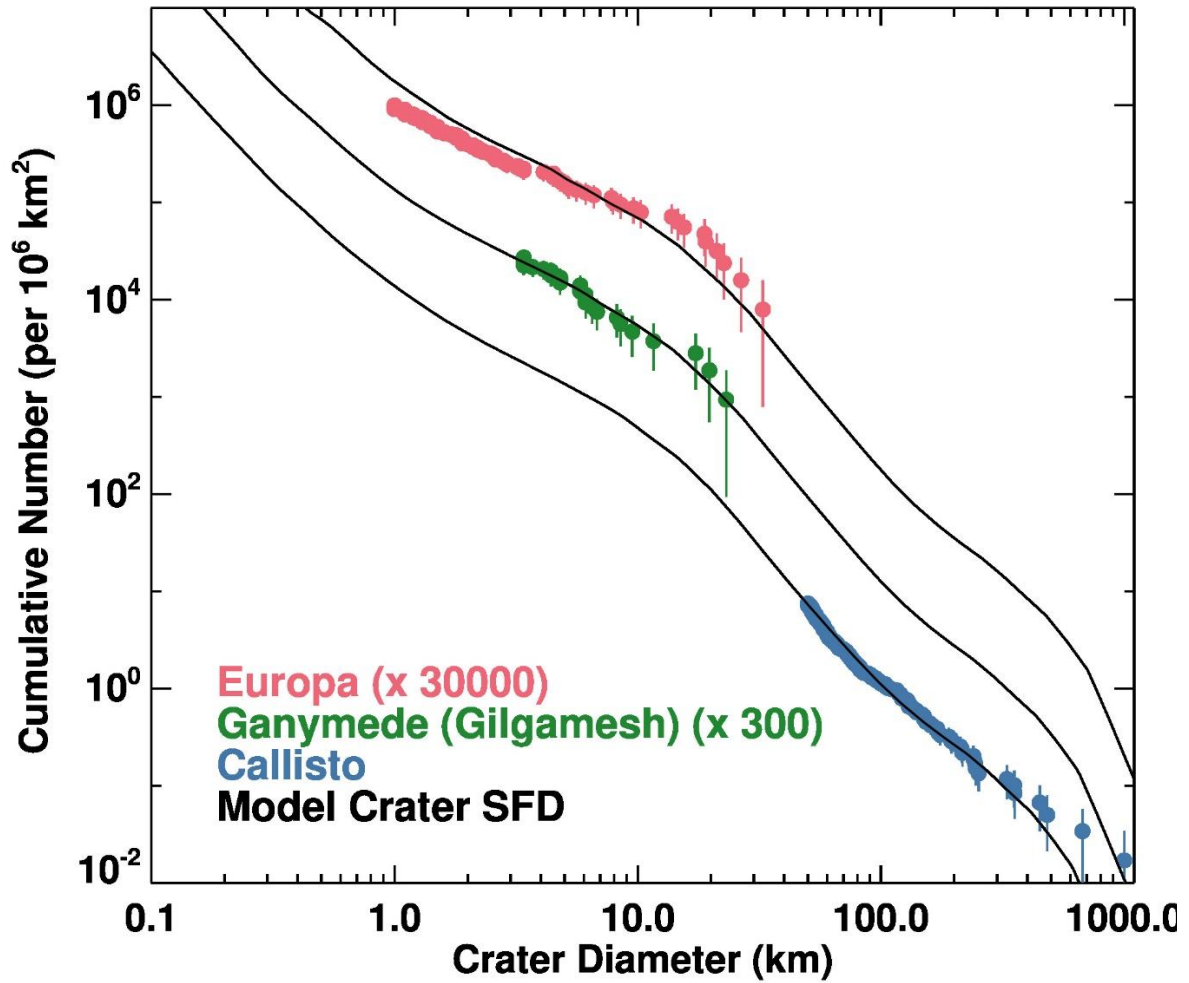

**Figure 13.** A comparison between the shapes of several observed crater SFDs found on Europa, Ganymede, and Callisto (Zahnle et al. 2003; Schenk et al. 2004) and versions of our model SFD from Table 3, Run #2 (see Fig. 10). The crater SFDs were multiplied by the factors in the legend so their shapes could be compared to the model without overlap. The Europa and Ganymede craters, the latter superposed on the Gilgamesh basin, are interpreted to be relatively young, while the Callisto craters are interpreted to be relatively old (see text). Accordingly, for Europa/Ganymede craters, we plot the model SFD from the last timestep in Fig. 10. For Callisto, to account for the fact that ancient surfaces are an integrated history of crater production, we integrated our model SFDs between 4500 Myr and 40 Myr once they had been multiplied by the impact rate curves from Fig. 8. The combined shape resembles the 40 Myr timestep in Fig. 10.

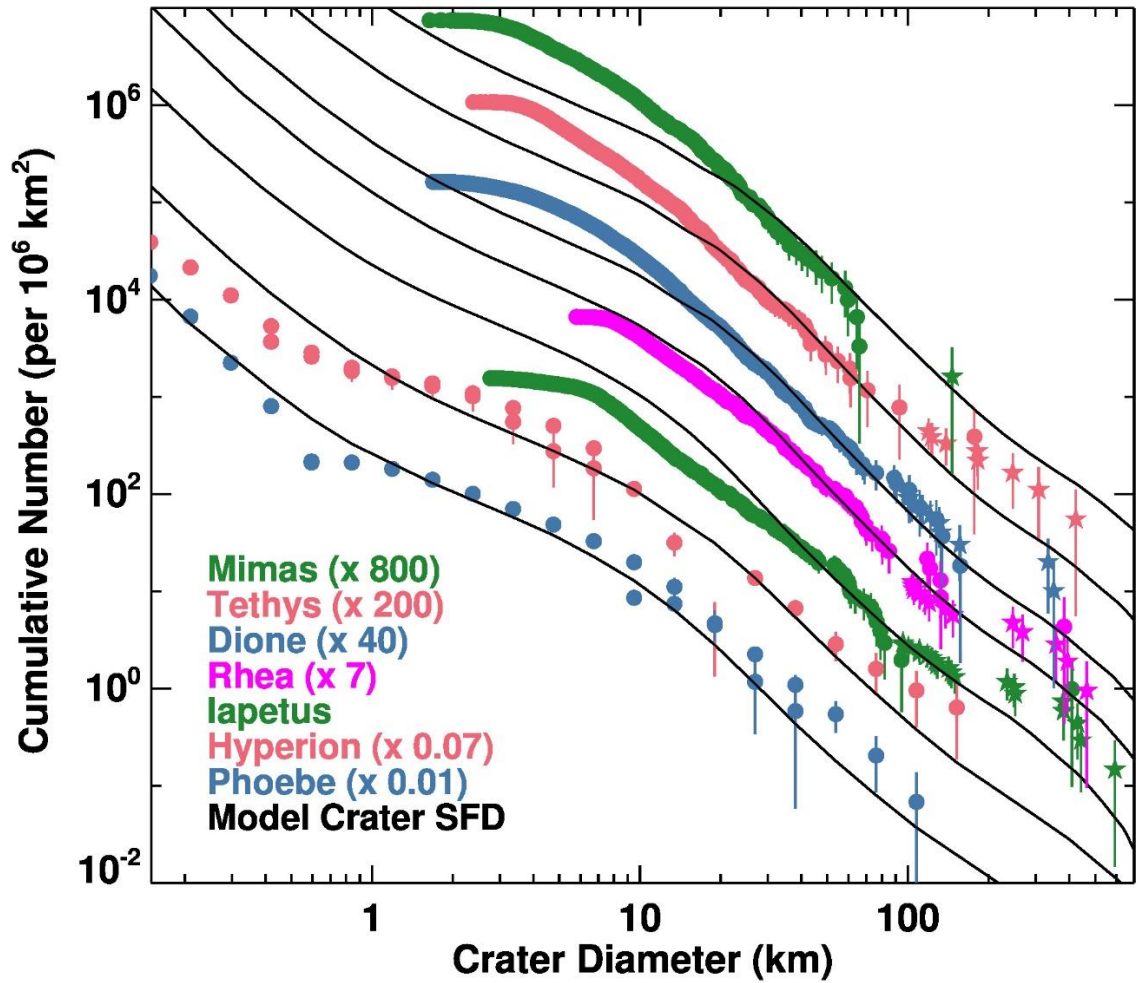

**Figure 14.** A comparison between the shapes of several observed crater SFDs found on the Saturnian satellites (Kirchoff and Schenk 2010; Thomas et al. 2013) and versions of our model SFD from Table 3, Run #2 (see Fig. 10). The crater SFDs were multiplied by arbitrary factors so their shapes could be easily compared to one another. The stars are the spatial densities of $D >$ 100 km basins found globally on a given satellite, while the points are craters derived from a selected terrain. The match is reasonable for all crater SFDs if we focus on $D_{\text{crat}} > 10\text{-}15$ km craters. For smaller craters, the match is more variable, possibly because secondary and sesquinary craters are dominating the shallow shape of the production SFD (Ferguson et al. 2022a,b).

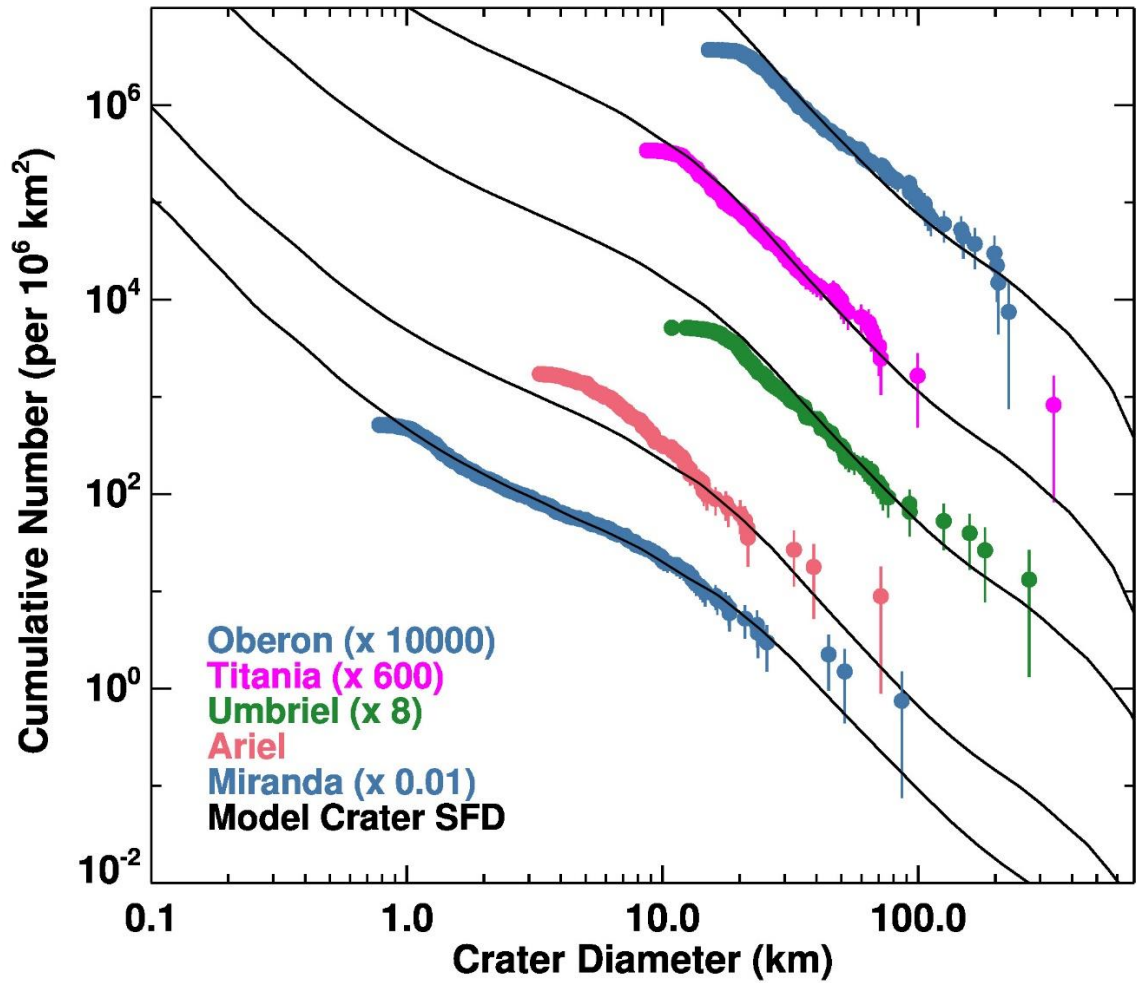

**Figure 15.** A comparison between the shapes of several observed crater SFDs found on the Uranian satellites (Kirchoff et al. 2022) and versions of our model SFD from Table 3, Run #2 (see Fig. 10). The crater SFDs were multiplied by arbitrary factors so their shapes could be easily compared to one another. For Miranda, we show the spatial densities of craters from the "Cratered D" terrain defined by Kirchoff et al. (2022). Overall, the shape of the model SFD is reproduced except for $D_{\mathrm{crat}} < 10$ km craters on Ariel. That mismatch may stem from secondary and/or sesquinary craters.

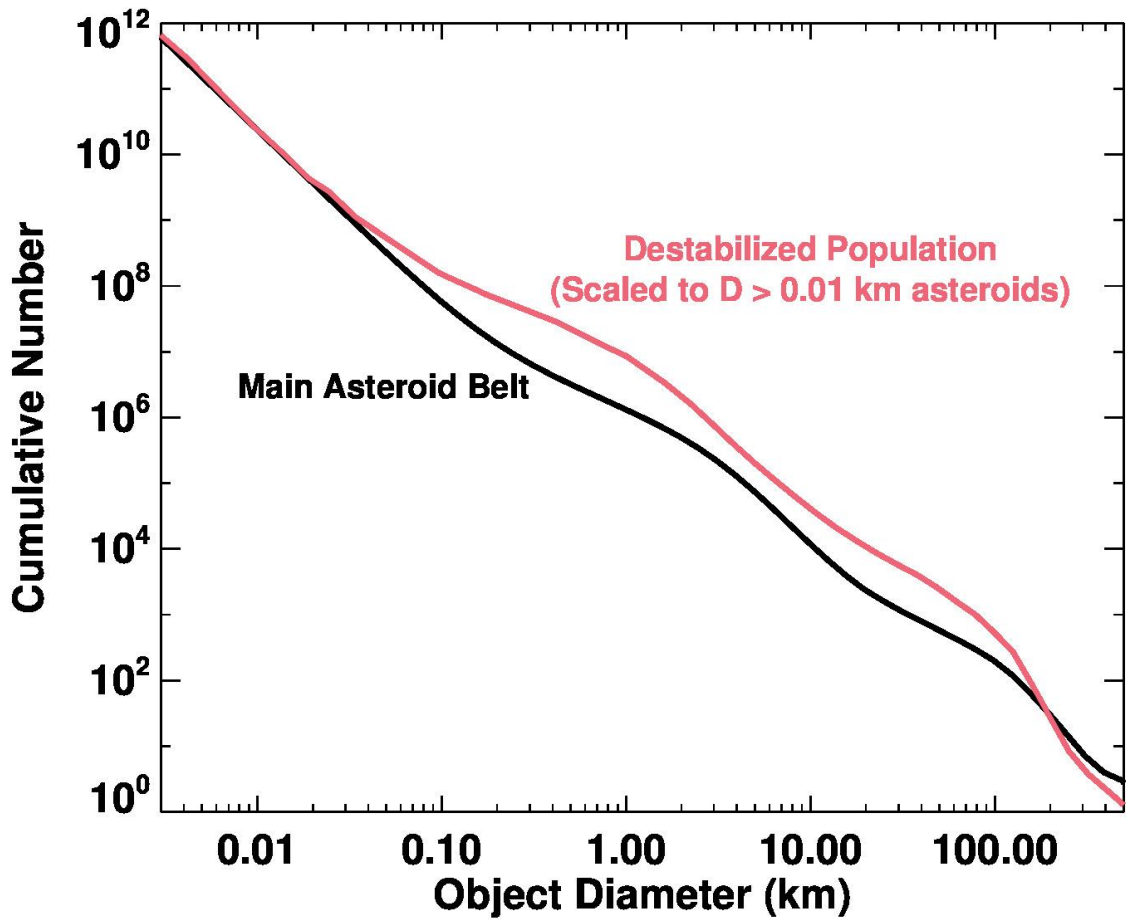

**Figure 16.** A comparison between the SFDs of the main asteroid belt from Bottke et al. (2020) (black curve) and destabilized population at 40 Myr shown in Fig. 10 (red curve). The destabilized population's SFD was normalized to match the number of $D > 0.01$ km bodies. Both have wavy shapes with a Dohnanyi-like SFD at small sizes that follows a cumulative power law slope of $q = -2.7$. Their differences stem from the fact that collisional evolution in the two populations follow different disruption functions, as shown in Fig. 9. The main belt has a bump at $D \sim 2$-$3$ km, while destabilized population has one at $D \sim 1$ km.

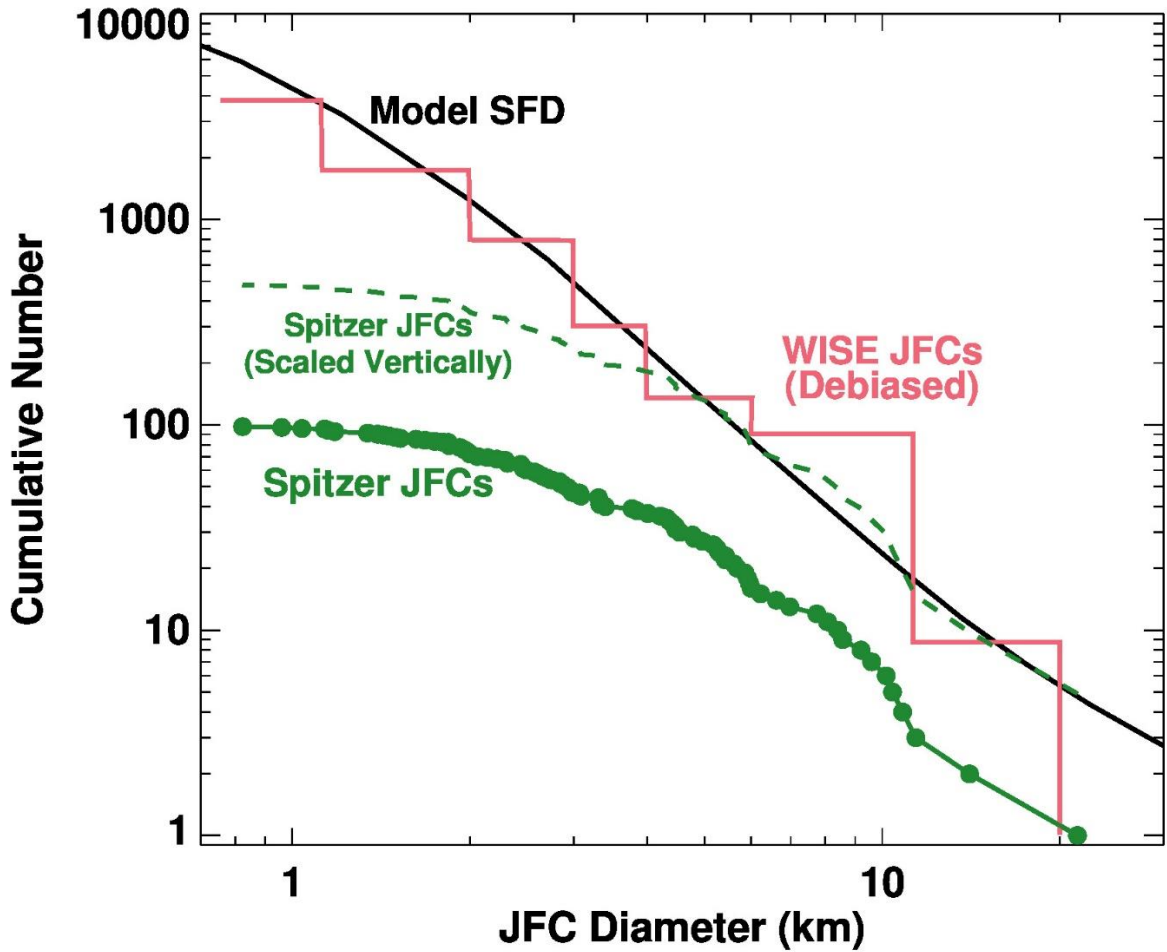

**Figure 17.** A comparison between the shapes of the deduced SFD for Jupiter family comets (JFCs) and our model SFD in Fig. 10 (Table 3, Run #2). The green dots are comet nuclei sizes deduced from a set of Spitzer observations (Lisse et al. 2020; see also Fernandez et al. 2013). The red histogram shows the debiased estimate of the JFC SFD calculated from WISE observations (Bauer et al. 2019). Our model SFD (black line) and the Spitzer SFD (dashed green line) have been vertically scaled to match the debiased WISE SFD at $D = 5$ km. The shape of our model SFD is reasonably consistent with that of the debiased JFC SFD from WISE for $D > 2$ km and with the shape of the JFC SFD from Spitzer for $D > 5$ km.

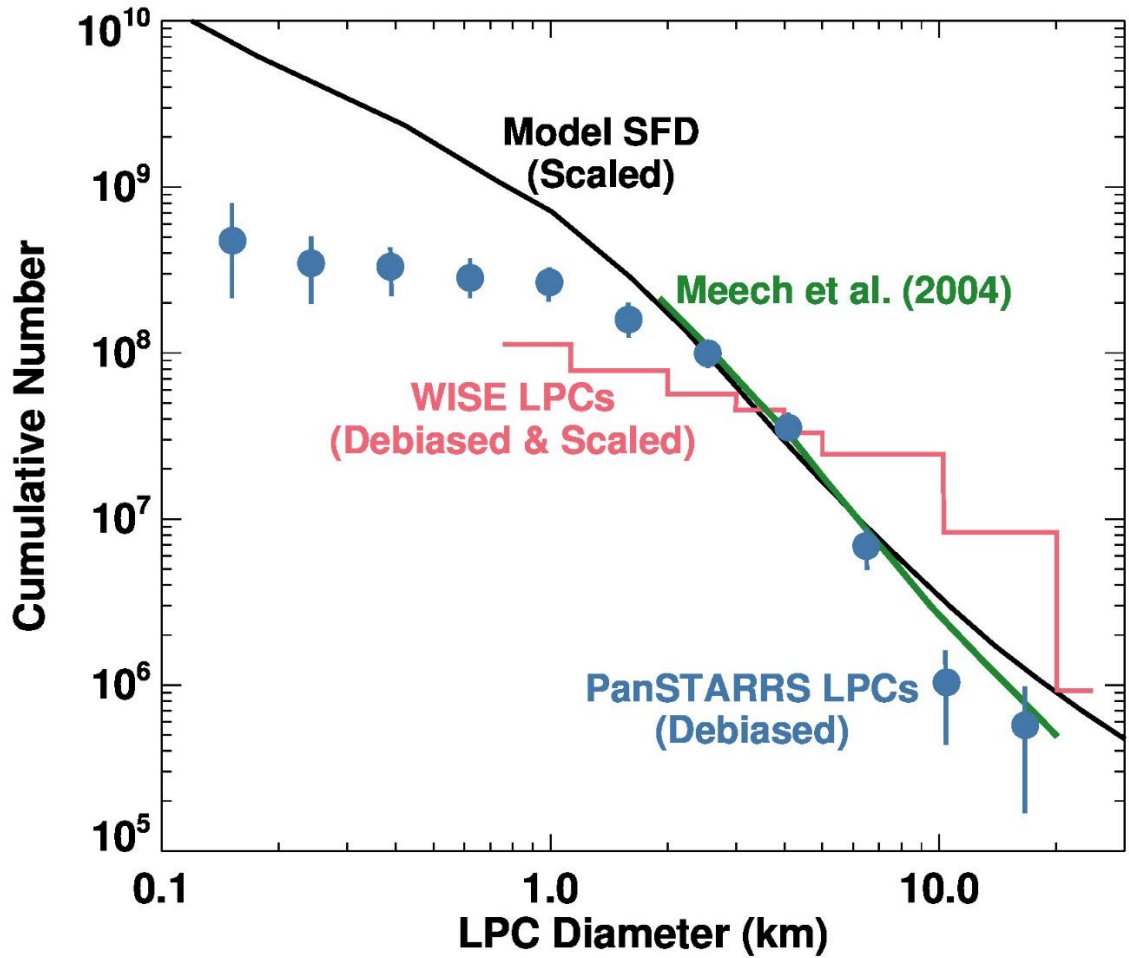

**Figure 18.** A comparison between the shapes of SFDs derived from long period comet (LPC) observations and our model SFD in Fig. 10 (Table 3, Run #2). Many LPCs were likely sent to the Oort cloud shortly after the giant planet instability, such that our model SFD was selected to be from an early timestep (i.e., the 40 Myr timestep from Fig. 10). The blue dots are the debiased SFD derived from PanSTARRS observations (Boe et al. 2019), while the red histogram is the debiased and scaled SFD from WISE observations (Bauer et al. 2019). Our model SFD is most consistent with the PanSTARRS SFD, but there is a mismatch for $D < 2$ km bodies. This difference may suggest that many small LPCs disrupt as they approach the Sun.

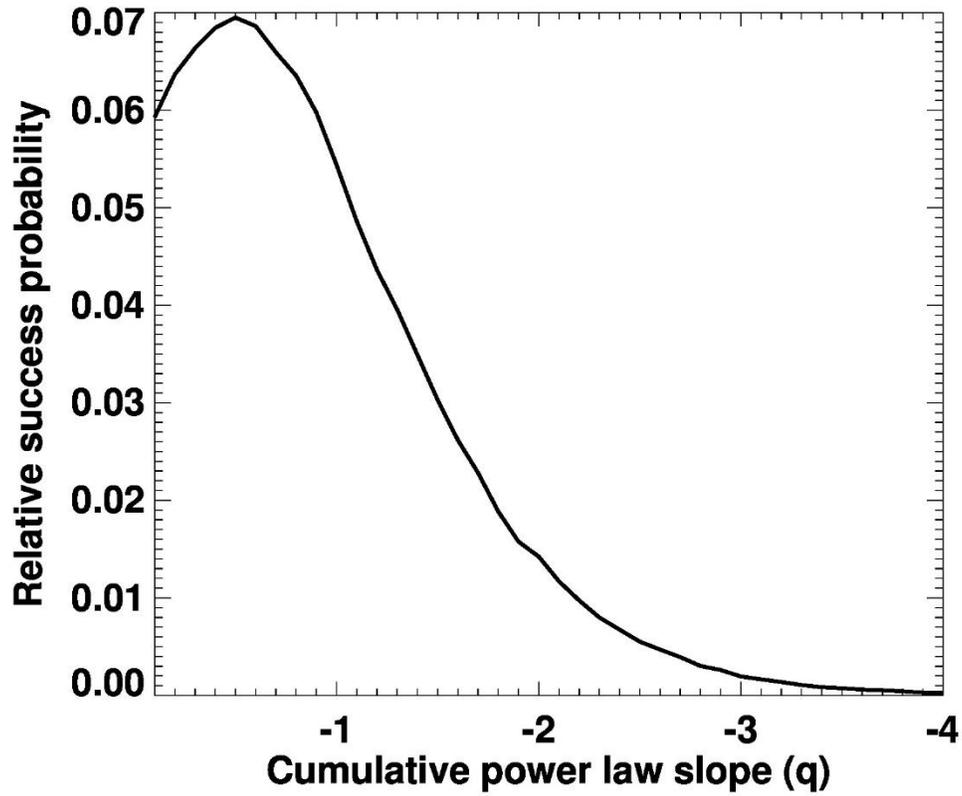

**Figure A1.** The relative success probability that two random draws from various power law SFDs will reproduce the sizes of interstellar comets 'Oumuamua and Borisov. We assumed the real size of 'Oumuamua was between 0.14 < *D* < 0.22 km, while the real size of Borisov was between 0.8 < *D* < 1.0 km. The cumulative power law slope of the comet SFD (*q*) is given on the y-axis. Note that this calculation does not account for the discovery circumstances of the two objects, and as such should be interpreted with caution.